\documentclass[aps,preprint,reprint,nofootinbib,superscriptaddress]{revtex4}

\usepackage{graphicx}
\usepackage{xcolor}
\usepackage{amssymb}
\usepackage{amsmath}
\usepackage{bm}
\usepackage{hyperref}
\usepackage{cleveref}

\usepackage{datetime}
\usepackage[T1]{fontenc}
\usepackage{ulem}
\usepackage{booktabs}

\makeatletter
\renewcommand\tableofcontents{%
    \begingroup
    {\Large \bfseries Contents\par}% Custom title
    \@starttoc{toc}% Start the table of contents
    \endgroup
}
\makeatother

\newcommand{\orcid}[1]{\,\href{https://orcid.org/#1}{\includegraphics[width=9pt]{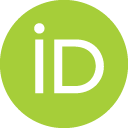}}\,}

\begin{document}

\title{
CJ26 Global QCD Analysis \\ with Large-$x$ Jefferson Lab 6 and 12 GeV Data
}

\author{Alberto Accardi
\orcid{0000-0002-2077-6557}}
\email{accardi@jlab.org}% \\ ORCID: 0000-0002-2077-6557}
\affiliation{Christopher Newport University, Newport News, Virginia 23606, USA}
\affiliation{Jefferson Lab, Newport News, Virginia 23606, USA}
\newcommand{\orcidAA}{0000-0002-2077-6557}

\author{Matteo Cerutti
\orcid{0000-0001-7238-5657}}
\email{matteo.cerutti@cea.fr}% \\ ORCID: 0000-0001-7238-5657}
\affiliation{Universit\'e Paris-Saclay – CEA – IRFU, 91191 Gif-sur-Yvette, France}
\newcommand{\orcidMC}{0000-0001-7238-5657}

\author{Cynthia E. Keppel
\orcid{0000-0002-7516-8292}}
\email{keppelc@gmail.com}
\affiliation{Jefferson Lab, Newport News, Virginia 23606, USA}
\affiliation{3q~Scientific~Consulting,~LLC, Newport News, Virginia 23602, USA}
\newcommand{\orcidCEK}{0000-0002-7516-8292}

\author{Shujie~Li\orcid{\orcidSL}}
\email{shujieli@lbl.gov}% \\ ORCID: 0000-0003-1252-5392}
\affiliation{Lawrence Berkeley National Laboratory, Berkeley, California 94720, USA}
\newcommand{\orcidSL}{0000-0003-1252-5392}

\author{J.~F.~Owens
\orcid{0000-0002-7351-0218\orcidJO}}
\email{owens@hep.fsu.edu}
\affiliation{Florida State University, Tallahassee, Florida 32306, USA}
\newcommand{\orcidJO}{0000-0002-7351-0218}

\author{Sanghwa Park
\orcid{0000-0002-8898-1231}}
\email{sanghwa@jlab.org}
\affiliation{Jefferson Lab, Newport News, Virginia 23606, USA}
\newcommand{\orcidSP}{0000-0002-8898-1231}

\author{Peter Risse
\orcid{0000-0002-8570-5506}}
\email{rissep@jlab.org}
\affiliation{Jefferson Lab, Newport News, Virginia 23606, USA}
\affiliation{Department of Physics, Southern Methodist University, Dallas, Texas 75275-0175, USA \\ 
        \vspace*{0.2cm}
        {\bf The CTEQ-JLab Collaboration}
        \vspace*{0.4cm} }
\newcommand{\orcidPR}{0000-0002-8570-5506}

\preprint{JLAB-THY-26-4768}
\preprint{SMU-PHY-25-07}

\begin{abstract}
We present CJ26, the new CTEQ-JLab global QCD analysis that incorporates for the first time the complete suite of JLab 6 GeV DIS measurements and the first published JLab 12 GeV measurements. Focused on the large-$x$ region, the analysis utilizes the increased $Q^2$ leverage of the 12 GeV data to robustly disentangle higher-twist effects from off-shell nucleon corrections. This leads to a highly accurate determination of the neutron-to-proton structure function ratio $n/p$ and the $d/u$ valence quark ratio, with uncertainties reduced by 30–50\% and 5–10\%, respectively. We highlight the critical role of experimental correlated systematic uncertainties in achieving this precision and provide the resulting NLO PDFs and structure functions in LHAPDF format for general use.
\end{abstract}

\maketitle
\tableofcontents
\newpage

\section{Introduction}

Parton Distribution Functions (PDFs) form the connection between observables in high energy collisions and theoretically calculable hard scattering quark and gluon processes. As such, PDFs can be constrained by fitting a wide variety of data from various types of high energy experiments. Ideally, such comparisons would yield knowledge of the different PDFs over the full range of momentum fractions $ 0\leq x \leq 1$ and values of the four-momentum $Q^2$ greater than some minimum value marking the boundary at which perturbative techniques may be applied. However, PDFs typically decrease as powers of $1-x$, leading to rather small values in the region as $x \rightarrow 1.$ Accordingly, one must employ observables that are sensitive to this large-$x$ region. One such class of observables includes high mass lepton pair or vector boson production measured at large values of rapidity. Such processes depend on products of PDFs depending on one small value of $x$ and one large value. Since PDFs at small values of $x$ are constrained by other processes such as deep inelastic scattering, the large rapidity measurements can provide constraints on the participating PDFs at large values of $x$. An alternative is to consider deep inelastic scattering measured at lower energies where the PDFs do not fall off as rapidly in the large $x$ region. High statistics data are available from the 6 GeV and 12 GeV runs at Jefferson Lab. One of the goals of the present analysis is to quantify the impact of these data sets on our knowledge of the large $x$ PDF behaviors.

The CTEQ-JLab (CJ) analysis program~\cite{Owens:2012bv,Accardi:2016qay,Accardi:2023gyr,Cerutti:2025yji} has as its goal the determination of PDFs over as wide a range of $x$ as possible. A wide variety of data are used, in particular including deep inelastic lepton proton and lepton deuteron scattering.
Utilizing data taken at lower energies means that the kinematic coverage includes regions where power suppressed $1/Q^2$ effects must be taken into account. Such contributions include target mass corrections (TMCs) and higher twist contributions (HT) from various nonperturbative and methodological effects. While the effects of the TMCs can be calculated~\cite{Georgi:1976ve,DeRujula:1976baf,Aivazis:1993kh,Kretzer:2002fr,Accardi:2008ne,Steffens:2012jx,Schienbein:2007gr,Brady:2011uy,Ruiz:2023ozv}, the HT contributions must at this point be parametrized~\cite{Virchaux:1991jc,Alekhin:2003qq,Blumlein:2006be,Alekhin:2008ua,Blumlein:2008kz,Accardi:2009br,Owens:2012bv,Blumlein:2012se,Accardi:2016qay,Accardi:2023gyr,Alekhin:2012ig,Cocuzza:2021rfn,Alekhin:2017kpj,Alekhin:2017fpf,Alekhin:2022tip,Alekhin:2022uwc}. In this analysis we present and discuss several different options for incorporating the effects of the HT contributions, following our previous work (see Ref.~\cite{Cerutti:2025yji}).

Deep inelastic scattering on a proton target at large values of $x$ is primarily sensitive to the $u$ quark PDF. In order to constrain the $d$ quark PDF it is often advantageous to use a deuteron target. It is therefore necessary to account for nuclear effects since the deuteron is a bound state of a proton and neutron~\cite{Kulagin:1989mu,Kulagin:1994fz,Kulagin:2004ie,Kulagin:2007ph,Kahn:2008nq,DelDotto:2016vkh,Fornetti:2023gvf}. The formalism used to treat this bound state in the CJ analysis will be discussed with particular attention to the correlations between the function accounting for the fact that the nucleons are off-shell and the functions parameterizing the HT contributions \cite{Cerutti:2025yji}.

The modeling of nuclear effects also influences the extraction of the $\bar d / \bar u$ sea quark ratio at moderate $x$ via proton-to-deuteron cross section ratio measurements in fixed target Drell-Yan measurements that correlate it to the $d/u$ PDF ratio at large $x$, see for example Ref.~\cite{Accardi:2023gyr}. In a global fit, it is then possible to benchmark the adopted nuclear modeling with data that is sensitive to either quark ratio but does not involve nuclear targets, for example measurements of the $W$ charge asymmetry in $p + \bar p$ collisions at the Fermilab's D\O\ experiment \cite{D0:2013lql} and measurements of the rapidity distribution of the $W^\pm$ decay lepton charge ratio from the STAR experiment at BNL's RHIC accelerator \cite{STAR:2020vuq}. On the one hand, these data show consistency with the adopted nuclear modeling and an asymmetric light quark sea with $\bar d / \bar u > 1$ \cite{Accardi:2023gyr,Cerutti:2024hrm} (see also Figures~\ref{f:Obs_Tevatron}, \ref{f:Obs_SQ-NS} and \ref{f:CJ26-extr_largex} in this paper).
On the other hand, the recent proton-target-only global analysis of Ref.~\cite{Ma:2025aga}, that profiles the CT18 PDFs using the forward-backward asymmetry in lepton-pair production at the Z boson pole measured at the LHC~\cite{Xie:2022tzo}, finds a $\bar d / \bar u$ ratio close to 1.
This is in tension with the extractions performed by the CJ and other collaborations \cite{Cocuzza:2021cbi,NNPDF:2021njg,Hou:2022ajg,Ablat:2024muy,Harland-Lang:2025wvm,Cocuzza:2026zoy}, including the CTEQ-TEA collaboration. In particular, they recently found that reweighting the newer proton-only CT25 PDFs with the STAR $W^+/W^-$ ratio~\cite{STAR:2020vuq} results in $\bar d / \bar u \approx 1.2$ at $x \approx 0.15$~\cite{Ponce-Chavez:2026lqh}. It will be important to understand the origin of this tension in future studies. 

Beyond hadron- and nuclear-structure-focused fits such as that presented here, higher-twist and nuclear corrections have recently received renewed attention in global QCD analyses that exploit high-precision data from the LHC and other high-energy facilities whilst re-examining lower kinematic cuts and the associated power corrections \cite{Ball:2025xtj,Harland-Lang:2025wvm}. While these effects are generally found to be small within the adopted kinematic cuts, they can significantly improve the perturbative convergence and stability of the fits and the precision that can be reached in the determination of Standard Model parameters such as $\alpha_S$.

Precise knowledge of PDFs at large $x$ is also essential for precision measurements of parity-violating processes at Jefferson Lab \cite{PVDIS:2014cmd,JeffersonLabSoLID:2022iod,Boughezal:2021kla,Bacchetta:2023hlw,Zaidi:2026rxo} and for reliable background estimates in searches for physics beyond the Standard Model (BSM).  
Indeed, the high invariant mass and large rapidity reach of LHC measurements probes momentum fractions $x\gtrsim 0.1$, so that uncertainties in the (anti-)quark and gluon sectors propagate directly, for example, into predictions for high-mass Drell-Yan (DY), dijet and $t\bar{t}$, and indirectly to smaller $x$ values through DGLAP evolution
\cite{Accardi:2015lcq,Accardi:2016ndt,Accardi:2021ysh,Ball:2022qtp,Ubiali:2024pyg,Brady:2011hb,Willis:2018yln,Fu:2025anw}.
A growing concern in recent studies is that, when high-mass LHC measurements are included in PDF fits under a Standard Model hypothesis, genuine signals of new physics can be partly absorbed into a sufficiently flexible large $x$ parametrization, blunting the sensitivity of indirect BSM searches \cite{Greljo:2021kvv,Hammou:2023heg,Hammou:2024xuj,Cole:2026eex}. This makes baseline PDF determinations that maximize their large-$x$ coverage without recourse to LHC data, like the one presented here, important not only for hadronic and nuclear structure studies but also for collider-driven physics analyses.

The outline of this paper is as follows. Section \ref{sec:formalism} contains a description of the formalism used while Section \ref{s:data} contains a description of the data employed. Results are presented in Section \ref{s:Results} while a discussion of the effects of systematic uncertainties are contained in Section \ref{sec:systematics}. Finally, in Section \ref{sec:Conclusions} the conclusions are summarized and information on how to access the PDF and structure functions in LHAPDF~\cite{Buckley:2014ana} format are provided. The appendices include tables of the best fit parameters, and a detailed discussion of the impact of different Jefferson Lab data subsets on the global fit. 

%%%%%%%%%%%%%%%%%%%%%%%%%%%%%%%%%%%%%%%%%%%%%%%%%%%%%%%
\section{Formalism}
\label{sec:formalism}

The \texttt{CJ26} global QCD analysis presented in this paper is based on the \texttt{CJ22} and \texttt{CJ22ht}~\cite{Accardi:2023gyr, Cerutti:2025yji} QCD setup, with a modest generalization of the PDF and higher twist parametrizations allowed by the substantially enlarged data set that now includes all available Jefferson Lab DIS data. We also adopt the \texttt{CJ22ht} implementation of nucleon motion and binding effects in the deuteron, treatment of Target Mass Corrections, and treatment of residual Higher Twist corrections. We will briefly review these elements here with a particular focus on their interplay. For more detail readers can consult the above mentioned references. The best fit parameters and their uncertainties are given in \Cref{tab:CJ26mHT_pdf_params,tab:CJ26mHT_ht_off_params,tab:CJ26aHT_pdf_params,tab:CJ26aHT_ht_off_params} in \Cref{app:parameter_tables}.

%%%%%%%%
\subsection{QCD setup and PDF parametrization}
\label{s:QCD_setup}

We calculate observables at next-to-leading order (NLO) perturbative accuracy, and for DIS observables we apply the ACOT-$\chi$ heavy-quark scheme~\cite{Kramer:2000hn}. Target mass corrections, nuclear corrections, and other power corrections will be discussed in detail in the following subsections.
For computational efficiency, we use the APPLgrid fast NLO interface~\cite{Carli:2010rw} to dynamically calculate the $W$ and $Z$ production cross sections during the fits. For more details, see Ref.~\cite{Accardi:2023gyr}.

For the fits, we use the same proton PDF parametrization considered in the \texttt{CJ22ht} release \cite{Cerutti:2025yji}, and obtain neutron PDFs from these by isospin symmetry. Namely, we adopt a standard five-parameter functional form at the initial scale $Q_0^2 = 1.69$ GeV$^2$ for most of the parton species
\begin{align}
   x \phi(x) = a_0 x^{a_1} (1-x)^{a_2} (1 + a_3\sqrt x + a_4 x)
\end{align}
with $\phi=u_\texttt{v}, \, \bar d + \bar u, \, \bar d - \bar u, \, g$. For ease of notation, we suppress the dependence of PDFs and observables on their parent hadron and only reintroduce them when necessary for the context.

Since the global dataset included in our analyses imposes only weak constraints on the strange and heavier quarks -- we do not use, for example, dimuon production in $\nu+A$ collision to avoid contamination from final state effects \cite{Accardi:2015lcq,Accardi:2009qv} --  we assume $s=\bar s =\kappa(\bar d + \bar u)/2$ with $\kappa=0.4$ at the initial $Q_0^2$ scale. As $Q^2$ increases, perturbative QCD evolution drives $\kappa$ towards 1 at small $x \lesssim 0.01$  values, qualitatively consistent with $W$ and $Z$ measurements at the LHC \cite{CMS:2011scv, CMS:2012mpr, CMS:2013mlg, ATLAS:2016fzp, CMS:2016iof, ATLAS:2012mjs} and with a suppressed strangeness at larger $x$ and lower $Q^2$ from HERMES \cite{HERMES:2008vug, HERMES:2013xef}.
We leave to future analyses a more flexible determination of the strange quark distribution, which would require the inclusion of additional data directly sensitive to the $s$ quark but free from nuclear modifications, such as $W$+charm production in $pp$ collisions at the LHC \cite{Anderson:2024evk}.
The charm and bottom quarks are generated perturbatively, with masses $m_c = 1.3$ GeV and $m_b=4.5$ GeV, respectively. We also fix $\alpha_s(M_Z)=0.118$ at the reference scale $M_Z=91.118$ GeV. 

In order to allow the $d/u$ PDF ratio to have a finite limit as $x$ approaches 1, we mix the $u_\texttt{v}$ distribution into the $d_v$ parametrization at the initial scale:
\begin{equation}
    \label{e:PDF_du}
xd_\texttt{v}^{\,\text{CJ}} (x, Q_0^2) = a_0^{d_\texttt{v}} \bigg ( x^{a^{d_\texttt{v}}_1} (1-x)^{a^{d_\texttt{v}}_2} (1 + a^{d_\texttt{v}}_3\sqrt x + a^{d_\texttt{v}}_4 x) + b x^c xu_\texttt{v} (x, Q_0^2) \bigg ).
\end{equation}
The $d$-quark normalization $a_0^{d_\texttt{v}}$ is determined by the valence sum rule, and $b$ and $c$ are free parameters. 
In previous CJ fits, we had fixed $c = 2$ due to the lack of strong constraints from the available experimental data, and $a_3^{u_\texttt{v}}=0$
to improve the stability of the fits. With the enlarged data set considered for the \texttt{CJ26} fit presented in this paper, the constraint on $a_3^{u_\texttt{v}}$ is removed. Consequently, we re-evaluate $c$ as a free parameter, and freeze it to its best-fit value to stabilize the uncertainty analysis.

We observed that the parametrization of $\bar d - \bar u$ exhibits strong correlations between the normalization parameter $a_0^{\bar d - \bar u}$ and the shape parameters $a_1$ and $a_2$. With the relatively weakly constraining lepton pair production data set available so far, this produces flat directions in the fit's $\chi^2$ profile which render the PDF uncertainty analysis unstable. 
To reduce this instability, for $\bar d - \bar u$ only we adopt a reparameterization in which the shape is normalized independently of $a_0$, which is then treated as an overall multiplicative factor.  In practice, we introduce a normalized form
\begin{align}
    x\tilde{\phi}^{\bar d - \bar u}(x; a_1,a_2,a_3,a_4) =
    \frac{x^{a_1} (1-x)^{a_2} \left(1 + a_3 \sqrt{x} + a_4 x \right)}
    {\mathcal{N}(a_1,a_2,a_3,a_4)},
\end{align}
which is analytically normalized to 1 by
\begin{align}
\mathcal{N}(a_1,a_2,a_3,a_4) &=
B(a_1+1, a_2+1)
+ a_3 \, B(a_1+\tfrac{3}{2}, a_2+1)
+ a_4 \, B(a_1+2, a_2+1),
\end{align}
with $B(p,q)$ the Euler Beta function. The final parametrization for $\bar d - \bar u$ is then written as
\begin{align}
    {\phi}^{\bar d - \bar u}(x) = a_0 \, \tilde{\phi}^{\bar d - \bar u}(x; a_1,a_2,a_3,a_4),
\end{align}
so that $a_0$ controls the overall normalization while $(a_1,a_2,a_3,a_4)$ determine the shape. This construction significantly reduces correlations between normalization and shape parameters and improves the stability of the fit. We furthermore fixed $a_4=0$ to avoid flat directions in the fit's $\chi^2$ profile.

%%%%%%%%
\subsection{Target mass corrections}

Target mass corrections (TMCs) are kinematic power corrections that scale as $M^2/Q^2$ with $M$ the nucleon's mass \cite{Schienbein:2007gr} and can be analytically calculated although with some degree of theoretical uncertainty. Trivial kinematic factors apart, the main effect is to substitute the Bjorken $x$ scaling variable with Nachtmann's $\xi = 2x/ (1+\gamma)$ in the evaluation of the DIS structure functions. For ease of notation, the mass corrections are bundled in the $\gamma=\sqrt{1+4x^2 M^2/Q^2}$ factor, with $\gamma \to 1$ corresponding to the massless limit $M^2/Q^2 \to 0$. Beyond this, TMC prescriptions vary, in particular regarding how to treat the violation of the $x \leq 1$ threshold engendered by the $\xi$ rescaling \cite{Brady:2011uy}.  For speed of calculation, in this paper we adopt the approximation of the Georgi-Politzer momentum space formula proposed by Schienbein et al. \cite{Schienbein:2007gr},
updated to reproduce the target mass corrections calculated with the CJ15 PDFs, 
namely,
\begin{align}
    F_2^{TMC}(x,Q^2;\gamma) = \frac{(1+\gamma)^2}{4\gamma^3} F_2^{(0)}(\xi,Q^2) 
    \left( 1 + 6 \frac{M^2}{Q^2} \frac{x\xi}{\gamma}(1-\xi)^2 \right)
\end{align}
with $F_2^{(0)} = C \otimes \phi\,|_x = \sum_\phi \int dz\, C_\phi(z)\, \phi(x/z)$ the nucleon-massless, factorized leading twist structure function calculated in perturbative QCD as a convolution of perturbative coefficients $C_\phi$ and PDFs $\phi$.  
In the DIS region, the differences with respect to the full Georgi-Politzer formula are modest and can be reabsorbed in the fitted phenomenological higher-twist term to be discussed below.

%%%%%%%%
\subsection{Residual power corrections} 
\label{ss:residual_power_corrections}

Other $1/Q^2$ corrections to the logarithmic evolution of the DIS structure functions dictated by the DGLAP evolution can have different origins. Here we distinguish \textit{dynamical power corrections}, that originate from parton rescattering on the target, and \textit{residual power corrections}.
The dynamical corrections can be calculated and are sensitive to twist-4 nonperturbative multiparton correlation functions~\cite{Jaffe:1982pm,Ellis:1982wd,Ellis:1982cd,Qiu:1988dn}. 
The residual power corrections are all remaining $1/Q^2$ corrections and originate, among other mechanisms, from phase space limitation near the $x=1$ threshold~\cite{Corcella:2005us,Accardi:2008ne,Accardi:2014qda,Bonvini:2015ira,Simonelli:2025xpm}, choice of TMC prescription, approximations in the off-shell mass corrections in nuclear targets, and other mechanisms (see Ref.~\cite{Cerutti:2025yji} for a detailed discussion).

The combination of all these sources is commonly referred to as ``higher-twist'' (HT) corrections, even though they do not only arise from twist-4 hadron matrix elements.
Including suitable, fitted power correction terms in a global QCD fit allows one to reliably extract the leading-twist dynamics (PDFs and offshell function) of the nucleon, provided the interplay and correlations of power corrections, nuclear corrections and leading-twist functions is properly analyzed and quantified as detailed in Ref.~\cite{Cerutti:2025yji} and discussed further in the next subsection. Notably, higher-order corrections to DIS calculations can also be in large part absorbed in a fitted higher twist term  leading to perturbative stability of the DIS data description \cite{Blumlein:2006be,Blumlein:2008kz,Harland-Lang:2025wvm}.

Following the \texttt{CJ22ht} analysis \cite{Cerutti:2025yji} we therefore parametrize the dynamical and residual power corrections by means of either a \textit{multiplicative} or an \textit{additive} modification of the target-mass-corrected (TMC) structure functions, 
\begin{align}
    F_{2,N}^\text{mult}(x,Q^2;\gamma) &= F_{2,N}^\text{TMC}(x,Q^2;\gamma) \bigg ( 1 + \frac{C_N(x)}{Q^2} \bigg ) \ , 
    \label{e:ht_mult}
    \\
      F_{2,N}^\text{add}(x,Q^2;\gamma) &= F_{2,N}^\text{TMC}(x,Q^2;\gamma) + \frac{H_N(x)}{Q^2} \ ,
    \label{e:ht_add}
\end{align}
and fit the $x$-dependent ``higher-twist'' functions $C_N(x)$ or $H_N(x)$ individually for each nucleon $N=p,n$. For this, we adopt a standard parametrizations \cite{Virchaux:1991jc,Blumlein:2012se,Alekhin:2012ig,Cerutti:2025yji}, namely the same parametrization as in \texttt{CJ22ht}, given by
\begin{align}
    C_N(x) = \alpha_N x^{\beta_N} (1 + \gamma_N\sqrt{x}+ \delta_N x)
    \label{e:ht_mult_par} \ ,\,\\
    H_N(x) = \alpha_N x^{\beta_N} (1-x)^{\gamma_N} (1 + \delta_N x) \ .
    \label{e:ht_add_par}
\end{align}

A priori there is no strong reason for choosing either implementation.
However, it is important to note that any given choice implies specific assumptions on the $Q^2$ scaling and isospin behavior of the HT functions. For instance, one can rewrite the multiplicative implementation~\eqref{e:ht_mult} in an additive way,
\begin{equation}
    F_{2,N}^\text{mult}(x,Q^2) = F_{2,N}^\text{TMC}(x,Q^2) + \frac{\tilde{H}_N(x,Q^2)}{Q^2} \ .
    \label{e:ht_tilde}
\end{equation}
However, the resulting additive $\tilde H$ coefficient inherits both isospin dependence and $Q^2$ evolution from the $F_2$ structure function, namely, 
\begin{equation}
    \tilde{H}_N(x,Q^2) = F_{2,N}^\text{TMC}(x,Q^2) C_N(x) \ ,
\label{e:Htilde}
\end{equation}
even if one decided to choose an isospin- and $Q^2$-independent $H$ function in Eq.~\eqref{e:ht_add}. Likewise, rewriting the additive implementation in a multiplicative way, the equivalent multiplicative factor $\tilde{C}_N = H_N(x)/F_{2,N}^{TMC}(x,Q^2)$ would acquire isospin and scale dependence from the structure function in the denominator at the right hand side independently of the approximations adopted in the additive implementation. 

To minimize model assumptions and biases, we therefore independently fit the HT coefficients for the proton and the neutron in either multiplicative and additive HT implementation, and interpret the difference between the two implementations as an estimate of the related theoretical systematic uncertainty \cite{Cerutti:2025yji}. In practice, we use the envelope of the additive and multiplicative fits to quote the extracted functions and the calculations performed using these. Finally, note that to avoid flat directions in the fit's $\chi^2$ profile we fixed $\gamma_n = 0$ in the multiplicative implementation of the neutron's higher-twist corrections. 

We emphasize that the isospin dependence of the HT terms obtained in the fit should not be interpreted as a direct determination of isospin symmetry violation in genuine twist-4 contributions, since the fitted terms effectively capture a variety of power-suppressed corrections. A simple example are truncations or neglected terms in the adopted HMC scheme, that operates differently on proton and neutron structure functions due to their different slope in $x$. A less obvious example is the modeling of mass corrections for bound nucleons, which is necessary to utilize deuteron DIS data for up and down quark separation and to which we turn next.

%%%%%%
\subsection{Nuclear binding and off-shell corrections} 

The partonic structure of a bound nucleon depends not only on $x$ and $Q^2$ but also on the nucleon's four-momentum squared $p^2$ which is in general smaller than its on-shell mass $M^2$. In the weakly-bound deuteron, one can Taylor-expand the bound nucleon structure around $p^2 \simeq M^2$ (for more details, see Refs~\cite{Kulagin:1994cj,Kulagin:1994fz}). Following the well established CJ and JAM methodology \cite{Owens:2012bv,Accardi:2016qay,Cerutti:2025yji,Cocuzza:2021rfn}, we perform this expansion at the PDF level and approximate the $\phi^*$ bound nucleon PDFs $\phi^*$ as
\begin{equation}
\phi^*(x,Q^2,p^2) = \phi(x,Q^2) \bigg ( 1 + \frac{p^2 - M^2}{M^2} \delta f (x) \bigg ) \, ,
\label{e:off_pdf}
\end{equation}
where $\phi$ is the free-nucleon PDFs, and $\delta f$ is an ``off-shell function'' that will be fitted to experimental data and quantifies the deformation of a parton belonging to bound nucleon. Since $\delta f$ is defined as a ratio of PDFs, we assume that the $Q^2$-evolution at the numerator compensates that at the denominator, and we only parametrize the $x$-dependence. Also, we utilize a flavor-independent $\delta f$ because the vast majority of nuclear target data included in our analysis consists of inclusive electron-deuteron scattering, which do not provide information about the difference between $u$- and $d$-quark deformations\footnote{In fact, we also include BONuS tagged DIS measurements but these are not precise enough to require separating the up and down quark off-shell deformations.}.

For the offshell deformation $\delta f$, we adopt a generic polynomial parametrization,
\begin{equation}
    \delta f (x) = \sum_{n=0}^{k} a_{\text{off}}^{(n)} x^n \, .
    \label{e:off_poly}
\end{equation}
With the current experimental data sets we obtain satisfactory fits with a second order polynomial as in the \texttt{CJ22ht} fit \cite{Cerutti:2025yji}. However, in the \text{CJ26} fits presented in this paper we use a third order polynomial, for a total of 4 free parameters, to provide a more conservative uncertainty estimate. We analyze the sensitivity of the fit to the chosen polynomial order in Section~\ref{ss:syst-corr}, which will also allow us to identify $x\lesssim 0.7$ as the momentum fraction region in which the \texttt{CJ26} dataset constrains $\delta f$.

We then evaluate the deuteron DIS structure function $F_{2,D}$ in the nuclear impulse approximation \cite{Kulagin:1989mu,Kulagin:1994fz,Kulagin:2004ie,Kulagin:2007ph,Kahn:2008nq,DelDotto:2016vkh,Fornetti:2023gvf}, under the assumption that the nucleon is weakly bound. In this approximation, the exchanged photon scatters off a bound, off-shell nucleon inside the deuteron and $F_{2,D}$ is given by
\begin{equation}
    F_{2,D} (x, Q^2) = \sum_{N=p,n} \int d y \,d p_T^2 \, f_{N/D}(y, p_T^2; \gamma) \, F^*_{2,N} \bigg (\frac{x}{y}, Q^2, p^2 ; \gamma^* \bigg ),
\label{e:F2D}
\end{equation}
where $x=M_D \, Q^2 / (M \, P_D\cdot q)$ is the \textit{per-nucleon} Bjorken invariant, with $q$ and $P_D$ being the photon and deuteron momenta, and $M$ and $M_D$ the nucleon and deuteron masses, respectively. $F_{2,N}^*$ is the target-mass- and power-corrected structure function of an off-shell nucleon $N$ (a proton $p$ or a neutron $n$) with momentum $p$ and Bjorken invariant $x/y = Q^2/(p\cdot q)$, and $y=(p\cdot q)/ (p_D \cdot q)$ is the bound nucleon momentum fraction with respect to the deuteron. The bound nucleon structure function is calculated as in Eq.~\eqref{e:ht_mult} or \eqref{e:ht_add} but substituting the onshell PDF $\phi$ with its offshell counterpart $\phi^*$. The deuteron smearing function $f_{N/D}$ can be interpreted as the probability of finding a nucleon of momentum fraction $y$ inside a deuteron $D$, and can be calculated using non-relativistic nuclear wave functions. For the deuteron, $f_{p/D}=f_{n/D}$ by energy momentum conservation. We however keep the subscript $N$ explicit in view of generalizations to heavier nuclei.

Note that the deuteron smearing function $f_{N/D}$ depends on the TMC parameter $\gamma$ because the deuteron is on shell, but the bound nucleon structure functions depends on $\gamma^*=\sqrt{1+4x^2p^2/Q^2}$ because the nucleon is off shell. In a lightly bound nucleon such as the deuteron, the nucleon's off-shellness $v=(p^2-M^2)/M^2$ is small, and $\gamma^*$ differs from the on-shell $\gamma$ by terms of order $O(v M^2/Q^2)$ that are further suppressed as $Q^2$ increases. 

The transverse momentum in Eq.~\eqref{e:F2D} can be explicitly integrated away by utilizing the offshell expansion \eqref{e:off_pdf} and furthermore approximating $\bar\gamma^* \approx \gamma$. The result is a 1D convolution formula that highlights in $\delta F_{2,D}$ the effect of the offshell PDF deformation \eqref{e:off_pdf}, and separates it from the dominant onshell contribution: 
\begin{align}
    F_{2,D}(x,Q^2) = \sum_{N=p,n} \int dy\, \mathcal{S}_N(y;\gamma) F_{2,N}\big({\textstyle \frac{x}{y}},Q^2;\gamma \big)
        + \sum_{N=p,n} \int dy\, \mathcal{S}_N^{(1)}(y;\gamma) \otimes \delta F_{2,N}\big({\textstyle \frac{x}{y}},Q^2;\gamma \big) \ ,
\label{eq:smearing_off}
\end{align}
where
$
    \mathcal{S}_N(y,\gamma) \equiv \int dp_T^2 \, f_{N/D}(y_D,p_T^2,\gamma) 
$
and
$
    \mathcal{S}_N^{(1)}(y;\gamma) \equiv \int dp_T^2 \, \frac{p^2-M^2}{M^2} \, f_{N/D}(y,p_T^2;\gamma) 
$
are $y$-dependent spectral functions, and the $\delta F_{2,N}$ off-shell contribution to the nucleon structure function is calculated as $F_2$ but with the $\phi \to \phi\delta f$ substitution, i.e. we multiply the PDFs with the offshell function (see Eq.~\ref{e:off_pdf}). With $\gamma^* = \sqrt{1+4x^2p^2/Q^2} = \gamma + O(v M^2/Q^2)$ the corrections to the on-shell approximation for the TMC parameter are small in lightly bound nuclei and can be effectively absorbed in the fitted HT terms discussed above.

%%%%%%%%%%
\subsection{Determination of fit uncertainties}
\label{sec:fit_uncertainties}

The fit uncertainties are determined with the local tolerance variant of the Hessian method~\cite{Pumplin:2001ct,Pumplin:2002vw} implemented in the \texttt{CJ22} fits~\cite{Accardi:2023gyr,Cerutti:2024hrm}. Before discussing this, we note that the $\chi^2$ function adopted in the CJ analysis and detailed in Ref.~\cite{Li:2023yda} is not quadratic in the experimental normalization factors, and cannot be analytically minimized as one can do with the other systematic errors~\cite{Stump:2001gu}.
Instead, we determine the covariance matrix of the physical parameters of the fit through a 2-step procedure. 
In the first step, the experimental normalization factors are treated as free parameters and fitted simultaneously with the physical parameters, as in previous CJ analyses. 
In the second step, the normalization factors are fixed to their best-fit values, and the physical parameters are refitted. As a consequence, the correlations between normalization and physical parameters are effectively incorporated into the covariance matrix of the latter, producing a more stable determination of uncertainties especially for observables that include non perturbative contributions beyond those included in the PDFs, namely, higher-twist and off-shell corrections. In the limit of a quadratic dependence of the $\chi^2$ on the normalization parameters, this approach reproduces the result of the standard analytic treatment. In a future dedicated paper we will explore this connection more broadly, including different error propagation methodologies like non-parametric bootstrap.

Finally, following \cite{Accardi:2023gyr,Hunt-Smith:2022ugn}, we diagonalize the Hessian matrix obtained in the second step of our fitting procedure and scan the $\chi^2$ function along each of its eigenvectors to determine in a dynamical way the parameters that produce an increase $\Delta\chi^2 = T^2$ above the minimum $\chi^2$ determined by the fit.
We then determine the uncertainties on calculated observables using these parameters, see \cite{Accardi:2023gyr} for details. Numerical results in this paper and in the accompanying LHAPDF grids will be presented with a tolerance factor $T^2=2.71$. Our choice of a tolerance value larger than 1 is partly motivated by the need of accounting for tensions in the global dataset, which in themselves spoil a straightforward sigma-level interpretation. More importantly, it allows us to better detect and adjust for deviations of the of the parameter's likelihood from a Gaussian distribution, especially in regions that are weekly constrained by the data \cite{Accardi:2023gyr,Hunt-Smith:2022ugn}. 
An interesting recent discussion of tolerance in QCD global analyses and a more comprehensive set of references can be found in \cite{Harland-Lang:2024kvt} and a comparative study of PDFs at equal $\Delta \chi^2$ value in \cite{Accardi:2021ysh}. 

%%%%%%
\subsection{Interplay of corrections and theoretical systematics}
\label{ss:Formalism-interplay}

We stress that the evolution prescription implicit in the choice of an additive or a multiplicative implementation of power corrections has consequence well beyond the determination of the HT correction terms themselves.

As a first example, we note that, according to the DGLAP equation, the $Q^2$ evolution of the leading twist (LT) $F_2$ structure functions is determined to a large extent by the the gluon PDF. Hence, the assumed $Q^2$ dependence in the HT terms can be in principle compensated by a change of the LT gluon distribution. This is particularly true at small $Q^2$ since the DGLAP evolution resemble a power law at small $Q^2$, and at large $x$ where the gluon is minimally constrained by the available data. 
We can therefore expect a correlation between the choice of HT implementation and the fitted gluon distribution at large $x$. 

As a second example \cite{Cerutti:2025yji}, with the relatively limited $Q^2$ range experimentally available in DIS at large $x$, the choice of HT correction implementation can bias the extraction of leading twist quantities such as the $d/u$ ratio and the offshell deformation function $\delta f$. Nonetheless this bias can be largely removed by independently fitting the HT corrections for the proton and the neutron HT functions, as we have explicitly demonstrated analytically and in a global QCD analysis in Ref.~\cite{Cerutti:2025yji}.

To understand where the bias originates, consider the neutron-to-proton $F_2$ structure function ratio, denoted as $n/p$, that can be determined from proton and deuteron target DIS data after removing nuclear corrections. However, the presence of both HT and off-shell corrections affect the determination of the deuteron structure function, and biases in either one can be compensated by the fitted parameters of the other. Consequently, the  $n/p$ ratios extracted with a multiplicative or an additive HT implementation may end up to be incompatible with each other. 

To see this analytically, consider first the isospin-independent \textit{multiplicative} HT functions, $C(x) \equiv C_p(x) = C_n(x)$, the correction simply cancels in the $n/p$ ratio and one obtains the same limit at large $x\rightarrow1$ as in a LT calculation: $n/p \simeq 1/4$.
With isospin-independent \textit{additive} HT corrections, $H(x) \equiv H_p(x) = H_n(x)$, we obtain instead $n/p \simeq 1/4 + \Delta$, with $\Delta =  \frac{27}{16} \frac{H/u}{Q^2}$. Therefore, a larger tail originates with additive HT than the multiplicative HT. This is a direct consequence of the implementation choice, potentially leading to an over- or underestimate of the $n/p$ ratio.
If instead we utilized isospin-dependent HT terms, the fit would be able to interpolate between these two extremes as warranted by the data.  For example, if the neutron correction was smaller than the proton's, for example $H_n \approx \frac12 H_p$ at large $x$ as in Ref.~\cite{Alekhin:2003qq}, one would obtain an intermediate value $n/p \simeq \frac{1}{4} + \Delta'$, with $ \Delta' = \frac{9}{16} \frac{{H}_p/u}{Q^2}$. The same results easily follow by using the equivalent additive representation ($H \rightarrow \tilde H$). Consequently, with isospin-dependent HT terms the fit can determine the appropriate $H_n/H_p$ value and the bias in the isospin-independent implementation is removed.

We therefore follow Ref.~\cite{Cerutti:2025yji} and will perform 2 fits with isospin-dependent HT terms, one with the multiplicative implementation \eqref{e:ht_mult} and one with the additive implementation \eqref{e:ht_add}.
The ensuing difference in the extracted quantities (especially the offshell function, $n/p$ ratio, and gluon distribution) can be interpreted as an estimate of the underlying theoretical uncertainty \cite{Cerutti:2025yji}. 
As a conservative procedure, we quote the average of \texttt{CJ26\_m} and \texttt{CJ26\_a} fits as the central value of any quantity calculated with our fit result, and the envelope of the additive and multiplicative Hessian error bands as the combined statistical and theoretical uncertainty band. This allows us to account for the spread between the distinct large-$x$ solutions, which is not fully captured by the $\Delta\chi^2=2.7$ tolerance alone.

%%%%%%%%%%%%%%%%%%%%%%%%%%%%%%%%%%%%%%%%%%%%%%%%%%%%%%%
\section{Data selection: the JLab program}
\label{s:data}

The CJ26 global analysis is based on a dataset that includes over 5000 data points from different high-energy scattering processes (see Table~\ref{tab:chi2-JLab} for details). We consider inclusive DIS data from fixed-target measurements of the proton and deuteron $F_2$ structure functions at Jefferson Lab~\cite{e06009,JeffersonLabHallCE94-110:2004nsn,e94110eric,Seely:2009gt,e03103thesis, Tvaskis:2006tv,Tvaskis:2010as,e99118_f2,Niculescu1999,JeffersonLabE00-115:2009jll,e00116_f2,CLAS:2011qvj,CLAS:2014jvt,JeffersonLabHallATritium:2021usd,Biswas:2024diw}, SLAC~\cite{Whitlow1992,e140x}, HERMES~\cite{HERMES:2011yno}, NMC~\cite{NewMuon:1996fwh,NewMuon:1996uwk} and BCDMS~\cite{BCDMS:1989qop,BCDMS:1990} and the DIS cross section from the HERA $ep$ collider~\cite{H1:2015ubc}. Direct sensitivity to the neutron structure function is provided by the BONuS tagged DIS measurement, in which a low-momentum proton is detected in coincidence with the scattered electron, and in the backward direction compared to the exchanged 4-momentum. Further flavor separation power is provided by $W$-boson~\cite{D0:2014kma,D0:2013xqc,CDF:2009cjw,D0:2013lql} and $Z$-boson~\cite{CDF:2010vek,D0:2007djv} asymmetries, and direct sensitivity to gluons by inclusive jet~\cite{CDF:2008hmn,D0:2000dzr,D0:2008nou} and $\gamma$+jet~\cite{D0:2008chx} cross sections from the CDF and DØ experiments at Tevatron. Finally, we included lepton pair production (LPP) from the E866 experiment at Fermilab \cite{NuSea:1998kqi,NuSea:2001idv} and from the E906/SeaQuest experiment~\cite{SeaQuest:2021zxb}, and the rapidity distribution of the $W^+/W^-$ ratio in proton-proton collisions by the STAR experiment at RHIC~\cite{STAR:2020vuq}.

Compared to the most recent analysis of the CJ Collaboration (\texttt{CJ22ht}~\cite{Cerutti:2025yji}), in this global fit we include 707 new data points from new inclusive DIS measurements of the proton and deuteron $F_2$ structure function from the JLab 6 GeV program~\cite{e06009,JeffersonLabHallCE94-110:2004nsn,e94110eric,Seely:2009gt,e03103thesis, Tvaskis:2006tv,Tvaskis:2010as,e99118_f2,Niculescu1999} and from the recent JLab 12 GeV program~\cite{JeffersonLabHallATritium:2021usd,Biswas:2024diw}, as well as the 8 high-statistics measurements from the E140X experiment at SLAC~\cite{e140x}. We passed the threshold of 5000 data points included in a global QCD fit.

In the PDF determination procedure, we include information on the experimental correlated systematic uncertainties when available, which is the case for some of the new Jefferson Lab data sets (see Table~\ref{tab:chi2-JLab}). For the SLAC data, we have taken the experimental correlations from the simultaneous analysis of 8 separate experiments performed by Whitlow and collaborators \cite{Whitlow1992, Whitlow1990,e140-web}. Note that they are included for the first time in a CTEQ-JLab QCD analysis. Remarkably, we find that the systematic shifts required to agree with the global data set included in the CJ26 fit are quite small, validating Whitlow's combination of SLAC data (see Appendix~\ref{ss:syst-corr} for more detail). We have similarly included correlated systematic errors for the lepton and $W$-boson asymmetries and for the jet cross sections measured at Tevatron. 

\begin{figure}[htb]
\centering
\includegraphics[width=0.495\textwidth]{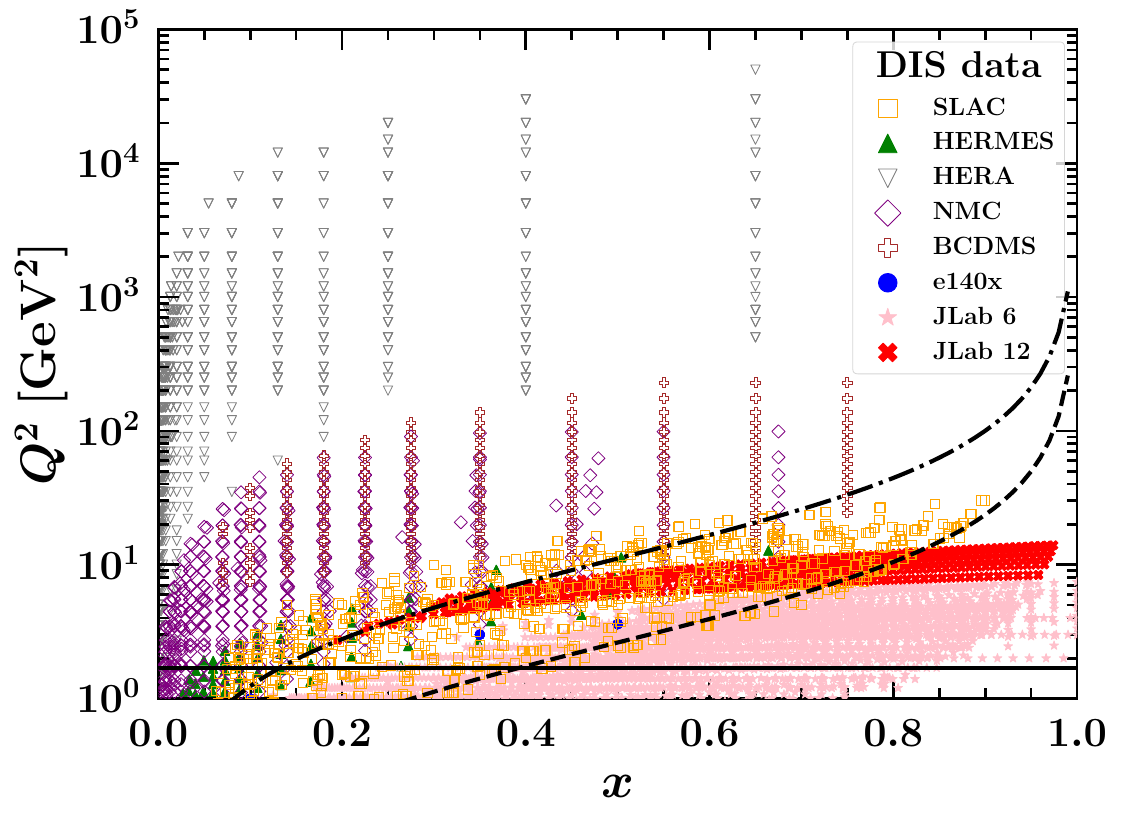}
\includegraphics[width=0.495\textwidth]{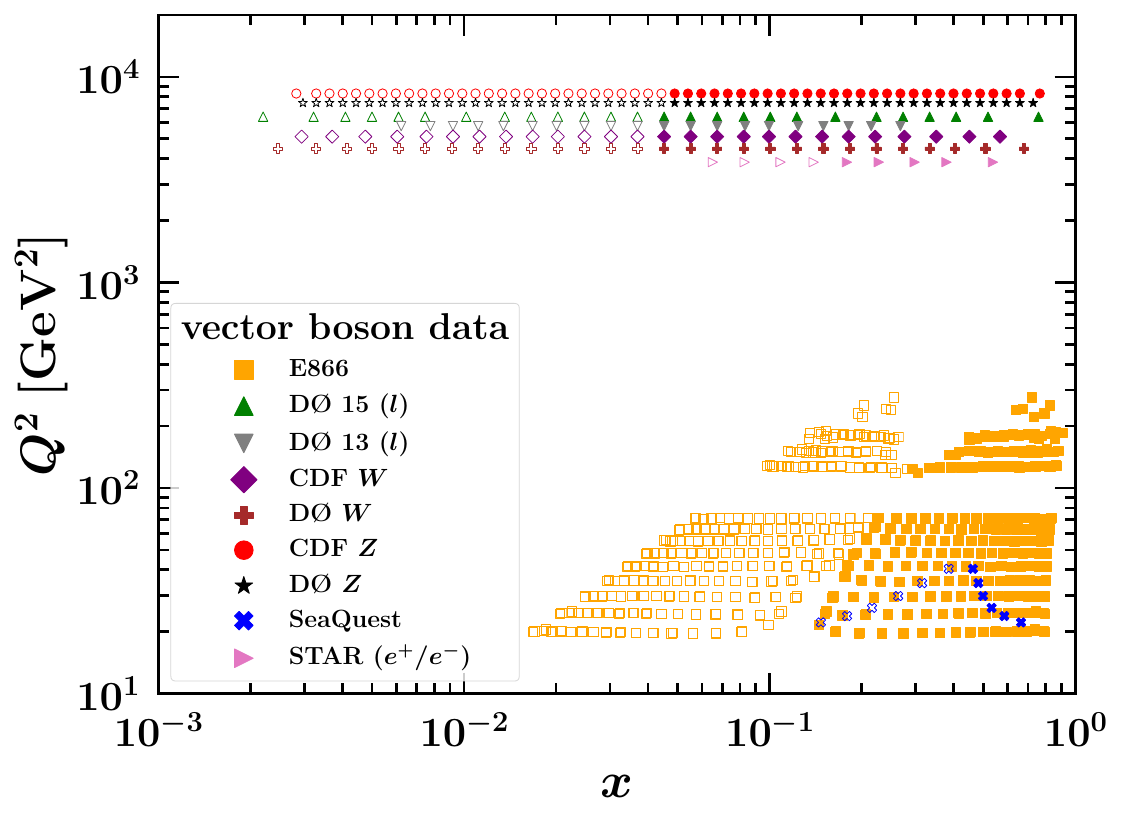}    
\includegraphics[width=0.495\textwidth]{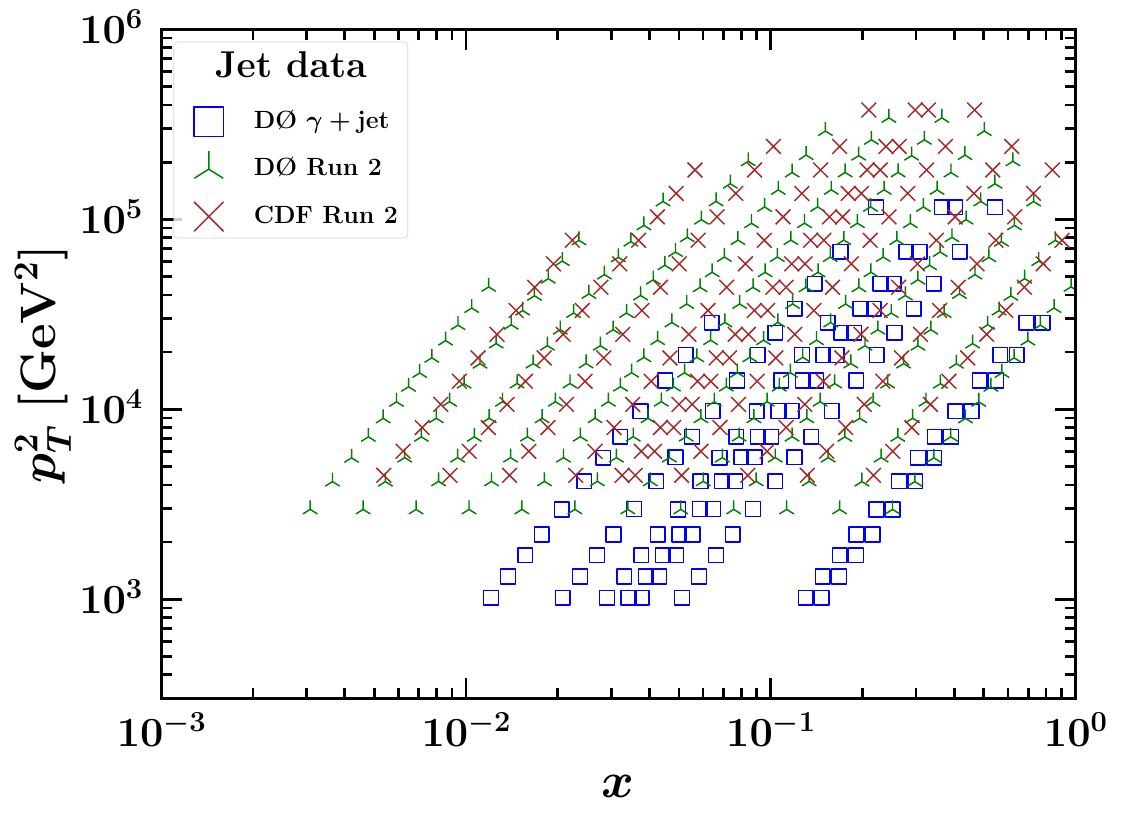}
\caption{Coverage of the data sets used in the CJ26 global fit. Left panel: DIS data and $W^2 = 12 \text{ GeV}^2$ (dot-dashed), $W^2 = 3.5 \text{ GeV}^2$ (dashed) and $Q^2 = 2 \text{ GeV}^2$ (plain) lines. Right panel: vector-boson data. To improve readability, a vertical offset is applied to the vector-boson data, and open (solid) markers are used to represent negative (positive) rapidity. Bottom panel: jet data.
}
\label{f:coverage}
\end{figure}

In Fig.~\ref{f:coverage}, we show the kinematic $x$ and $Q^2$ coverage of the DIS data (left panel), the vector-boson data (right panel) and of the jet data (bottom panel) included in our analysis. We display the leading order kinematics according to Refs.~\cite{Ball:2010de,D0:2008chx}.
%%%%%%%%%%%%%%
%%%%%%%%%%%%%%%%%%%%

We note that the DIS data from the JLab 6 GeV program cover a good slice of the $x,Q^2$ plane, above all at large-$x$ and small-$Q^2$. In our analysis, we impose the cuts $Q^2 > 1.69$ GeV$^2$ and $W^2 > 3.5$ GeV$^2$ to exclude the kinematic region dominated by resonances, which cannot be described in the standard perturbative QCD approach. Such a cut significantly limits the number of JLab 6 data points included in our analysis to a narrow slice in $Q^2$ at $0.25 \lesssim x \lesssim 0.7$, overlapping with SLAC data at the boundary of its phase space. However, the JLab 12 measurements cover a thicker, higher-$Q^2$ slice of the phase space, and most of the data point survive the imposed kinematic cuts reaching past $x\approx 0.8$. As they providing high-precision measurements at a $Q^2$ which is intermediate between JLab 6 GEV and SLAC, the JLab 12 data are fundamental to the determination of HT corrections, in particular at $0.7\lesssim x \lesssim 0.8$, where they are the only other source of information apart from SLAC data.

\begin{figure}[htb]
\centering
\includegraphics[width=0.495\textwidth]{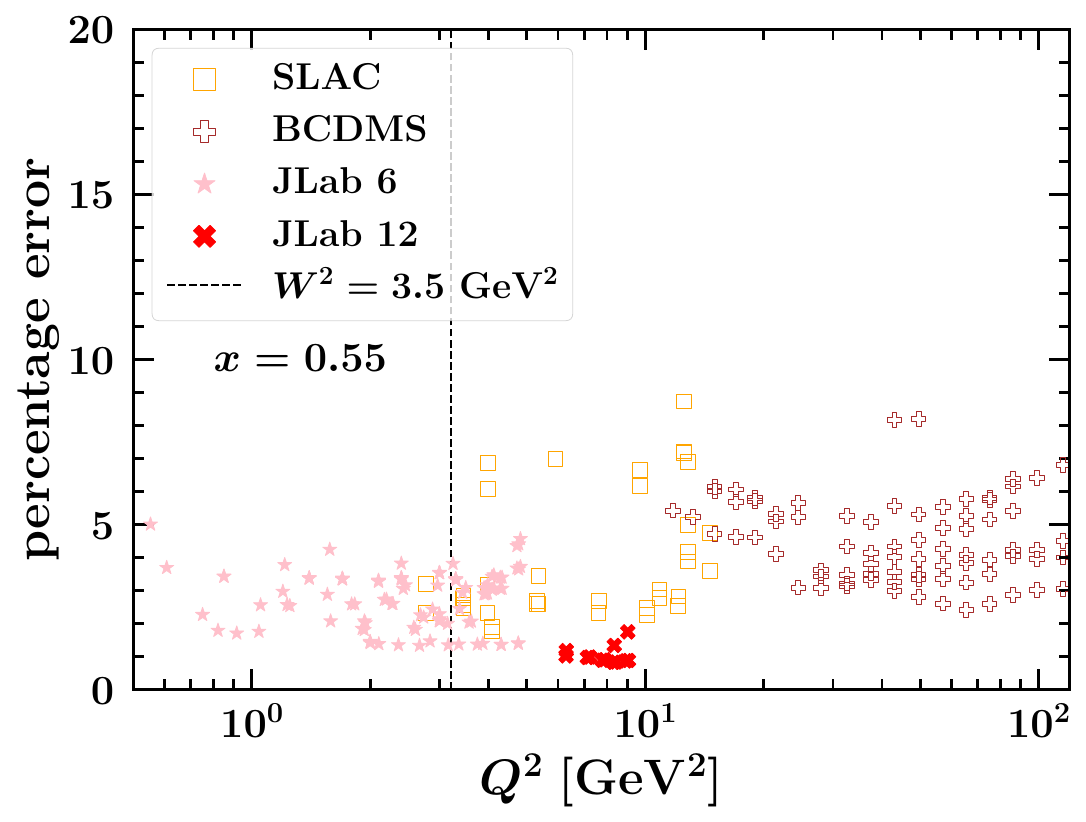}
\includegraphics[width=0.495\textwidth]{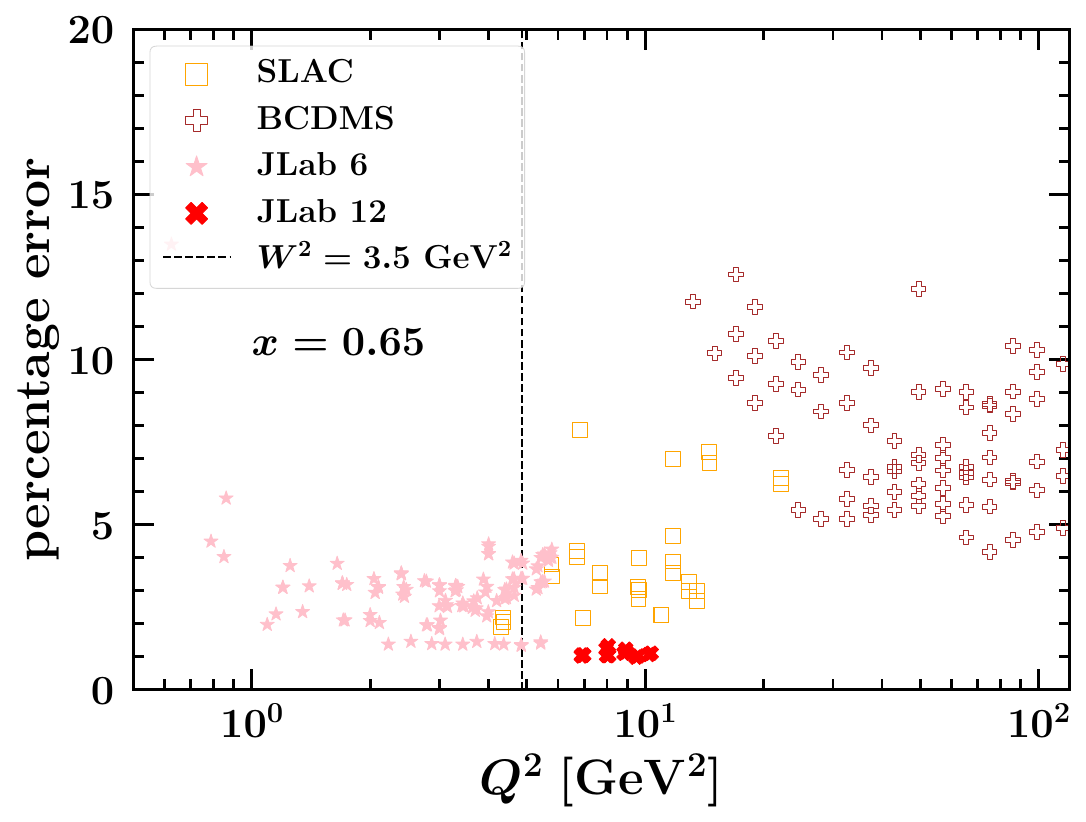}
\includegraphics[width=0.495\textwidth]{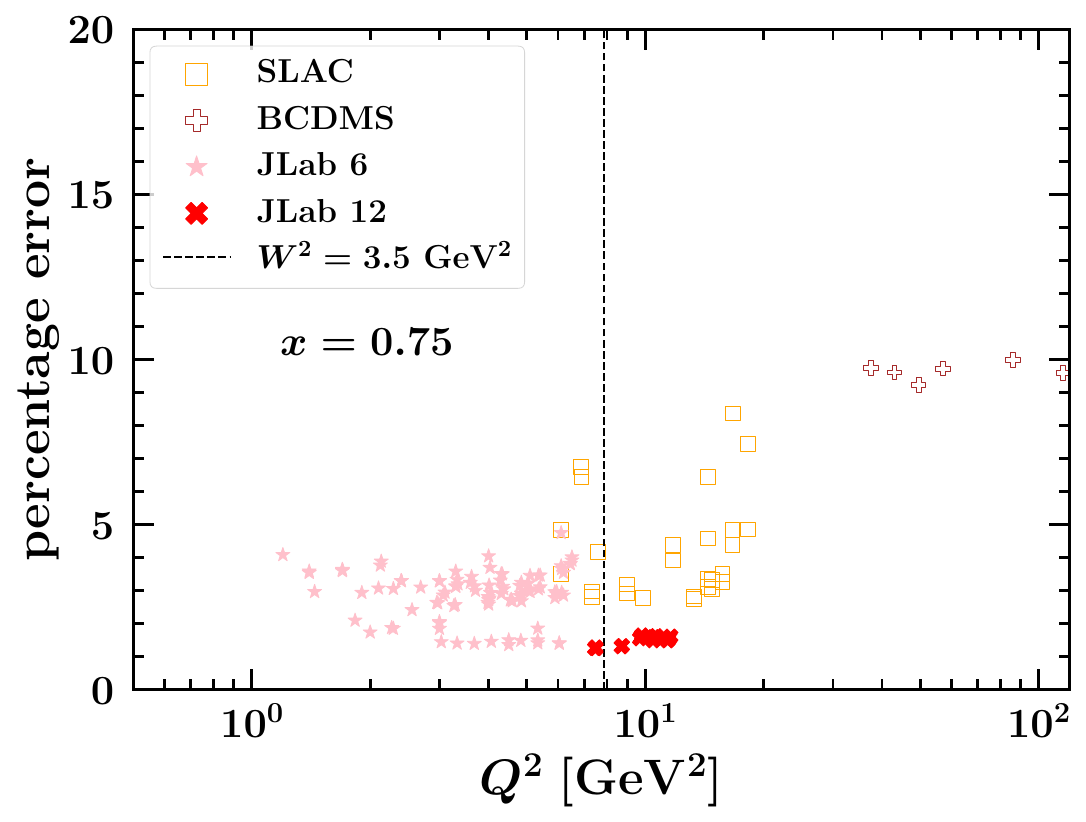}
\includegraphics[width=0.495\textwidth]{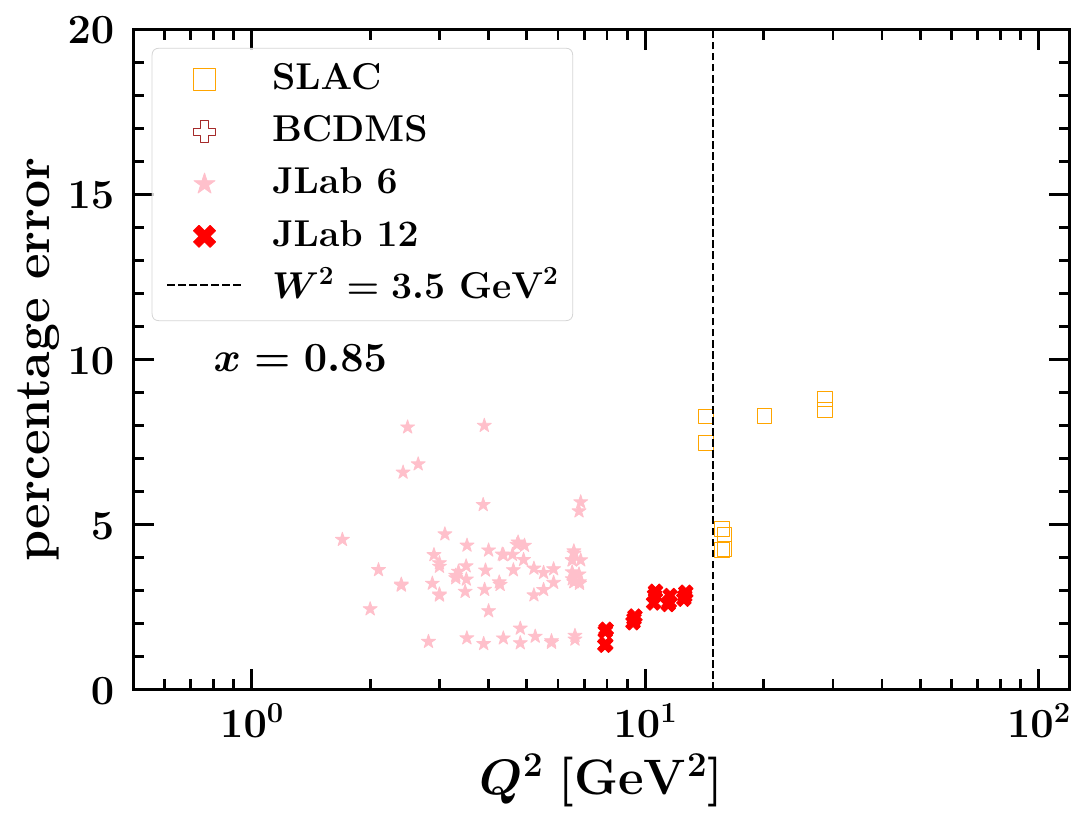}
\caption{Comparison of percent statistical experimental error as a function of the energy scale $Q^2$ of the DIS data included in our analysis. The four panels show data falling within $\pm 0.01$ of $x=0.55$, 0.65, 0.75, 0.85. The vertical, dashed show the position of the $W^2=3.5$ GeV$^2$ cut imposed in the \texttt{CJ26} analysis.}
\label{f:error_DIS_vs_Q}
\end{figure}

The high precision of the JLab 6 Gev data and even more of the JLab 12 GeV data is highlighted in Figure~\ref{f:error_DIS_vs_Q} where their realtive statistical errors are compared to those of previous experiments. The high precision and increased $Q^2$ leverage offered by the combined JLab 6 and 12 GeV data is fundamental to disentangle power correction, from the leading-twist dynamical offshell corrections in deuteron targets which was only partially achieved in the previous CJ.
%%%%%%%%%%%%%%
%%%%%%%%%%%%%%%%%%%%

\newpage

%%%%%%%%%%%%%%%%%%%%%%%%%%%%%%%%%%%%%%%%%%%%%%%%%%%%%%%
\section{Results: the CJ26 fit}
\label{s:Results}

In this section, we present the main results of the CJ26 NLO global analysis of unpolarized hard scattering data discussed in the previous section. This work represents a major upgrade with respect to our previous extractions due to the extent of the DIS dataset with JLab data and to the improved control of the model uncertainties. We will first discuss the quality of the fit in Sec.~\ref{ss:Results-quality}; then in Sec.~\ref{ss:Results-quantities} we will show the extracted physical quantities, including parton distributions, off-shell corrections, and power corrections; in Sec.~\ref{s:Results-JLab} we will analyze the impact of the JLab data; finally, in  Sec.~\ref{ss:Results-compHE} we will compare our results with other modern global QCD analyses.

%%%%%%%%%%
\subsection{Fit quality}
\label{ss:Results-quality}

In this section, we discuss the quality of the CJ26 global fit to the experimental datasets with kinematical cuts discussed in Sec.~\ref{s:data}. The PDF error analysis is performed through the variant of the the Hessian method discussed in Section~\ref{sec:fit_uncertainties}.

Table~\ref{tab:chi2_CJ26} reports summative statistical quantifiers for the 2 fit variants discussed in this paper, namely using the multiplicative and additive implementations of the HT corrections. Note that we have grouped measurements by reaction or final state, so that for example the the ``JLab (p)'' row groups 5 DIS experiments on proton target from JLab. For each group of data we report the $\chi^2$ value obtained in the multiplicative and additive fits ($\chi^2_\text{mult}$ and $\chi^2_\text{add}$, respectively, as defined in Eq.~(5) of Ref.~\cite{Li:2023yda}). We also include the value of the Gaussian variable $S_E$ defined in Eq.~(157) of Ref.~\cite{Kovarik:2019xvh}. This variable quantifies the fluctuation of a given data set from the calculated theoretical values in terms of standard deviations. We observe that most of the included datasets are satisfactorily described in our global fit. In fact, the global $\chi^2$ per data point is $1.1$ and it is very similar in the two fit configurations. This stability is due to the isospin dependence ($p \neq n$) introduced in the parameterization of the higher-twist as discussed in Sec.~\ref{ss:Formalism-interplay} and, more extensively, in Ref.~\cite{Cerutti:2025yji}. Given that the description of the global dataset shows independence of the HT implementation choice, in the plots we will include in this paper we will display the envelope between the results in the two configurations, which then includes this source of theoretical uncertainty on top of the statistical fit uncertainty.

\begin{table}%[h!]
\renewcommand{\arraystretch}{0.6}
\begin{tabular}{|c|l|c|c|cc|cc|}
\hline
Obs. & Experiment & Ref & \# Points & $\chi^2_{\rm mult}$ & $S_{\rm E, mult}$ & $\chi^2_{\rm add}$ & $S_{\rm E, add}$\\
\hline
DIS & JLab 12 GeV (D/p)${}^\dagger$ & \cite{JeffersonLabHallATritium:2021usd,Biswas:2024diw} & 339 & 284.1 & -2.2 & 284.4 & -2.2 \\
 & JLab 6 GeV (p) & \cite{e06009,JeffersonLabHallCE94-110:2004nsn,e94110eric,Seely:2009gt,e03103thesis, Tvaskis:2006tv,Tvaskis:2010as,e99118_f2,Niculescu1999,JeffersonLabE00-115:2009jll,e00116_f2} & 269 & 247.1 & -0.9 & 243.9 & -1.1 \\
 & JLab 6 GeV (D)${}^\dagger$ & \cite{e06009,JeffersonLabHallCE94-110:2004nsn,e94110eric,Seely:2009gt,e03103thesis, Tvaskis:2006tv,Tvaskis:2010as,e99118_f2,Niculescu1999,JeffersonLabE00-115:2009jll,e00116_f2} & 279 & 205.3 & -3.3 & 204.7 & -3.4 \\
 & BoNUS6 (n/D) & \cite{CLAS:2011qvj,CLAS:2014jvt} & 137 & 158.4 & 1.3 & 152.8 & 1.0 \\
 & HERMES (p) & \cite{HERMES:2011yno} & 37 & 42.4 & 0.7 & 45.5 & 1.0 \\
 & HERMES (D) & \cite{HERMES:2011yno} & 37 & 36.6 & 0.0 & 37.9 & 0.2 \\
 & SLAC (p)${}^\dagger$ & \cite{Whitlow1992} & 530 & 615.2 & 2.5 & 604.7 & 2.2 \\
 & SLAC (D)${}^\dagger$ & \cite{Whitlow1992} & 541 & 556.7 & 0.5 & 553.4 & 0.4 \\
 & E140X (p) & \cite{e140x} & 9 & 5.6 & -0.8 & 6.6 & -0.5 \\
 & E140X (D) & \cite{e140x} & 13 & 5.6 & -1.7 & 5.4 & -1.7 \\
 & BCDMS (p)${}^\dagger$ & \cite{BCDMS:1989qop} & 351 & 441.4 & 3.2 & 451.7 & 3.6 \\
 & BCDMS (D)${}^\dagger$ & \cite{BCDMS:1990} & 254 & 291.6 & 1.6 & 294.7 & 1.8 \\
 & NMC (p)${}^\dagger$ & \cite{NewMuon:1996fwh} & 275 & 404.7 & 5.0 & 403.7 & 5.0 \\
 & NMC (D/p)${}^\dagger$ & \cite{NewMuon:1996uwk} & 189 & 169.9 & -1.0 & 170.5 & -1.0 \\
 & HERA (NC $e^-p$)${}^\dagger$ & \cite{H1:2015ubc} & 159 & 247.5 & 4.4 & 239.1 & 4.1 \\
 & HERA (NC $e^+p$)${}^\dagger$ & \cite{H1:2015ubc} & 945 & 1091.7 & 3.3 & 1155.6 & 4.6 \\
 & HERA (CC $e^-p$)${}^\dagger$ & \cite{H1:2015ubc} & 42 & 47.4 & 0.6 & 49.0 & 0.8 \\
 & HERA (CC $e^+p$)${}^\dagger$ & \cite{H1:2015ubc} & 39 & 48.4 & 1.1 & 48.7 & 1.1 \\
LPP & E866 ($pp$) & \cite{NuSea:1998kqi} & 121 & 140.5 & 1.2 & 138.2 & 1.1 \\
 & E866 ($pD$) & \cite{NuSea:1998kqi} & 129 & 139.9 & 0.7 & 131.1 & 0.2 \\
 & SeaQuest ($D/p$) & \cite{SeaQuest:2021zxb} & 6 & 12.2 & 1.6 & 8.8 & 0.9 \\
W & D\O\ ($e$)${}^\dagger$ & \cite{D0:2014kma} & 13 & 18.2 & 1.0 & 20.4 & 1.4 \\
 & D\O\ ($\mu$)${}^\dagger$ & \cite{D0:2013xqc} & 10 & 16.2 & 1.3 & 16.2 & 1.3 \\
 & CDF ($W$) & \cite{CDF:2009cjw} & 13 & 14.3 & 0.3 & 15.4 & 0.5 \\
 & D\O\ ($W$)${}^\dagger$ & \cite{D0:2013lql} & 14 & 10.4 & -0.6 & 8.3 & -1.1 \\
 & STAR ($e^\pm$) & \cite{STAR:2020vuq} & 9 & 22.9 & 2.6 & 22.3 & 2.6 \\
Z & CDF (Z) & \cite{CDF:2010vek} & 28 & 28.9 & 0.2 & 29.8 & 0.3 \\
 & D\O\ (Z) & \cite{D0:2007djv} & 28 & 16.4 & -1.7 & 16.5 & -1.7 \\
jet & CDF (jet)${}^\dagger$ & \cite{CDF:2008hmn} & 72 & 97.0 & 2.0 & 91.5 & 1.6 \\
 & D\O\ (jet)${}^\dagger$ & \cite{D0:2008nou} & 110 & 102.4 & -0.5 & 98.0 & -0.8 \\
$\gamma$+jet & D\O\ (1, 2, 3, 4) & \cite{D0:2008chx} & 44 & 50.7 & 0.7 & 51.5 & 0.8 \\
\hline
 & total & & 5054 & 5582.8 & 5.1 & 5612.8 & 5.4 \\
 & total + norm & &  & 5602.4 & 5.3 & 5633.2 & 5.6 \\
\hline
\end{tabular}
\caption{Breakdown of the $\chi^2$ and $S_E$ values for the  data included in the \texttt{CJ26} analysis, grouped by facility and process. The values obtained with the multiplicative (mult) and additive (add) higher-twist implementations are separately displayed along with the number of fitted points. \mbox{${}^\dagger$For} these datasets correlated uncertainties were taken into account.
}
\label{tab:chi2_CJ26} 
\end{table}

\begin{figure}[h!]
    \centering
    \includegraphics[width=0.5\linewidth]
    {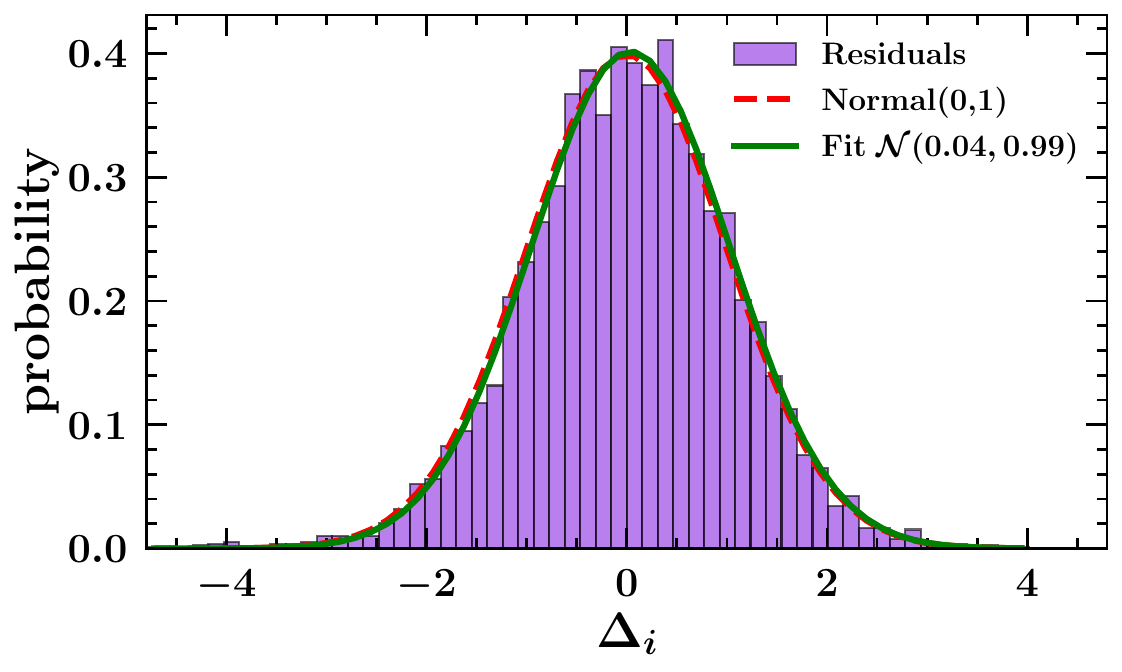}%
    \includegraphics[width=0.5\linewidth]{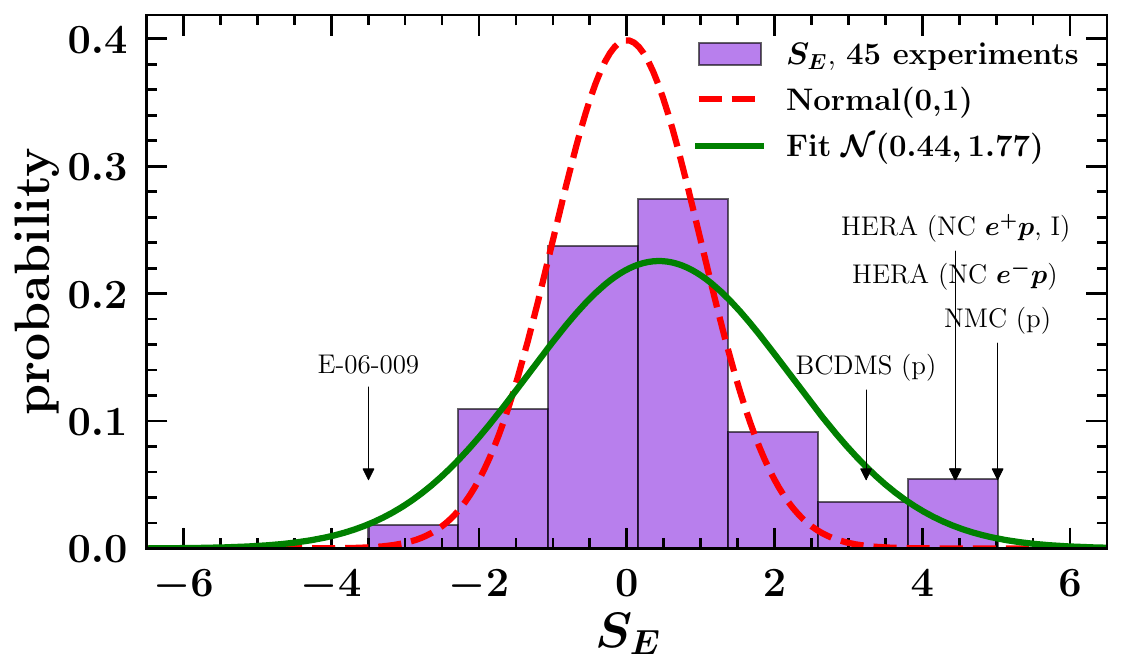}
    \caption{The pull distribution for the global dataset on the left and the Gaussian variable $S_E$ per experiment on the right. The statistical outliers with $|S_E|\geq 3$ are labeled individually.}
    \label{f:pull_global}
\end{figure}

For a more granular evaluation of the CJ26 fit quality, 
in the left panel of Fig.~\ref{f:pull_global} we display the histogram of the pull distribution $\Delta_i$ of each data point (that is, the difference between data and theory in units of the statistical error shifted by the fit's best estimate of the correlated systematic uncertainties when available) as defined in Eq.~(13) of Ref.~\cite{Pumplin:2002vw}. In the right panel we show the distribution of the Gaussian-equivalent $\chi^2$ variable $S_E$ of each individual data set in the multiplicative HT fit; the distribution is similar also in the additive HT fit. 
Both the pull and $S_E$ histograms are expected to follow a normal distribution ${\cal N}(0,1)$ with mean 0 and standard deviation equal to 1, displayed by the dashed red curve, while the green curve is a normal distribution fitted to the histograms.

The pull distribution is very close to the statistical expectation. Considering that we included 45 different datasets, we believe that this result clearly indicates that our global fit is statistically consistent. Nevertheless, the $S_E$ distribution in the right panel shows evidence of tensions in the global dataset. While the distribution's peak is close to 0, its width is significantly larger than 1, being driven by outliers with large and positive values of $S_E$. For concreteness, we have highlighted datasets with $|S_E|\geq 3$, but it is the HERA neutral current reduced cross sections and the $F_{2,p}$ measurements at NMC with $S_E = 4$-$5$ that drive the deviation from the normal behavior. With all known sources of correlated errors being consistently taken into account, the larger than expected $\chi^2$ values do not originate from easily identifiable regions in phase space, disfavoring a simple interpretation in terms of underestimated experimental uncertainties. We rather think that the large values of $S_E$ may originate from other sources; for example, from an intrinsically non-Gaussian distribution of statistical fluctuations, or missing elements in the theoretical calculations such as small-$x$ resummation~\cite{xFitterDevelopersTeam:2018hym} or other higher-twist corrections~\cite{Abt:2016vjh}. In fact, as already observed in the comparative study of Ref.~\cite{Kovarik:2019xvh}, the HERA, NMC, and BCDMS data display large $S_E$ values also in other global analysis. Recent examples include MSHT26 \cite{Harland-Lang:2025wvm}, NNPDF4.0~\cite{NNPDF:2021njg}, CT18~\cite{Hou:2019efy} and ABMP16~\cite{Alekhin:2017kpj}, which reinforces our view.

We now turn to the new experimental dataset included in this analysis, namely the JLab 6 and 12 GeV measurements, that are listed at the top of Tab.~\ref{tab:chi2_CJ26}. We observe that the theory/data agreement is excellent for each of the DIS datasets on proton ($p$) or deuteron ($D$) targets, as well as $D/p$ structure function ratios and the tagged $n/D$ measurements from BONuS targets, indicating that our theoretical models and phenomenological setup are robust. (If anything, the typically negative $S_E$ values indicate a tendency to overestimate the experimental uncertainties.) No tensions are observed between the new datasets and the rest of the data. This result supports the reliability of the extracted quantities.

We can now discuss comparison plots of the CJ26 fitted results with a representative selection of experimental data sensitive to valence quarks and the light-antiquark sea.

\begin{figure}[h!]
    \centering
    \includegraphics[width=0.51\linewidth]{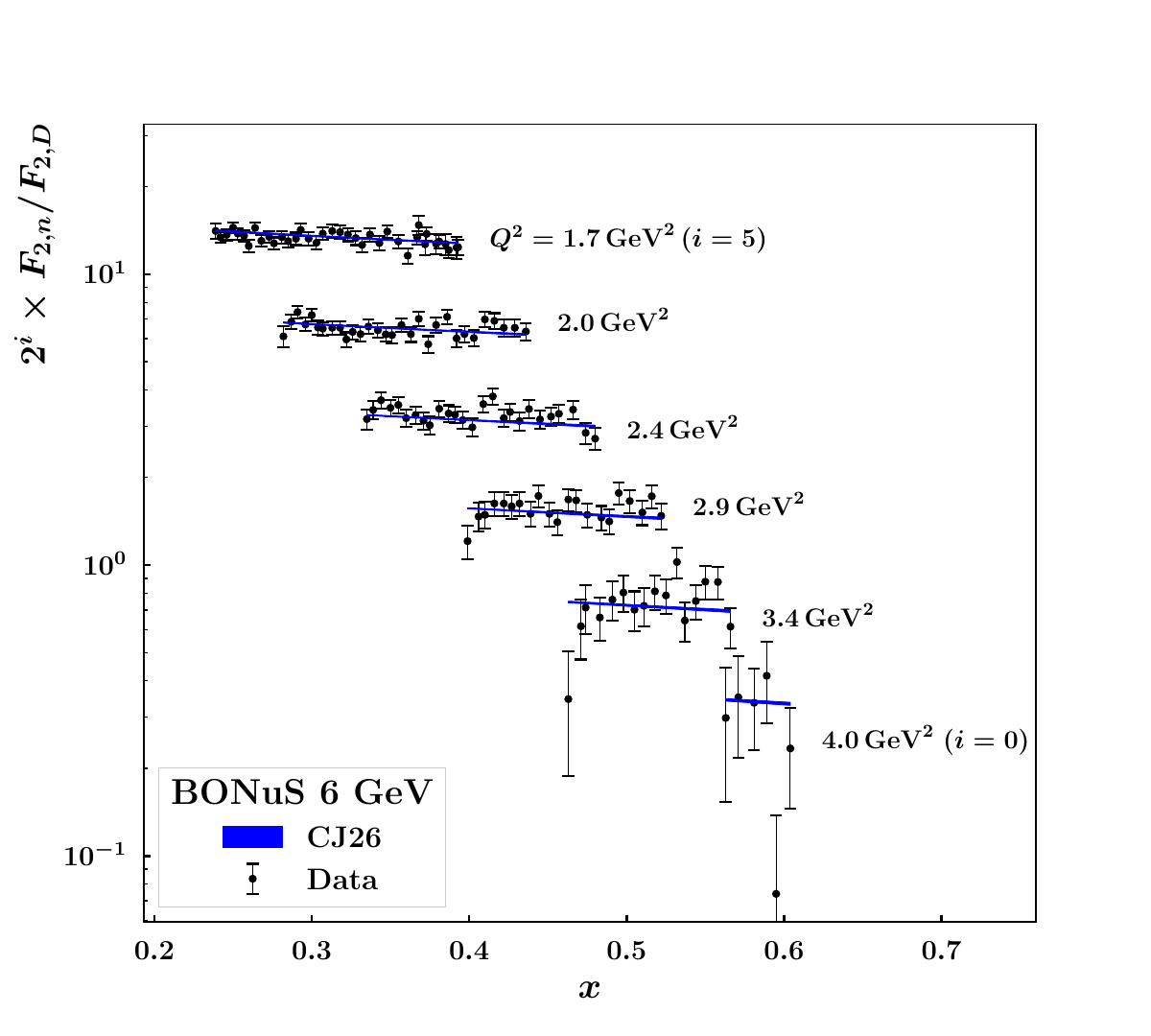}
    \includegraphics[width=0.46\linewidth]{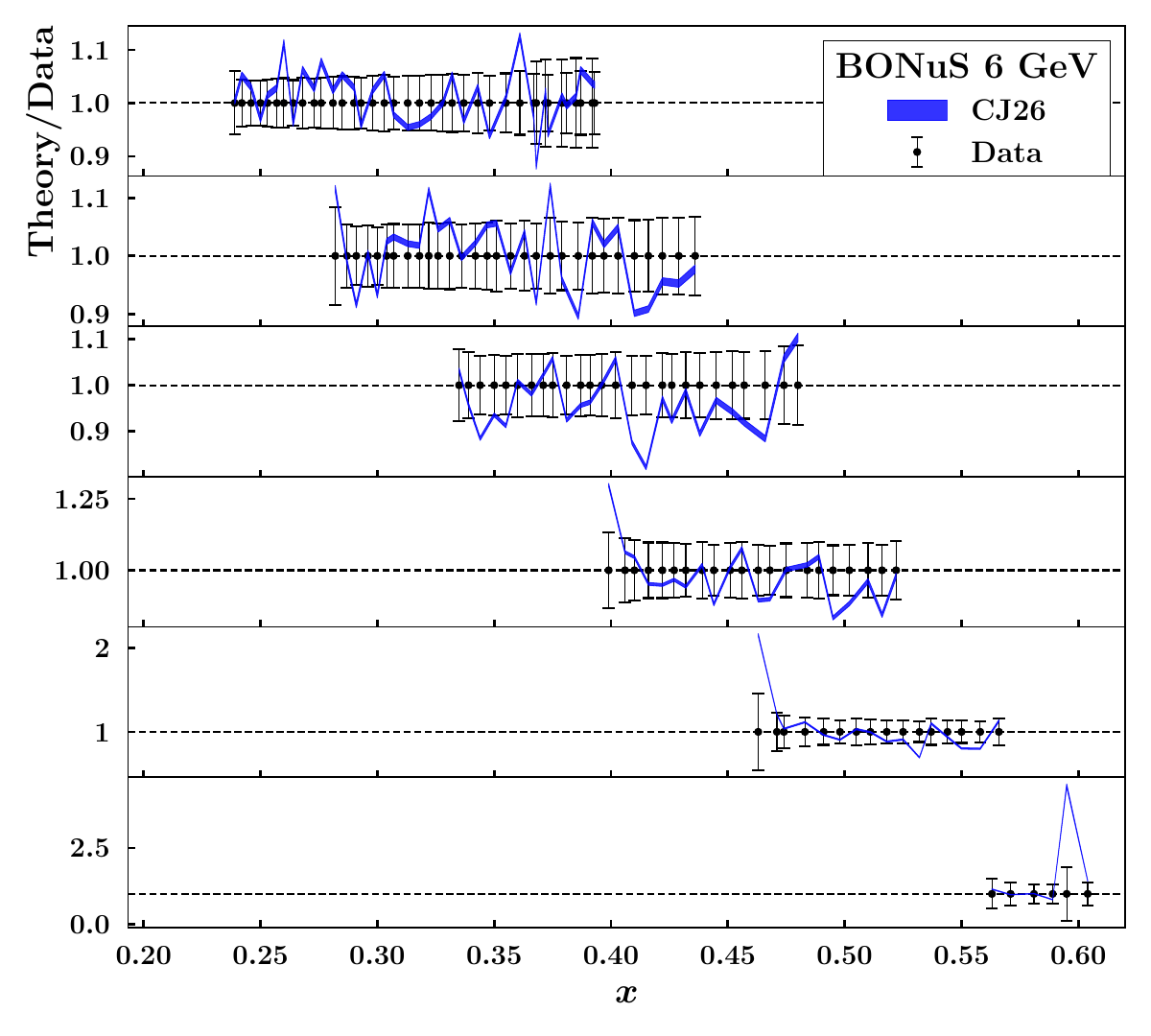}
    \caption{Data-theory comparison for $F_{2,n}/F_{2,D}$ ratio from the 6 GeV BONuS experiment. The data and corresponding CJ26 results at each kinematics are scaled for visibility on the left panel. CJ26 to data deviations are shown on the right.  Uncertainty are the envelope of $T=1.645$ errors of multiplicative and additive HT fits.}
    \label{f:bonus6}
\end{figure}

In Fig. ~\ref{f:bonus6} we show the comparison for the $n/D$ ratio of $F_2$ structure functions measured by the BONuS experiment, where the neutron structure function is obtained by tagging low-momentum spectator protons and reconstructing the active neutron's kinematics. With the numerator sensitive to off-shell corrections and the denominator to the full nuclear smearing, this plot validates the nuclear correction model we have adopted.

\begin{figure}[h!]
    \centering
    \includegraphics[width=0.49\linewidth]{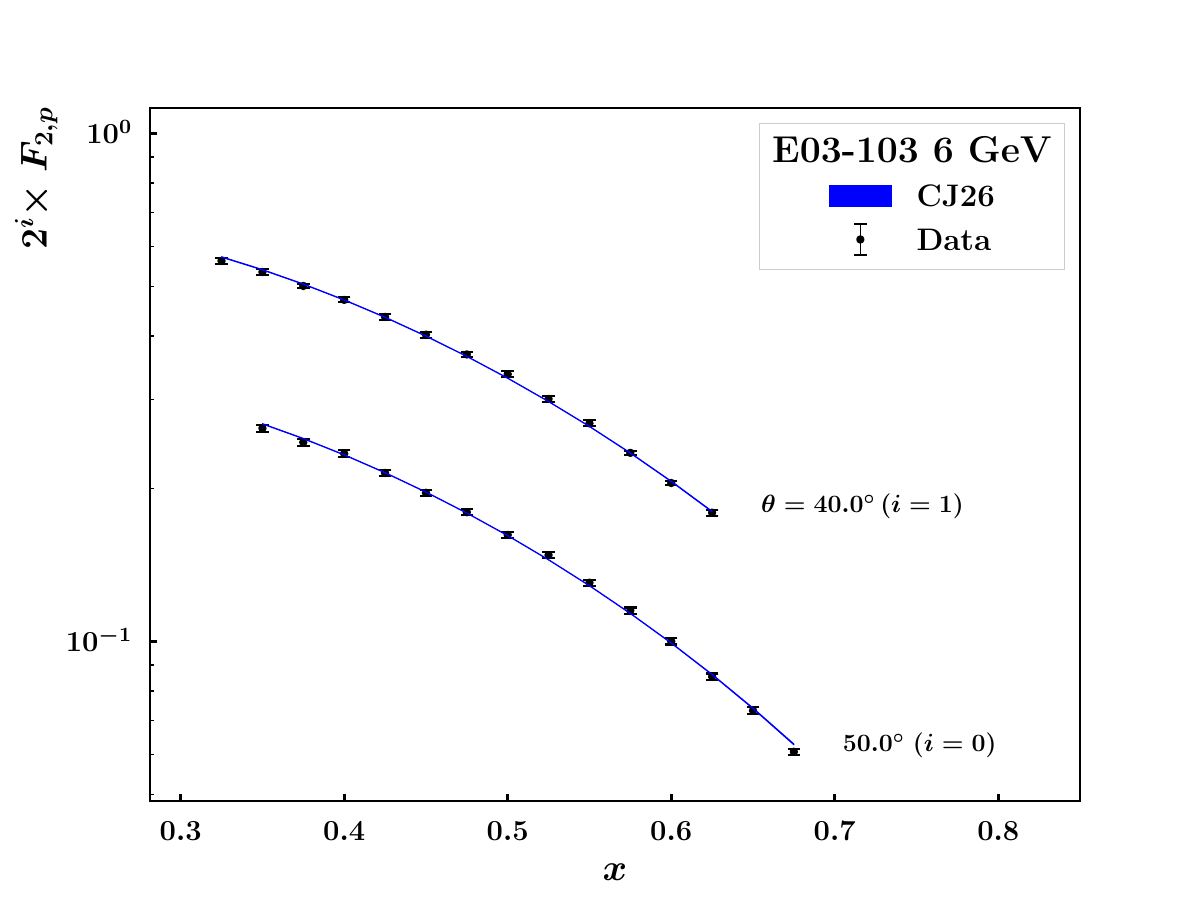}
    \includegraphics[width=0.48\linewidth]{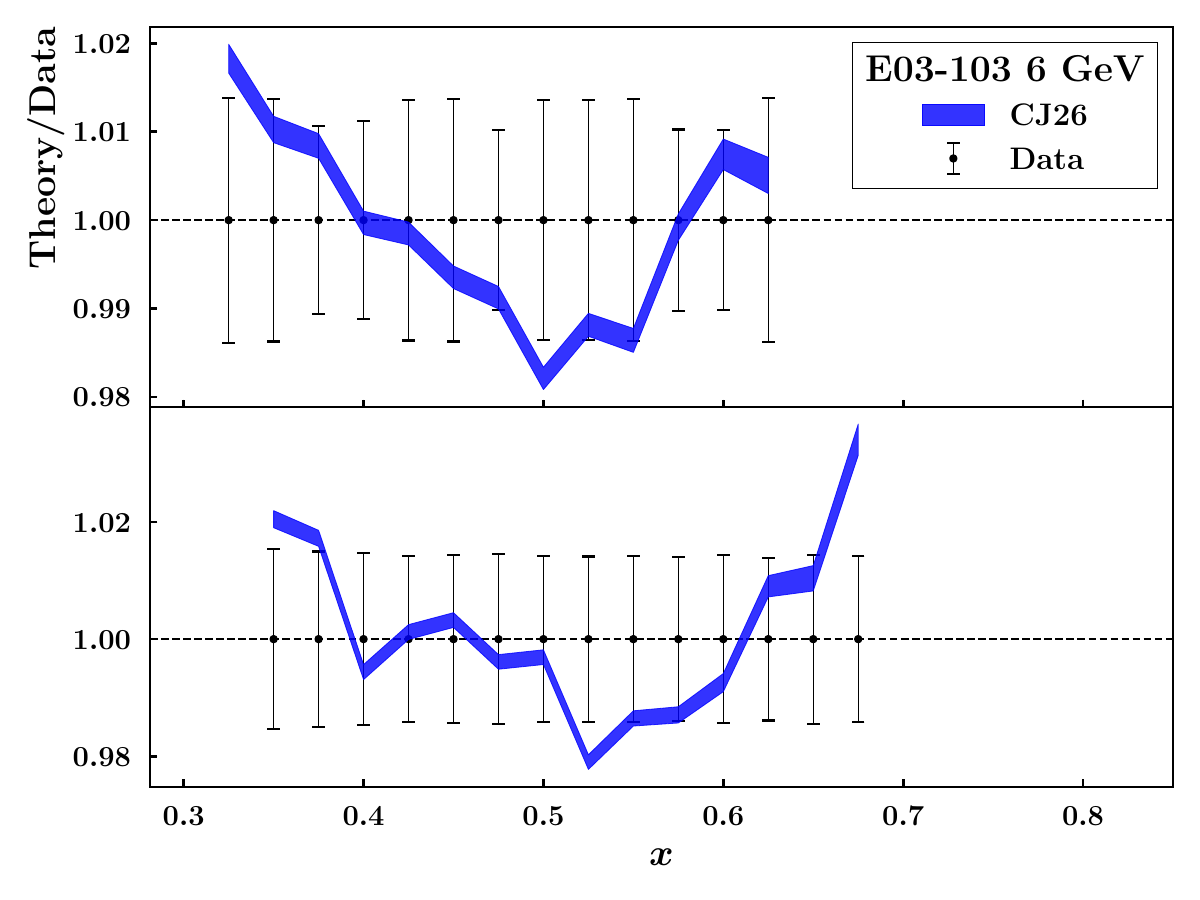}
    \includegraphics[width=0.49\linewidth]{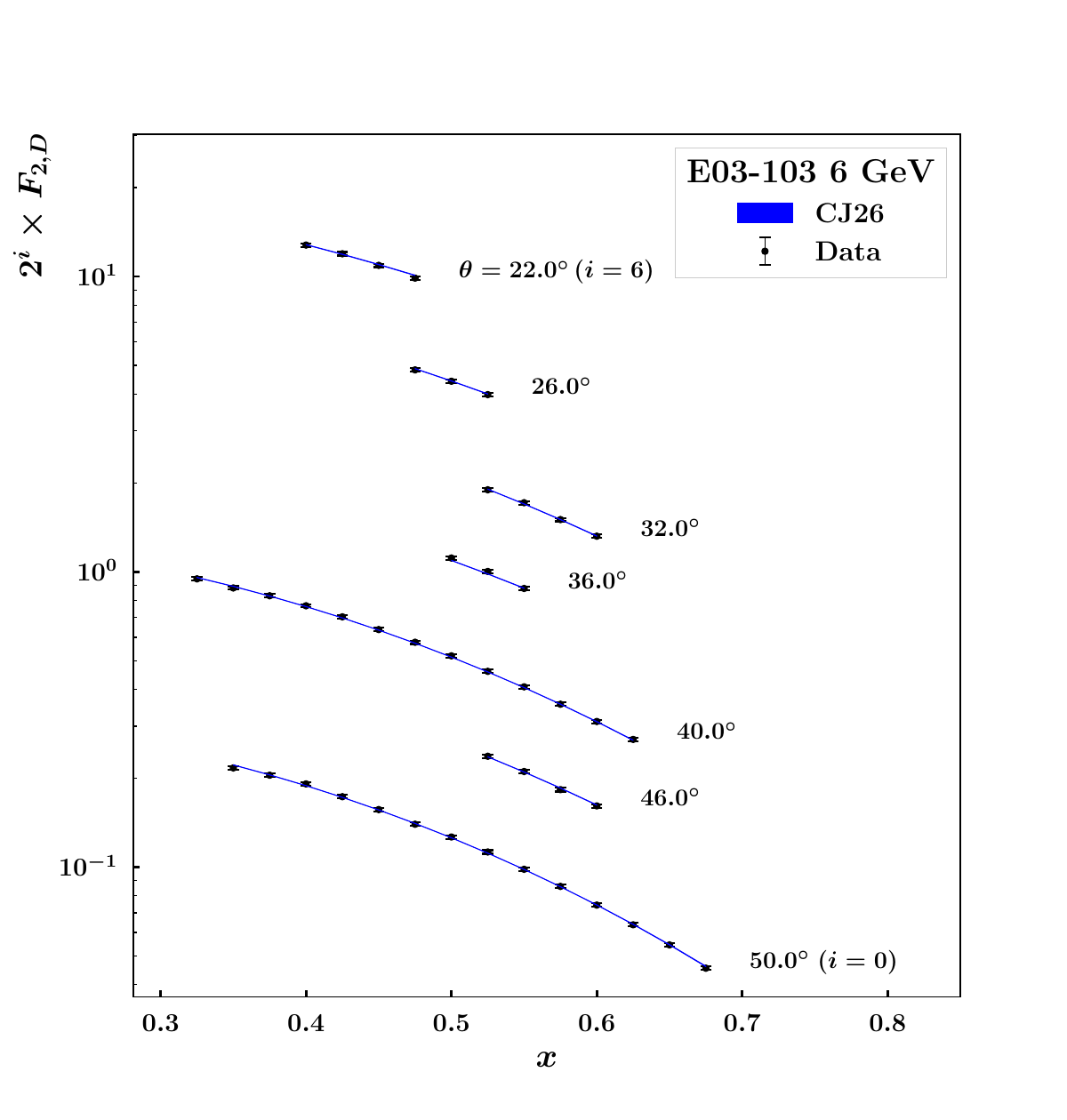}
    \includegraphics[width=0.47\linewidth]{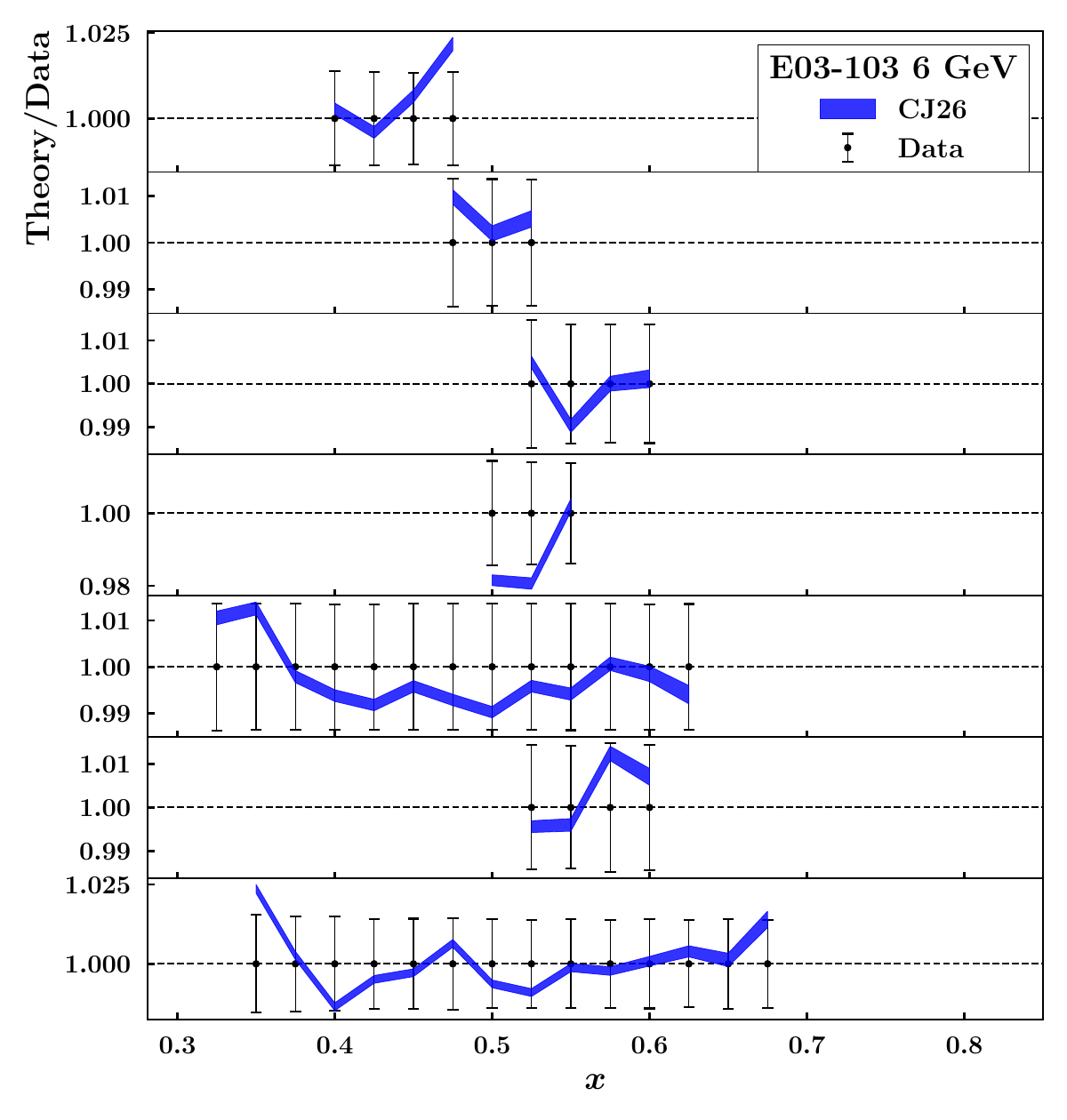}
    \caption{Data-theory comparison for $F_{2,p}$ (upper panels) and $F_{2,D}$ (lower panels) structure functions from the JLab 6 GeV experiment E03-103. The data and corresponding CJ26 results at each spectrometer angle are shown in the left panels and are scaled vertically for clarity. The corresponding deviation of the CJ26 calculations from the data are shown on the right. The data points from the same $x$ and $Q^2$ bins are combined for clarity. The theoretical uncertainties (blue bands) are the envelope of $T=1.646$ error bands of the multiplicative and additive HT fits.}
    \label{f:e03103}
\end{figure}

As a representative of the newly included JLab 6 GeV measurements, 
in Fig.~\ref{f:e03103} we show the data and theory comparison for the $F_{2,p}$ and $F_{2,D}$ structure functions measured by the E03-103 experiment. The agreement is excellent.

\begin{figure}[h!]
    \centering
    \includegraphics[width=0.49\linewidth]{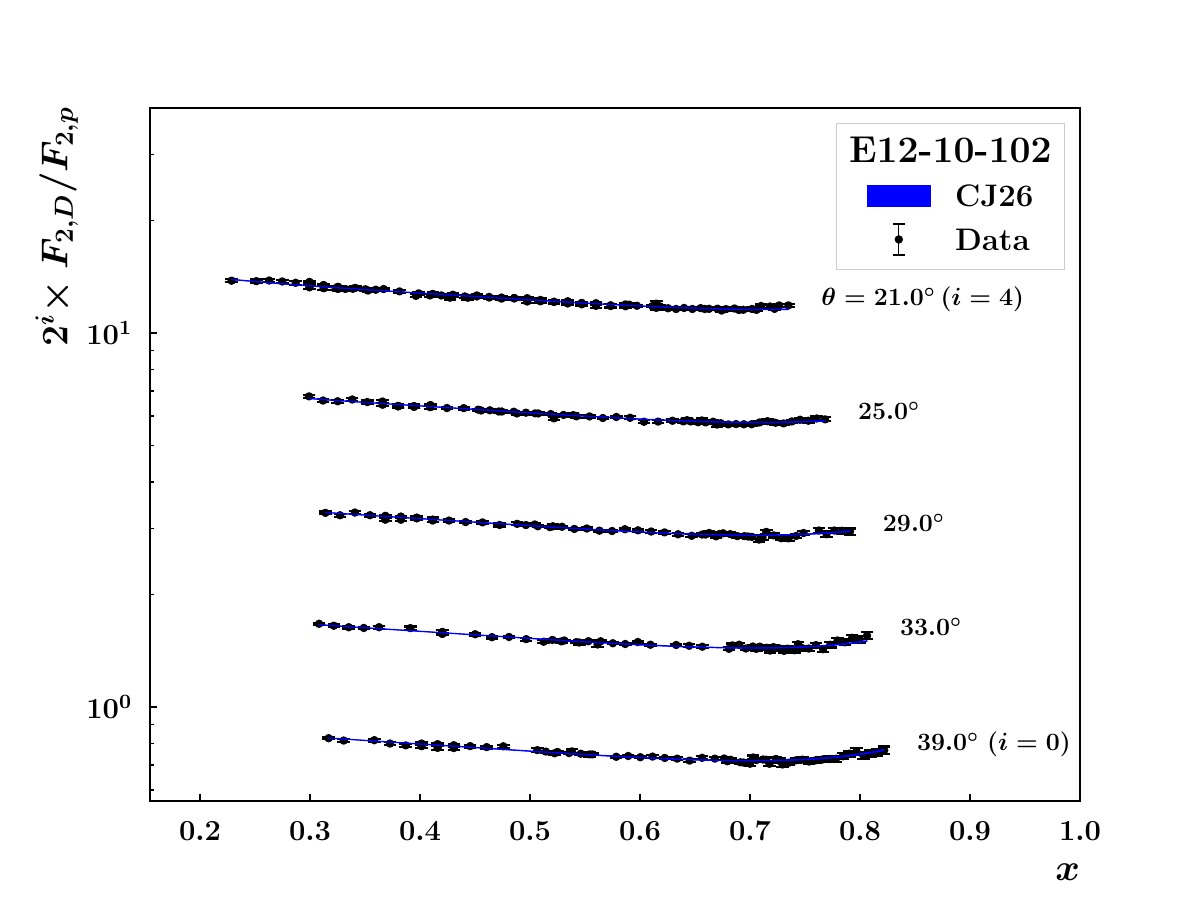}
    \includegraphics[width=0.49\linewidth]{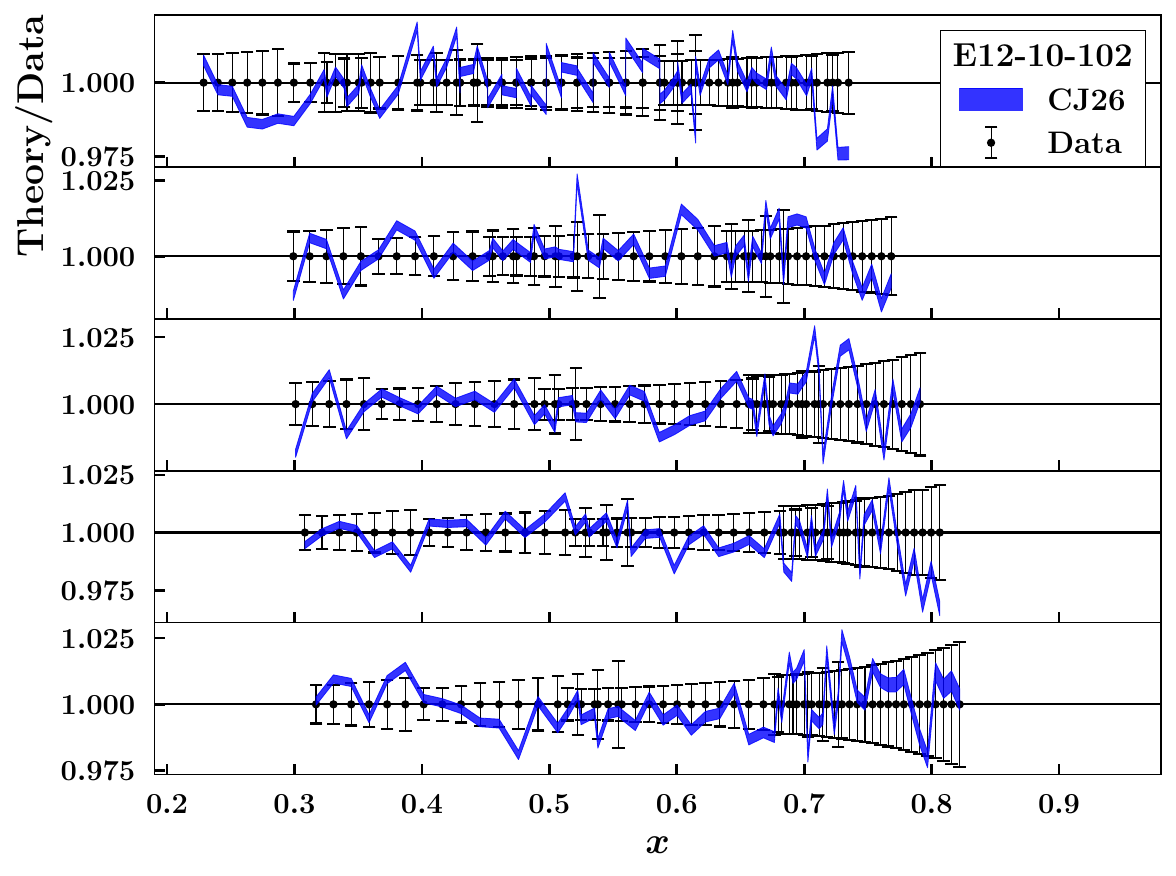}
    \caption{Data-theory comparison for the $D/p$ ratio of $F_2$ from the JLab 12 GeV experiment E12-10-102. The data and corresponding CJ26 calculations at several electron spectrometer angles are shown. The deviations of theory from data are shown on the right. The data points from the same $x$ and $Q^2$ bins are combined for clarity. The fit uncertainties are the envelopes of the $T=1.645$ errors of the CJ26 multiplicative and additive HT fits.
    }
    \label{f:e1210102}
\end{figure}

In Fig.~\ref{f:e1210102} we address the recent high statistics measurement of the $D/p$ ratio of $F_2$ structure functions by the JLab 12 GeV E12-10-102 experiment. The high statistics data are accompanied by a careful evaluation of correlated systematic errors, making it an important stress test of our global QCD analysis. The excellent agreement between data and theory is in part a validation of the CJ26 fit and a testament to the E12-10-102 experimental analysis.

\begin{figure}[h!]
    \centering
    \includegraphics[width=0.49\linewidth]{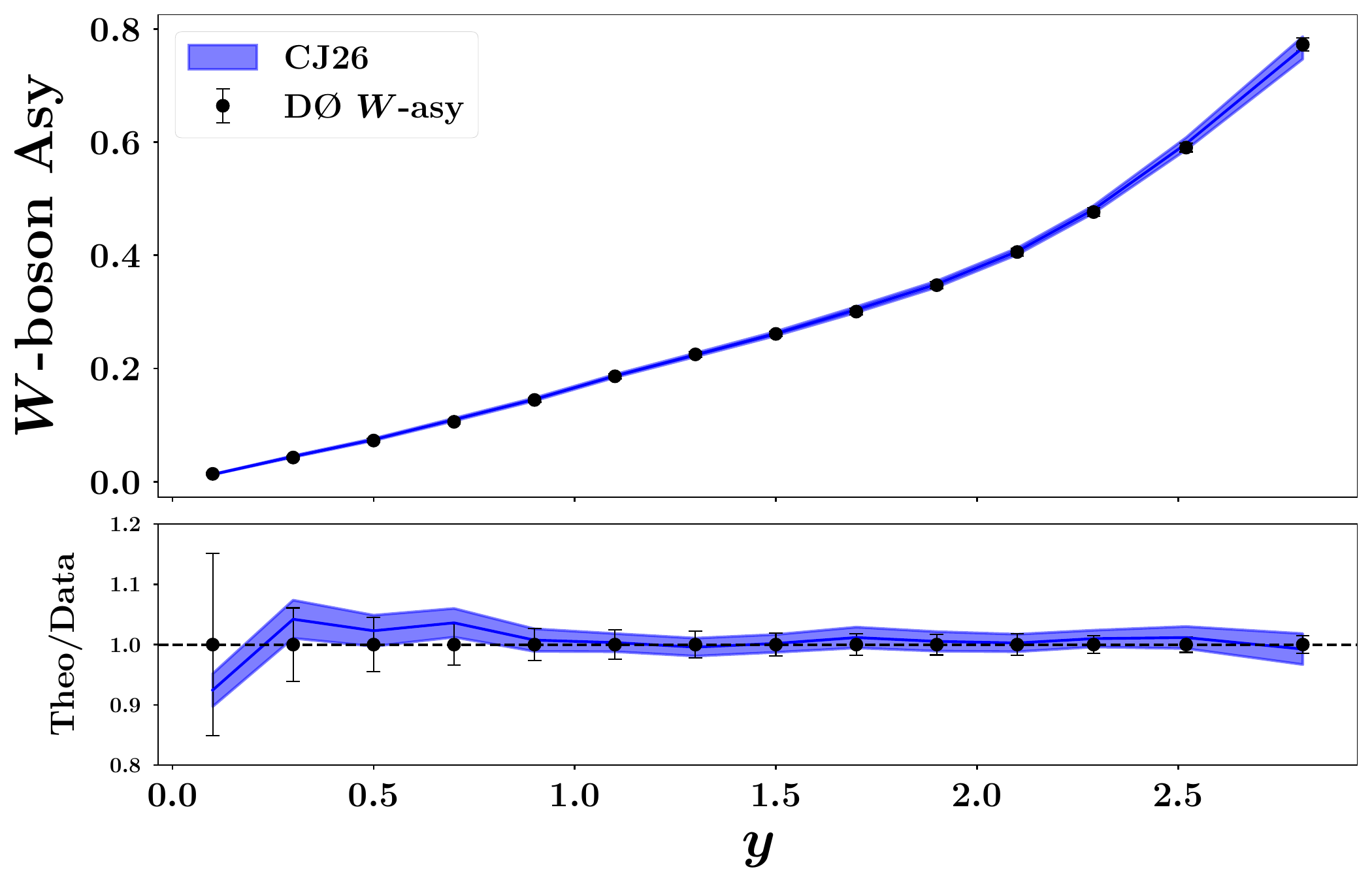}
    \includegraphics[width=0.49\linewidth]{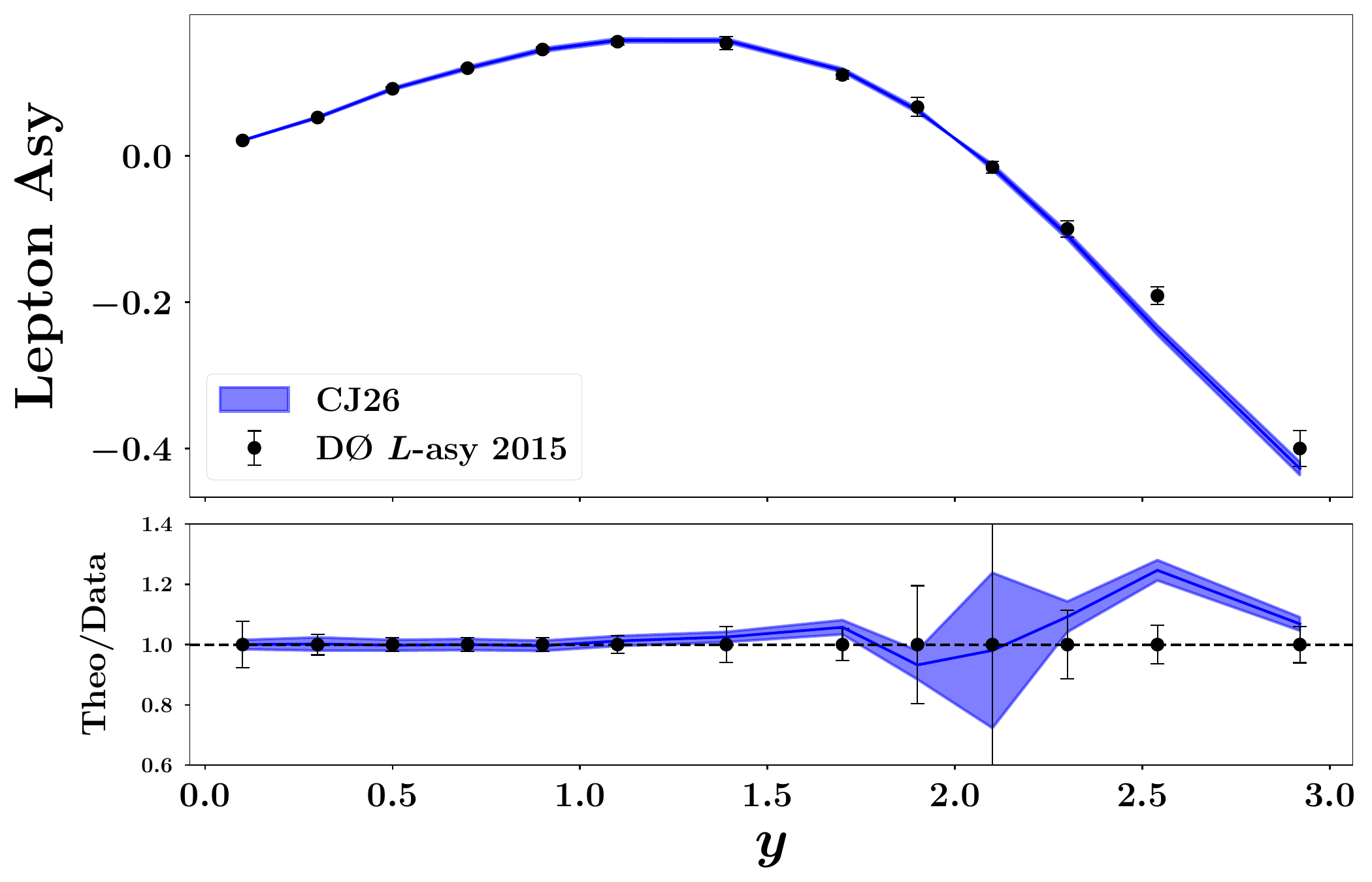}
    \caption{Data-theory comparison for D\O\  W-boson asymmetry (left panel) and D\O\ 2015 lepton asymmetry (right panel). The fit uncertainties (blue bands) are the envelopes of the $T=1.645$ errors of the multiplicative and additive HT fits.}
    \label{f:Obs_Tevatron} 
\end{figure}

Figure~\ref{f:Obs_Tevatron} compares the data and CJ26 fit for $W$-boson (left panel) and lepton rapidity asymmetries in $pp$ collisions measured by the D\O\ experiment at Tevatron.
%%%%%%%%%%%%%%%%%%
%%%%%%%%%%%%%%%%%%
The agreement between data and theory is striking, especially considering that the D\O\ measurements have very small statistical errors. Furthermore, the description of the first few bins in $y$ is improved compared to previous CJ fits~\cite{Accardi:2023gyr,Accardi:2016qay} due to the complete treatment of the systematic error point-to-point correlations.
As demonstrated in Ref.~\cite{Cerutti:2025yji}, the $W$-boson asymmetry dataset places a strong constraint on the extracted $d/u$ ratio at large $x$, without any sizable double counting when included in the fit along with the lepton-asymmetry measurement, which is instead sensitive to smaller $x$ values. The $W$ asymmetry data is thus a cornerstone of our analysis: anchoring the extraction of the $d/u$ ratio at $x \lesssim 0.7$ it allows lower energy data such as DIS from JLab to identify the isospin dependence of HT corrections, as well as constraining quark deformations in deuteron targets. In the rest of the paper we will justify these claims in more detail.

\begin{figure}[h!]
    \centering
    \includegraphics[width=0.7\linewidth]{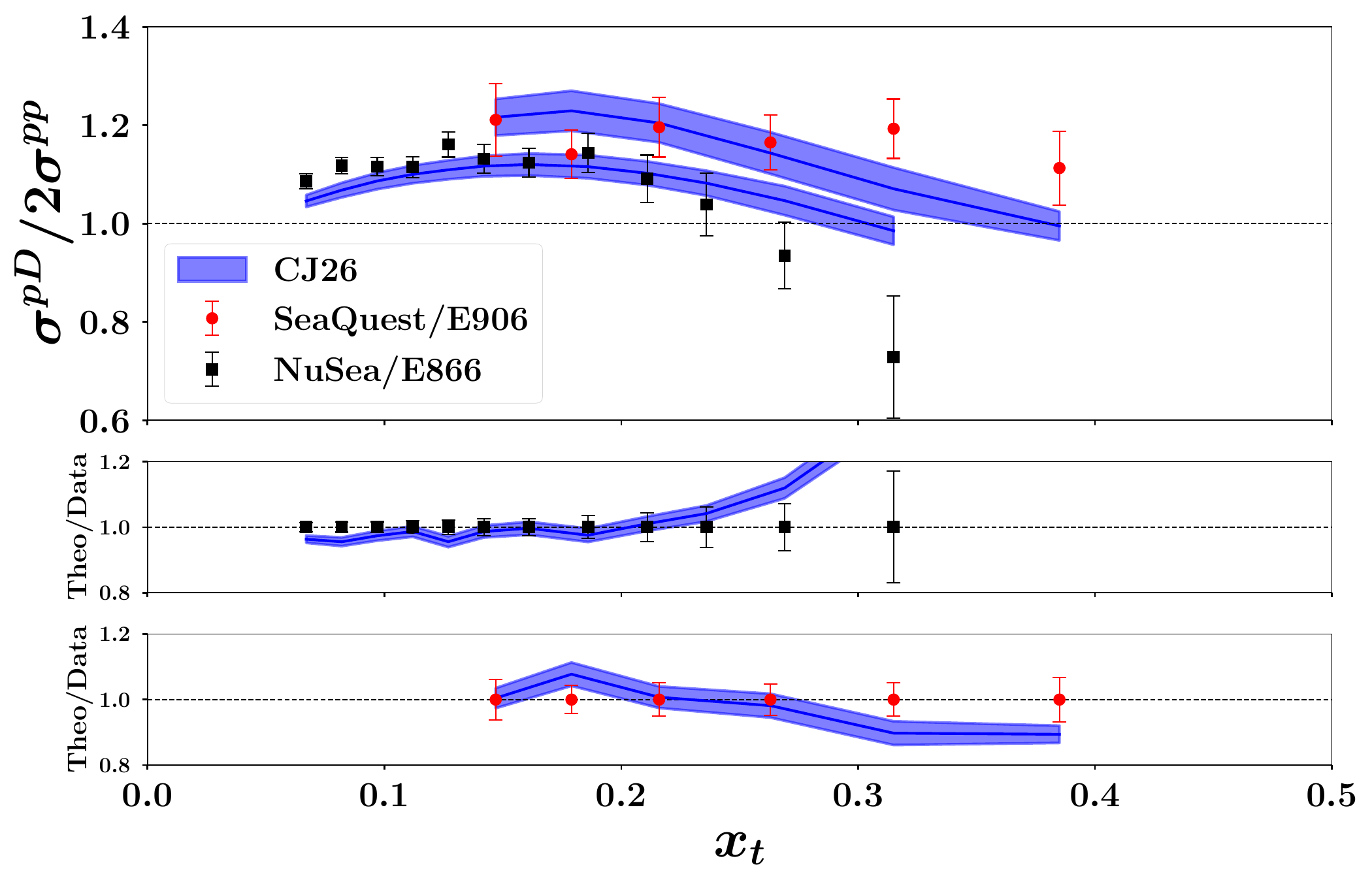}
    \caption{Data-theory comparison for SeaQuest and NuSea measurements. The uncertainty bands are the envelopes of $T=1.645$ errors of multiplicative and additive HT fits. 
    }
    \label{f:Obs_SQ-NS}
\end{figure}

Finally, in Fig.~\ref{f:Obs_SQ-NS} we compare the theoretical predictions from the CJ26 fit and the lepton pair production measurements by the SeaQuest/E906 (red points) and NuSea/E866 (black points) experiments as functions of the parton longitudinal momentum fraction $x_t$ of the target. While primarily constraining the $\bar d/\bar u$ ratio at large $x$, these data non-trivially correlate the light antiquark ratio with the $d/u$ ratio~\cite{Accardi:2023gyr}.
In the upper panel we show the direct comparison of data and theory calculations, while the bottom two panels display the theory-to-data ratios for the NuSea/E866 and SeaQuest/E906 data, respectively.
The comparison is performed separately for each experiment, with E866 (800 GeV proton beam) and SeaQuest (120 GeV proton beam) probing different ranges of the lepton-pair mass $M$.
%%%%%%%%%%%%%%%%%%
%%%%%%%%%%%%%%%%%%
The fitted cross section ratios are similar to those obtained in the \texttt{CJ22} fits~\cite{Accardi:2023gyr}. Small differences can be observed at large $x_t$, where the \texttt{CJ26} fit is slightly smaller, because of larger extracted $d/u$ and smaller $\bar{d} / \bar{u}$ ratios at large $x$ that will be discussed in the next section (see Fig.~\ref{f:CJ26-extr_largex}).

%%%%%%%%
\subsection{Extracted quantities}
\label{ss:Results-quantities}

In this subsection we present the main physical quantities extracted from the CJ26 fit. These include the resulting parton distribution functions, the HT and the off-shell functions, and other observables of phenomenological interest. 

\begin{figure}[h!]
    \centering
    \includegraphics[width=0.49\linewidth]{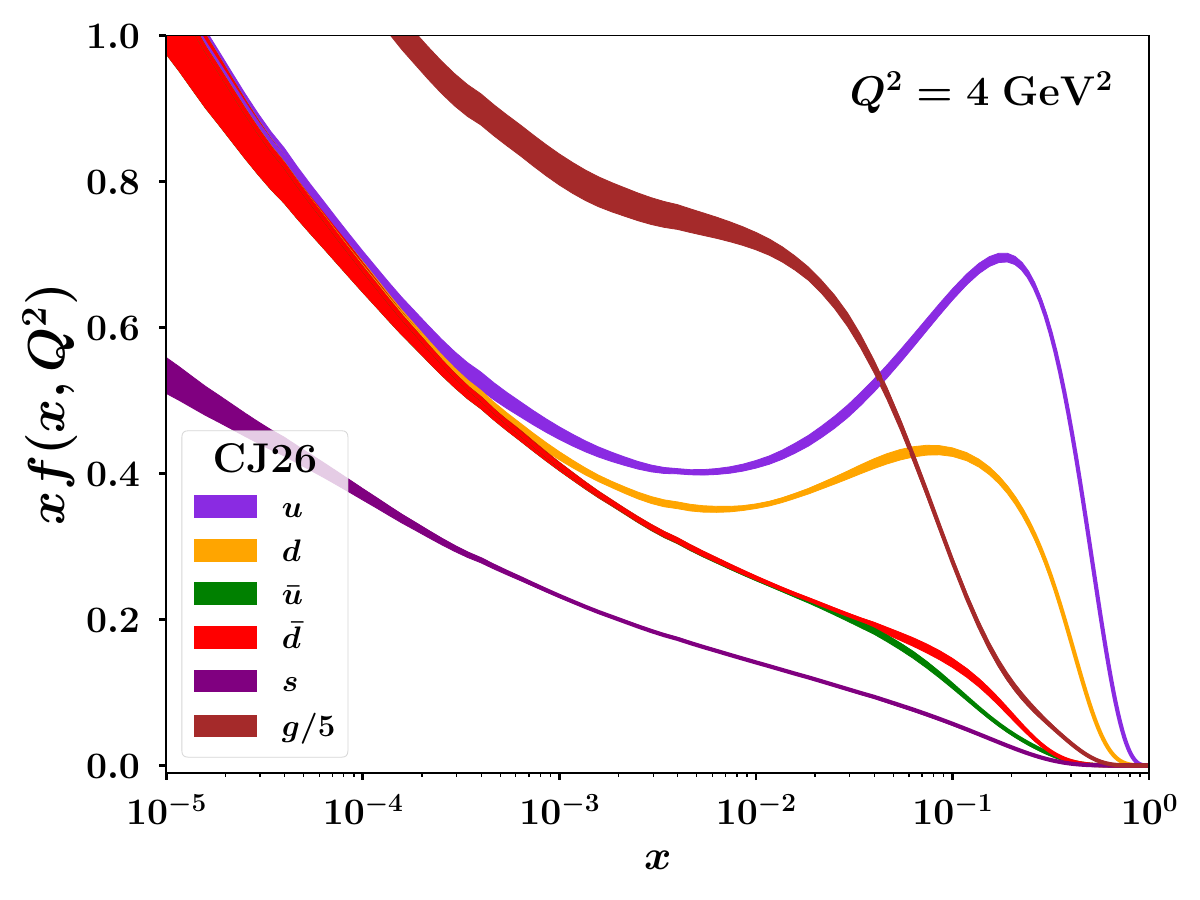}
    \includegraphics[width=0.49\linewidth]{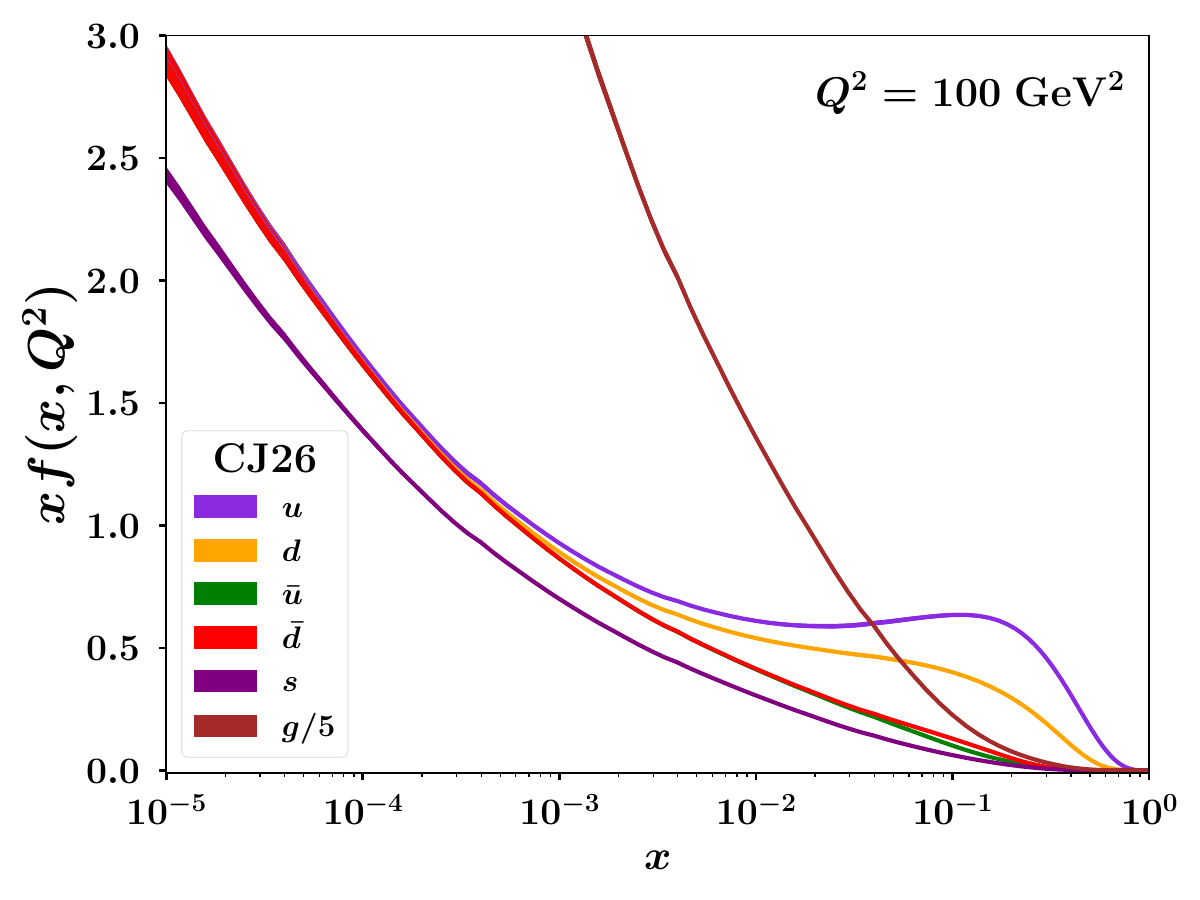}
    \caption{Breakdown of PDFs at $Q^{2}$ of 4 (left panel) and 100 (right panels) GeV$^2$. Uncertainty bands are envelope of additive and multiplicative HT results. Note that the gluon PDF is scaled by a factor $1/5$ in both panels.}
    \label{f:CJ26_pdfs}
\end{figure}

The CJ26 PDFs are shown at $Q^2 = 4$ and  100 GeV$^2$ in the left and right panels of Fig.~\ref{f:CJ26_pdfs}, respectively.   
At the lower scale, the valence $u$ and $d$ distributions dominate at large $x$ while the sea quarks and gluon contribute mainly at small $x$. Upon evolution to $Q^2=100$~GeV$^2$ (right panel), the gluon distribution grows rapidly at small $x$, driving the increase of the sea quark densities, while the valence quark distributions are shifted toward lower $x$. Note that, for better visibility, the gluon distribution is scaled by a factor $1/5$ in both panels.
The plotted uncertainty bands are the the envelope of the $\Delta\chi^2=2.71$ statistical uncertainty bands of the PDFs obtained in the multiplicative and the additive higher-twist correction fits, which in themselves differ by $\Delta\chi^2 = 31$, see Table~\ref{tab:chi2_CJ26}. As we will show in more detail later, the fitted PDFs are rather stable against the choice of HT implementation so that the envelope of the two fits allows us to take into account the associated epistemic uncertainty.

%%%%%%%%%%%%%%%%%%
%%%%%%%%%%%%%%%%%%

\begin{figure}[h!]
    \centering
    \includegraphics[width=0.7
        \linewidth]{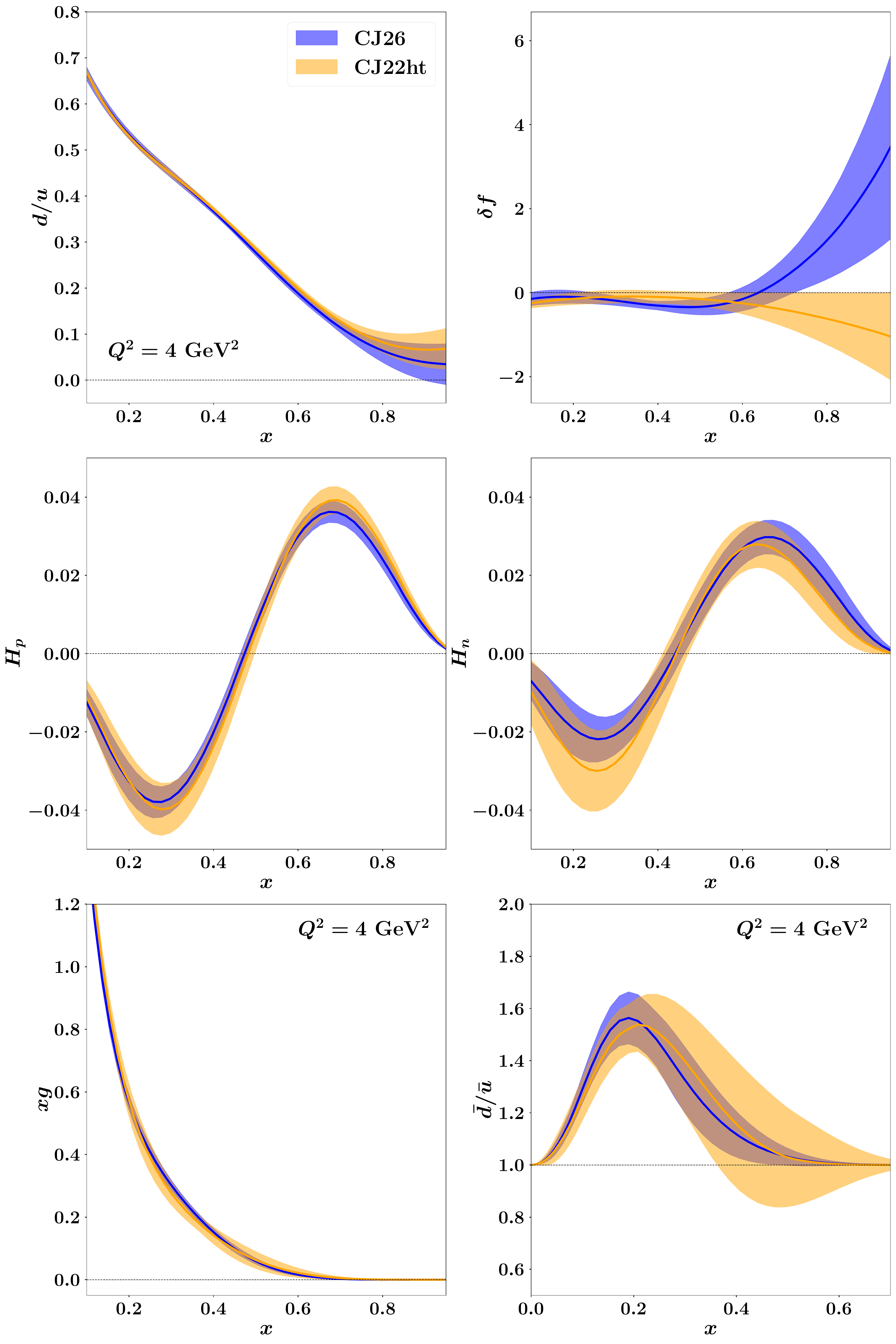}
    \caption{Comparison of selected extracted quantities from the CJ26 fit (blue bands) with those from CJ22ht~\cite{Cerutti:2025yji} (orange bands). From top to bottom: the $d/u$ ratio at $Q^2=4$~GeV$^2$ and the offshell function $\delta f$; the proton and neutron higher-twist functions; and the gluon PDF and the $\bar{d}/\bar{u}$ ratio at $Q^2=4$~GeV$^2$. Uncertainty bands represent the envelope of the multiplicative and additive HT implementations.}
    \label{f:CJ26-extr_largex} 
\end{figure}

\begin{figure}[h!]
    \centering
    \includegraphics[width=0.4\linewidth]{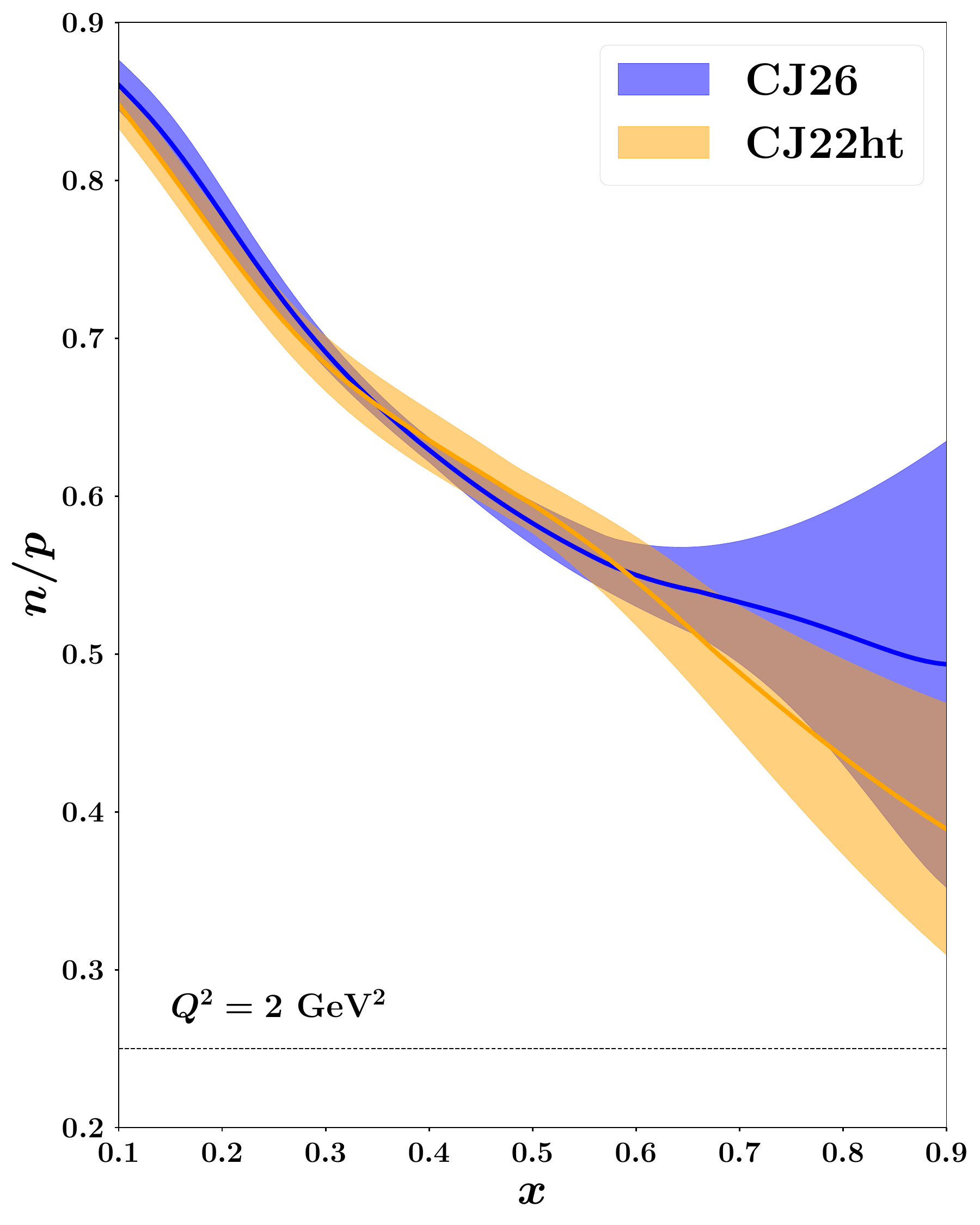}
    \caption{Neutron-to-proton structure function ratio $F_{2,n}/F_{2,p}$ at $Q^2=2$~GeV$^2$, comparing the CJ26 (blue bands) and CJ22ht~\cite{Cerutti:2025yji} (orange bands) fits. Uncertainty bands represent the envelope of the multiplicative and additive HT implementations.}
    \label{f:CJ26-nop}
\end{figure}

We now turn to the comparison between the \texttt{CJ26} and \texttt{CJ22ht} fits for the selection of extracted quantities displayed in Fig.~\ref{f:CJ26-extr_largex} and \ref{f:CJ26-nop}. Blue bands represent the results of the \texttt{CJ26} and orange bands the results of the previous \texttt{CJ22ht} fit~\cite{Cerutti:2025yji}. Overall, the comparison indicates that the new PDFs, corrections functions and $n/p$ ratio are compatible with those obtained in the previous fit. As we will discuss in detail below, the differences are localized to specific kinematic regions and well understood in terms of model correlations and effect of new datasets.

%%%%%%%%%%%%%%%%%%
%%%%%%%%%%%%%%%%%%
In the first row we can see that the CJ26 $d/u$ ratio at $Q^2=4$~GeV$^2$ (left panel) is smaller than in the CJ22ht fit. This results from a combined effect of the inclusion of new JLab data, and the updated treatment of correlated errors for the SLAC dataset (and for the D\O\ $W$-boson asymmetries).
As already discussed in Ref.~\cite{Cerutti:2025yji}, it is clear that the large-$x$ tail of the $d/u$ ratio correlates to the behavior of the off-shell function $\delta f$ (in the right panel). Indeed, the off-shell becomes large and positive as soon as $d/u$ gets smaller. The CJ26 off-shell functions remains compatible to the CJ22ht extraction up to $x\simeq 0.75$, which is at the limit of the kinematic coverage of the experimental data included in this analysis. The change in concavity is due to the increased flexibility in its parametrization which is now a polynomial of third degree, and the new data from Jefferson Lab at $x\lesssim 0.6$. We checked that the shape remains stable also when using a fourth degree polynomial, see the discussion of Fig.~\ref{f:syst_offs} in Section~\ref{ss:syst-offshell}.
The extracted $\delta f$ function is now compatible with recent results from AKP~\cite{Alekhin:2023uqx,Alekhin:2022uwc,Alekhin:2022tip,Alekhin:2017kpj,Alekhin:2017fpf} also in the very large-$x$ region. However, we stress that the result at $x \gtrsim 0.75$ should be interpreted carefully, since it is an extrapolation of the underlying phenomenological model which is not reflected in our method for estimating the uncertainty.

In the second row of Fig.~\ref{f:CJ26-extr_largex}, the higher-twist functions are shown in their additive form. We observe that the CJ26 results are well-compatible with the CJ22ht, with the proton HT being slightly smaller and the neutron HT being slightly larger at high $x$. This leads to an enhancement of the $n/p$ structure function ratio in the corresponding region (see Fig.~\ref{f:CJ26-nop} for more details). As discused in Section~\ref{ss:residual_power_corrections}, since the fitted HT corrections can absorb part of the missing higher-order effects, they also contribute to the robustness of our results against the truncation of the perturbative expansion.

In the third row of Fig.~\ref{f:CJ26-extr_largex}, both the gluon distribution and the $\bar d/\bar u$ ratio at $Q^2=4$~GeV$^2$ are in good agreement with CJ22ht. A small shift of the $\bar d/\bar u$ peak toward smaller $x$ is observed in CJ26. The uncertainty in both distributions is reduced compared to the previous analysis.  This is mainly due to the reparameterization of the $\bar d -\bar u$ distribution discussed in Section \ref{s:QCD_setup}, and furthermore having fixed $a_3^{\bar d - \bar u} =0$ which was necessary to stabilize the PDF uncertainty analysis. Due to DGLAP evolution, the gluon parameters are strongly correlated to those of the sea quarks, and are subject to a corresponding reduction in uncertainty at large $x$. 

Finally, we show in Fig.~\ref{f:CJ26-nop} the comparison of the $n/p$ structure function ratio at energy scale $Q^2=2$ GeV$^2$ obtained in the CJ26 (blue bands) and CJ22ht~\cite{Cerutti:2025yji} (orange bands) fits.
%%%%%%%%%%%%%%%%%%
%%%%%%%%%%%%%%%%%%
We note that, although compatible within uncertainties, the $n/p$ ratio from the CJ26 fit exhibits a more pronounced tail in the high-$x$ region compared to CJ22ht. At first sight, this behavior may appear in contrast with the smaller $d/u$ ratio observed in Fig.~\ref{f:CJ26-extr_largex}. In practice, however, it is driven by the combined effect of the enhancement of the neutron HT and the reduction of the proton HT in the same kinematic region (see second row of Fig.~\ref{f:CJ26-extr_largex}).
This illustrates the strong correlation between the extracted HT functions and the large-$x$ behavior of the structure function ratio. Importantly, the effect remains consistent with the data constraints at $x \lesssim 0.75$. The extrapolation region, where our result should be interpreted with caution, is signaled by the increase in the uncertainty bands, that can span a larger interval due to the additional open parameters we allowed in the CJ26 analyses, which in turn was possible due to the additional data from JLab 6 GeV and JLab 12 GeV. We will discuss their impact in more detail in the following Section and in Appendix~\ref{app:JLab_impact}. 

%%%%%%%%%%%%%%%%%%%%%%%%%%%%%%%%%%%%%%%%%%%%%%%%%%%%%%%
\subsection{Impact of JLab data}
\label{s:Results-JLab}

In this Section, we discuss in more detail the effect of the JLab experimental datasets referenced in Sec.~\ref{s:data} on our global QCD analysis. We stress that we included 685 data points from the Jefferson Lab 6 GeV program and 339 recently published data points from the 12 GeV program. In total, they represent $20\%$ of the full global dataset consisting of more than 5000 points. In contrast our previous \texttt{CJ22ht} analysis \cite{Cerutti:2024hrm} used only 272 point from the 6 GeV E00-106 experiment at JLab, which covered a reduced range in $x$ and $Q^2$ compared to the full JLab data set included in the \texttt{CJ26} fit. In Appendix~\ref{app:JLab_impact} we will also discuss the individual impact of JLab data subsets (tagged DIS from JLab 6 GeV, inclusive DIS at 6 GeV, inclusive DIS at 12 GeV). For definiteness, in this subsection, we only report fits with multiplicative HT corrections; we obtained analogous results with additive higher twists terms.

In order to better assess the specific impact of the Jefferson Lab measurements on the extracted PDFs and higher-twist terms, we first performed a dedicated baseline fit (``no JLab'') under the same settings as the nominal CJ26 analysis, but excluding all data sets from Jefferson Lab.
This allows us to directly quantify the role of the high-precision large-$x$ data from Hall C, BONuS, and other JLab experiments in constraining the valence region and testing the stability of the higher-twist and off-shell parametrizations. 

\begin{figure}[h!]
    \centering
    \includegraphics[width=0.95\linewidth]{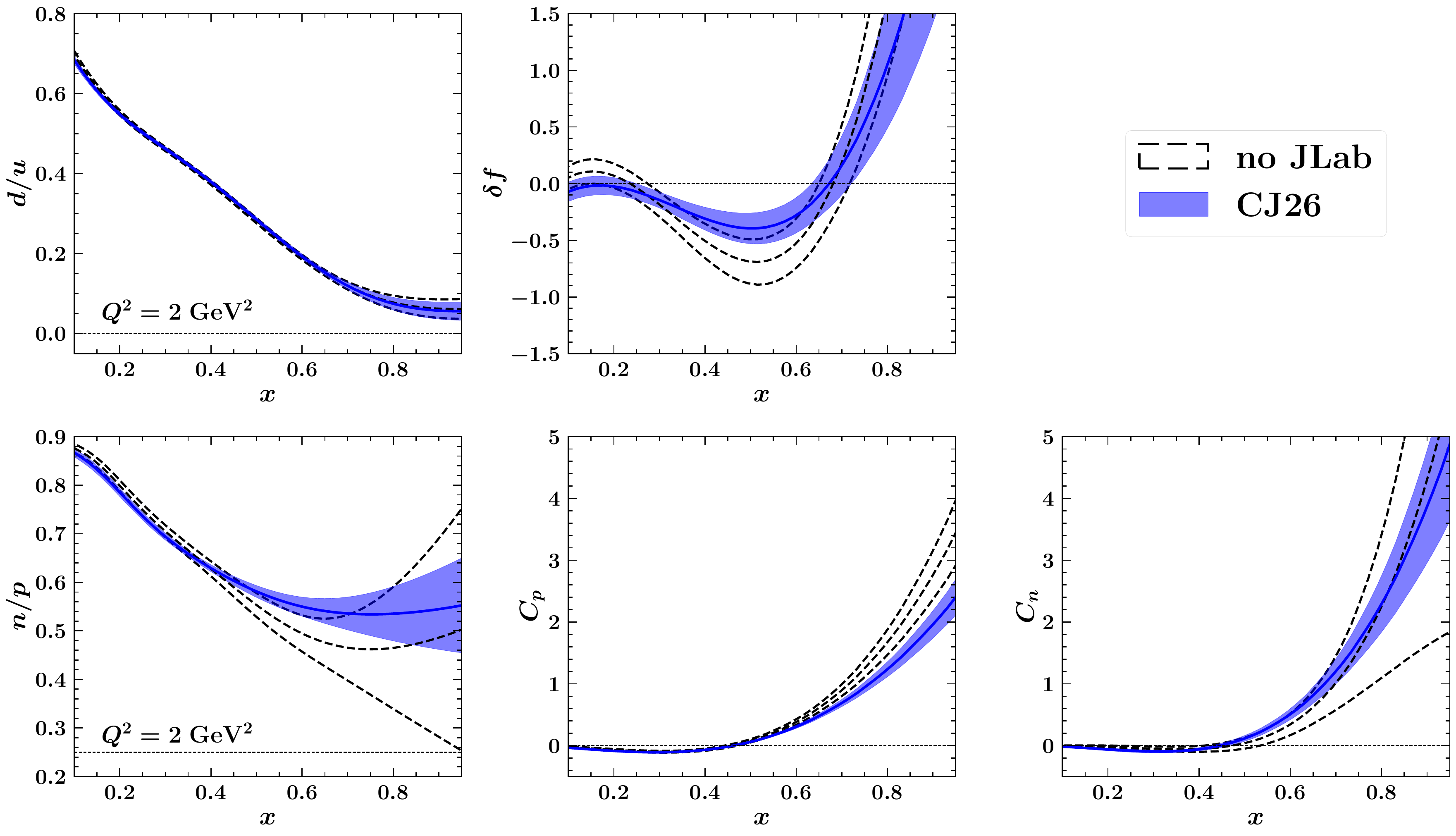}
    \caption{Comparison of a subset of extracted quantities between the baseline fit (``no JLab'', dashed bands) and the fit including JLab data (``CJ26'', blue bands).  
    For clarity, we show the results only in the multiplicative HT scenario.
    Upper row: $d/u$ quark ratio as a function of $x$ at $Q^2 = 2$ GeV$^2$ (left panel); off-shell function (right panel). Lower row: ratio $n/p$ of neutron and proton $F_2$ structure functions at $Q^2 = 2$ GeV$^2$ (left panel); proton higher-twist correction (central panel); neutron higher-twist correction (right panel). Bands represent $T^2$ = 2.7 uncertainties.}
    \label{f:imp-JLab}
\end{figure}

The baseline fit shows general agreement between data and theory, with a global $\chi^2$ per data point close to 1.17. It may be argued that the parameterizations of the fitted quantities are too flexible for such a reduced dataset. However, we do not observe clear evidence of instabilities in the result. Thus we can confidently discuss the modification in the fitted quantities generated by the introduction of JLab data.
In Fig.~\ref{f:imp-JLab}, we show the comparison of a selection of extracted quantities from the baseline fit (``no JLab'', dashed-bands) and the fit including JLab data (``CJ26'', blue bands). A similar result has been obtained for additive HTs.
In the upper row, we display the $d/u$ PDF ratio as a function of $x$ at $Q^2 = 2$ GeV$^2$ (left panel) and the fitted off-shell function (right panel). In the lower row, we display the $n/p$
ratio of $F_2$ structure functions at $Q^2 = 2$ GeV$^2$ (left panel), the proton higher-twist term (central panel), and the corresponding quantity for neutron (right panel). The error bands are obtained with a tolerance $T^2 = 2.7$ in the fit’s uncertainty as customary in the CJ fits~\cite{Owens:2012bv}.

It is immediately evident that the inclusion of JLab data results in a sizeable reduction of the uncertainty bands for most of the quantities shown in Fig.~\ref{f:imp-JLab}, notably the neutron HT correction ($C_n$) and the $n/p$ structure function ratio. %We also obtained a small but not negligible uncertainty reduction for the $d/u$ ratio. 

\begin{figure}[h!]
    \centering
    \includegraphics[width=0.55\linewidth]{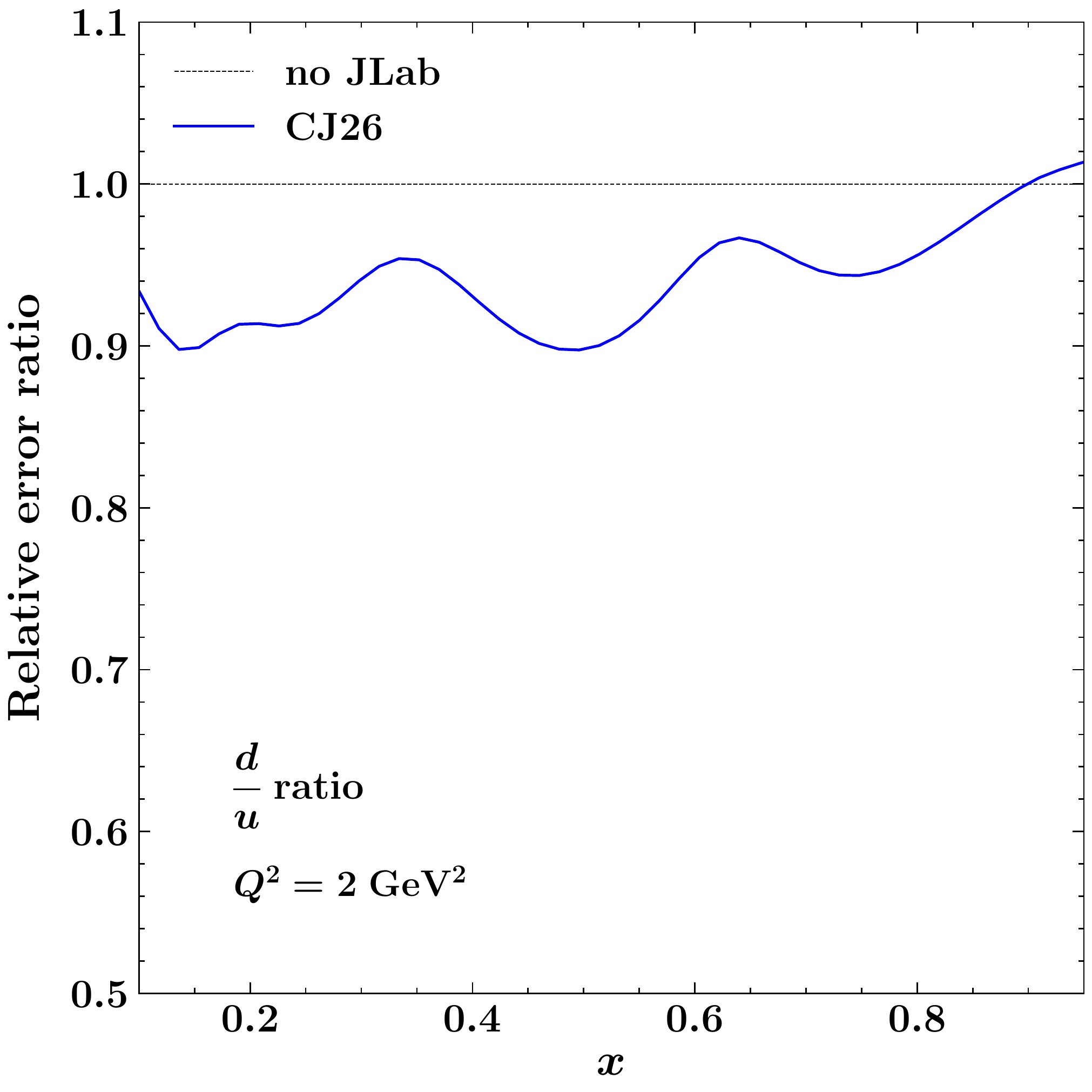}
    \caption{Relative error ratio between baseline and CJ26 fits of the $d/u$ ratio as a function of $x$ at $Q^2 = 2$ GeV$^2$.}
    \label{f:imp-JLab_du}
\end{figure}

The central value of the $d/u$ ratio is only marginally below the baseline fit, while its uncertainty bands is reduced by a 5-10\% factor. In order to better visualize the statistical impact of the JLab data on the $d/u$ ratio, we show in Fig~\ref{f:imp-JLab_du} the ratio of the $\delta(d/u)/(d/u)$ relative error in the \texttt{CJ26} determination of the $d/u$ ratio and the same quantity obtained in the baseline fit. 
We observe a reduction of about $10\%$ up to $x \simeq 0.6$, which decreases to approximately $5\%$ up to $x \simeq 0.8$, while the small-$x$ region remains essentially unaffected. For larger values of $x$, the behavior is increasingly driven by extrapolation effects rather than direct experimental constraints, and should therefore be interpreted with appropriate caution. Most of the statistical power of the 1024 JLab data points is utilized by the fit to constrain isospin-separated higher-twist corrections, as well as extending the $x$ range where off-shell effects can be constrained, as we will discuss further.

Beyond the $d/u$ ratio, the central value of the off-shell function $\delta f$ becomes smaller and negative around $x \simeq 0.2$, increases toward zero at intermediate $x \simeq 0.5$, and decreases again beyond $x \simeq 0.7$. The latter behavior is likely reflecting an extrapolation effect (see Sec.~\ref{ss:syst-offshell} for a more detailed discussion).
Regarding power corrections, the neutron higher-twist correction factor $C_n$ remains stable in its central value (slightly larger for $x>0.4$), but its uncertainty substantially decreases. However, the proton HT factor $C_p$ becomes markedly smaller at $x>0.4$, and also features a reduced uncertainty.
Altogether, these modifications have a pronounced impact on the $n/p$ ratio of the $F_2$ structure functions, which gets a markedly higher tail compared to the baseline (no JLab) fit. As discussed in Ref.~\cite{Cerutti:2025yji}, it is noteworthy that the trend observed in the $n/p$ ratio is opposite to that of the $d/u$ PDF ratio, indicating that the large-$x$ behavior of the $n/p$ ratio is primarily driven by higher-twist dynamics. Indeed, the enhanced $n/p$ tail directly reflects the combined effects of $C_n$ and $C_p$ increasing and decreasing at large $x$, respectively.

Since there are no differences in the fit setup of the two analyses compared in Fig.~\ref{f:imp-JLab}, we stress that the observed effects originate genuinely from the inclusion of JLab data, which thus represent a fundamental source of information on the partonic dynamics at large~$x$. In particular, with the increased $Q^2$ range provided by the 12 GeV data, they provide unique sensitivity to disentangle leading-twist contributions such as off-shell PDF deformations from higher-twist effects, highlighting the essential role of JLab measurements in constraining the nonperturbative structure of the nucleon.

%%%%%%%
\subsection{Comparison with other modern PDF fits}
\label{ss:Results-compHE}
It is also instructive to compare the CJ26 results with modern global analyses, such as NNPDF4.0~\cite{NNPDF:2021njg}, MSHT20~\cite{Bailey:2020ooq}, CT18~\cite{Hou:2019efy}, and ABMP16~\cite{Alekhin:2017kpj}. These fits make extensive use of collider data, in particular from the LHC, and -- with the exception of ABMPS analyses -- typically apply more conservative kinematic cuts in $W^2$ to avoid regions where power or nuclear corrections are required. (In more recent studies, the NNPDF and MSHT collaboration have re-evaluated the impact of lower cuts and the needed theoretical corrections \cite{Harland-Lang:2025wvm,Ball:2025xtj}.) In contrast, the CJ approach has consistently included such corrections since the beginning \cite{Accardi:2009br}, thereby enabling the use of a wider range of DIS and fixed-target data, but has not yet include the LHC data. The comparison therefore provides a valuable test of the consistency of the extracted PDFs across different methodologies and datasets. 
%%%%%%%%%%%%%%%%%%
\begin{figure}[h!]
    \centering
    \includegraphics[width=0.8\linewidth]{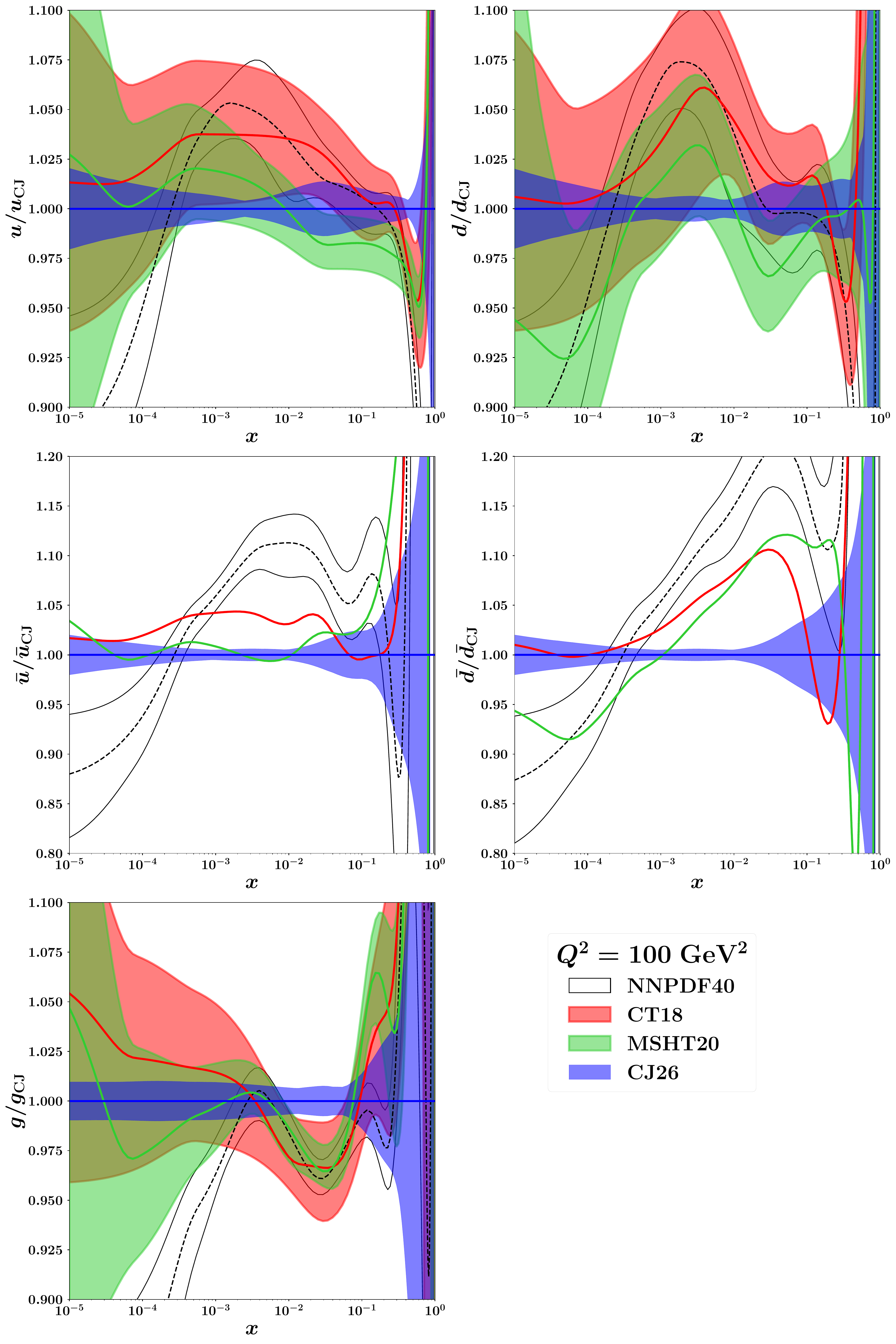}
    \caption{Ratios of the CJ26 PDFs to those from CT18 (red), MSHT20 (green), and NNPDF4.0 (dashed) at $Q^2=100$~GeV$^2$ for $u$, $d$, $\bar{u}$, $\bar{d}$, and $g$ distributions.
    % The CJ26 uncertainties (blue bands) represent the envelope of the multiplicative and additive HT treatments, while the uncertainties for the global fits are taken from the corresponding \texttt{LHAPDF} grids.
    }
    \label{f:CJ26-compHE}
\end{figure}
%%%%%%%%%%%%%%%%%%

In Fig.~\ref{f:CJ26-compHE}, we show the ratio of selected NLO unpolarized PDFs at scale $Q^2 =100$ GeV$^2$ from the global analyses discussed above and the PDFs extracted in the CJ26 analysis. 
Overall, CJ26 is in reasonable agreement with the high-energy extractions, either within their quoted PDF uncertainties or within their systematic differences (see Ref.~\cite{PDF4LHCWorkingGroup:2022cjn} for a discussion of these). 

Significant deviations appear, as expected, at $x \gtrsim 0.1$. In fact, this is driven by the inclusion in the CJ26 global analysis of the SeaQuest lepton-pair production data, as well as of a larger set of fixed-target DIS data, together with the deuteron and power corrections discussed in Sec.~\ref{sec:formalism}, that, similarly to the ABMP16 dataset, probe the large Bjorken-$x$ region. This additional data provide the CJ26 PDF determination with stronger constraints on $d/u$ and $\bar d/\bar u$ at large $x$. As a result, with the careful systematic studies presented in this paper, we believe that the CJ26 fit provides a more accurate and precise valence and light-sea quark determination in this region compared to the other global analyses.  

At intermediate $x$ down to $x \simeq 10^{-3}$, the $u$ and $\bar{u}$ distributions are consistent with the modern fits of Fig.~\ref{f:CJ26-compHE}, and mild differences appear for the $d$ quark. In contrast, the gluon and the $\bar d$ quark show a marked difference between CJ26 on one side and the other PDF sets on the other, with a smaller CJ26 $\bar d$ quark and a larger CJ26 gluon. This difference may reflect the absence of LHC data directly constraining this region or, as suggested in Ref.~\cite{Accardi:2021ysh}, the propagated effect of the nuclear corrections for deuteron target DIS calculations which are included in the \texttt{CJ26} fit. 

Below $x \simeq 10^{-3}$, all PDFs display large statistical and systematic uncertainties, with CJ26 remaining broadly consistent within errors. These comparisons underline the robustness of the CJ26 fit also at intermediate and small $x$, despite its primary scientific focus on fixed-target data and the explicit treatment of power and nuclear corrections.

\begin{figure}[h!]
    \centering
    \includegraphics[width=0.49\linewidth]
    {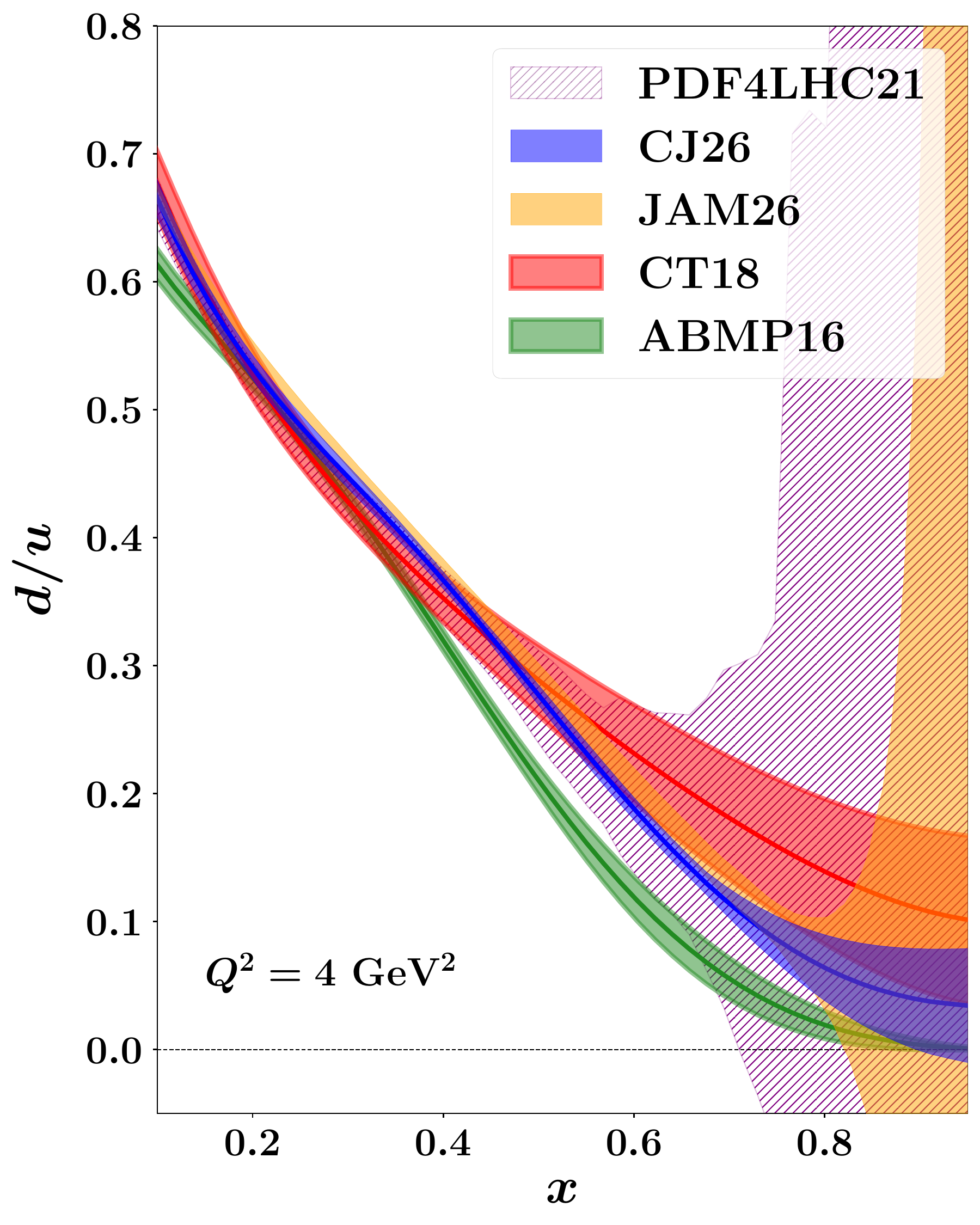}
    \caption{Comparison of CJ26 (blue band) fit to JAM26 (orange) CT18 (red), ABMP16 (green), PDF4LHC21 (purple) sets. The CJ26 uncertainty bands are the envelope of additive and multiplicative HT fits with tolerance T=1.645; the other bands utilize the nominal tolerance included in their LHAPDF grids.}
    \label{f:CJ26-du_comp}
\end{figure}

In order to further discuss the large $x$ region, in Fig.~\ref{f:CJ26-du_comp} we show the comparison of the CJ26 $d/u$ PDF ratio (blue band) at $Q^2 =  4$ GeV$^2$ with the same quantity extracted from the JAM26~\cite{Cocuzza:2026zoy}, CT18~\cite{Hou:2019efy}, ABMP16~\cite{Alekhin:2017kpj} and PDF4LHC21~\cite{PDF4LHCWorkingGroup:2022cjn} analyses. We selected these PDF sets because they introduce different specific treatments of the large-$x$ physics. CJ26, JAM26 and ABMP16 include nuclear corrections for deuteron targets and power corrections to lower their $W^2$ cut on DIS data; while CT18 and PDF4LHC21 use high $W^2$ cuts and no nuclear or power corrections. The CJ26 and CT18 analyses are the only ones to implement a $d$-quark parametrization that allows the $d/u$ ratio to have a finite limit as $x \to 1$. In the CJ26 case, this is obtained by enlarging the $d$-quark parametrization and mixing it with the $u$ quark, see Eq.~\eqref{e:PDF_du}; in the CT18 case, the parametrization is reduced by constraining $a_2^d=a_2^u$. In all other analyses the $d/u$ quark is forced to go to 0 or $\pm\infty$ when $x \to 1$ by the chosen parametrization. Finally, the PDF4LHC21 set provides a representative combination of the CT18, NNPDF3.1, and MSHT20 analyses, that include higher-energy data from the LHC. 

%%%%%%%%%%%%%%%%%%
%%%%%%%%%%%%%%%%%%

We observe good compatibility between CJ26 and JAM26 up to $x=0.8$ where the latter's uncertainty band diverges to infinity. This is a direct result of their choice of parametrization of the $u$ and $d$ quarks, and can be used to estimate the region in $x$ where the datasets lose constraining power on the $d/u$ ratio. In CJ26, instead, the $u$ quark is allowed to mix with the $d$ quark, allowing one to extrapolate the $d/u$ ratio to $x=1$ and gain insight on the non-perturbative nature of the forces confining the quarks in the proton \cite{Holt:2010vj}. However, the CJ26 band can only provide an underestimate of the extrapolation uncertainty due to the restriction introduced on the parametrization to avoid a flat direction in the fit's $\chi^2$ profile. New data and a more robust treatment of PDF uncertainty will be needed for a more realistic estimate of the uncertainties. 

The PDF4LHC21 band is fully compatible with CJ26 and JAM26, although with very large uncertainties at high $x$. These are induced by the high $W^2$ cut adopted to avoid power correction, that however severely limits the $x$ range probed by the fitted data. Among the fits included in the PDF4LHC21 combination, we highlight CT18 as the only one to allow a finite $d/u$ limit as $x\to1$. The obtained shape of the $d/u$ ratio is however different from CJ26 and JAM26, due likely in equal parts to the reduced flexibilty of the CT18 parametrization and its smaller kinematic coverage.
The ABMP16 ratio deviates substantially from all others, in part because it is driven to 0 by the adopted valence quark parametrization, with the $u$ quark dropping to 0 faster than the $d$ quark as $x\to1$ by a $(1-x)^{b}$ factor. The origin of the differences at moderate $x$ is less clear \cite{Alekhin:2022tip,Alekhin:2022uwc}.

In summary, CJ26 represents one of the most flexible PDF analyses at high $x$ currently available, with a simultaneous fit of higher-twist and nuclear target corrections for DIS data and realistic treatment of both statistical and systematic uncertainties. Future data from dedicated large-$x$ measurements, such as BONuS12, inclusive DIS at JLab 12~GeV, SoLID, the planned JLab 22~GeV program, and ultimately the EIC, will allow further refining the fit and will provide more stringent tests of offshell physics, due to their wider coverage in $Q^2$.
Moreover, with a comprehensive data set that maximizes the kinematic coverage at large $x$ without recourse to LHC data, CJ26 provides a solid baseline for the study of high-rapidity and high-mass observables at the LHC and for searching for signals of physics beyond the standard model. We will explore these aspects in the future.

%%%%%%%%%%%%%%%%%%%%%%%%%%%%%%%%%%%%%%%%%%%%%%%%%%%%%%%
\section{Investigating systematic uncertainties}
\label{sec:systematics}

In this section, we discuss the main sources of systematic uncertainties affecting the CJ26 fit and their impact on the extracted PDFs and related quantities: in Sec.~\ref{ss:syst-corr} we study the effect of the treatment of experimental systematic errors (either uncorrelated or correlated) on a specific subset of the fitted observables; in Sec.~\ref{ss:syst-add-mult} we examine the differences between the results obtained using the additive and multiplicative implementations of higher-twist corrections, assessing their influence on the fit stability; in Sec.~\ref{ss:syst-offshell} we investigate the sensitivity of the fit to the polynomial degree used to parameterize the off-shell function. Complementing the analysis of the HT implementation systematics discussed in earlier sections, and the interplay of HT and off-shell corrections discussed in \cite{Accardi:2023gyr,Cerutti:2024hrm}, these studies provide a quantitative assessment of the robustness of the CJ26 framework with respect to both phenomenological and experimental assumptions.

\subsection{Effect of correlated experimental uncertainties}
\label{ss:syst-corr}

The experimental measurements included in the CJ26 analysis are affected by statistical and systematic errors. The treatment of systematic uncertainties plays a crucial role in determining the overall precision and accuracy of global PDF fits.

Most of the datasets used in our analysis provide only the sum in quadrature of their systematic errors, whether correlated or uncorrelated, which may lead to an overestimation of the total experimental uncertainty and a distortion of the residual distributions. 

\begin{figure}[h!]
    \centering
    \includegraphics[width=1.0\linewidth]{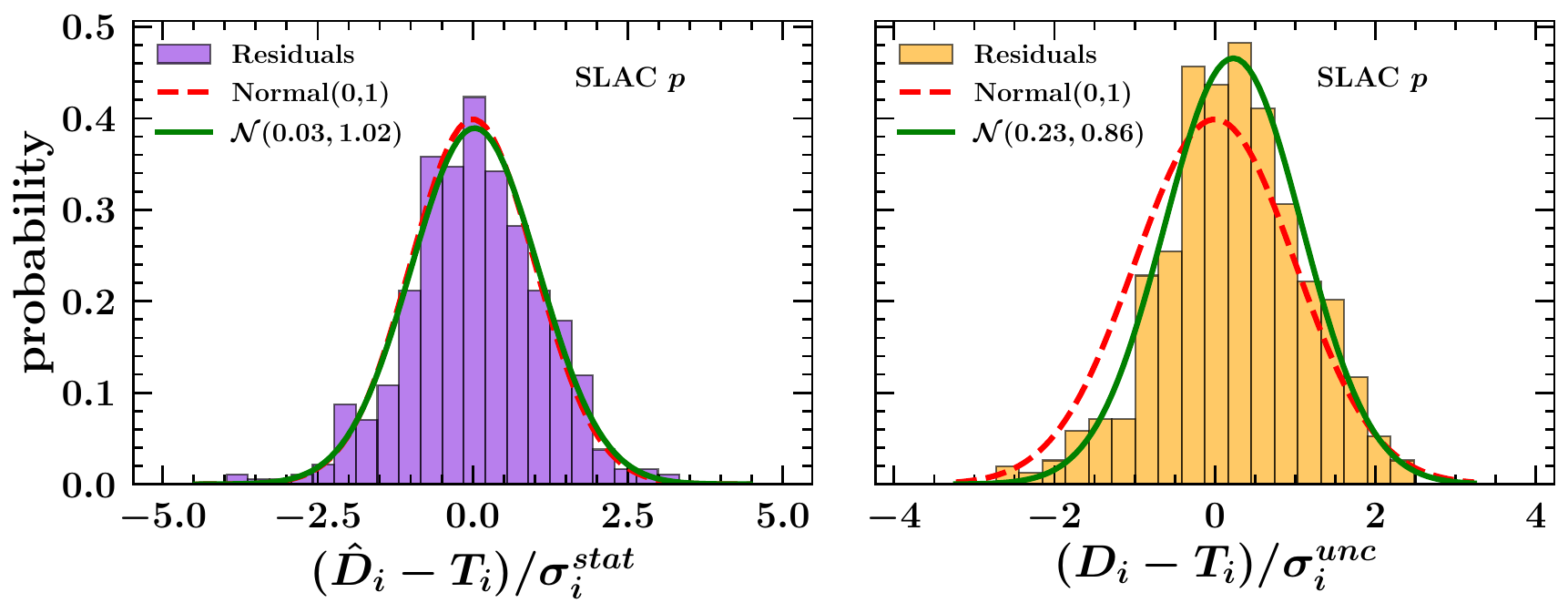}
    \includegraphics[width=1.0\linewidth]{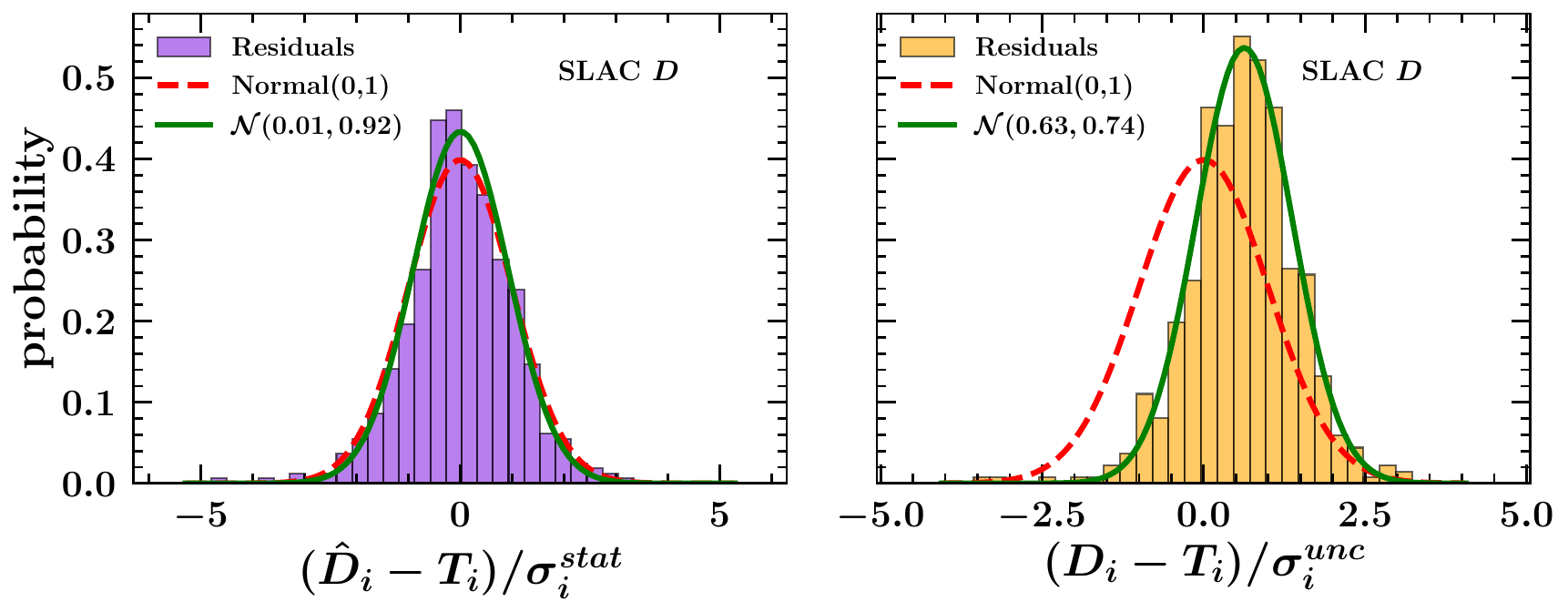}
    \caption{Comparison of pull distributions for the \textsc{SLAC} DIS data on proton ($p$) and deuteron ($D$) targets (upper and lower panels, respectively). Left panels include correlated systematic uncertainties, while right panels correspond to the decorrelated treatment $\Bigl(\hat{D}\rightarrow D$ and $\sigma^{unc}_i \to \sigma^{stat}_i=\sqrt{(\sigma^{stat}_i)^2+\sum_j(\sigma^{cor}_{i,j})^2}\Bigr)$. The red dashed line represents a standard Gaussian distribution, and the green line is a Gaussian fit to the histogram.}
    \label{f:pull_slac}
\end{figure}

\begin{figure}[h!]
    \centering
    \includegraphics[width=1.0\linewidth]{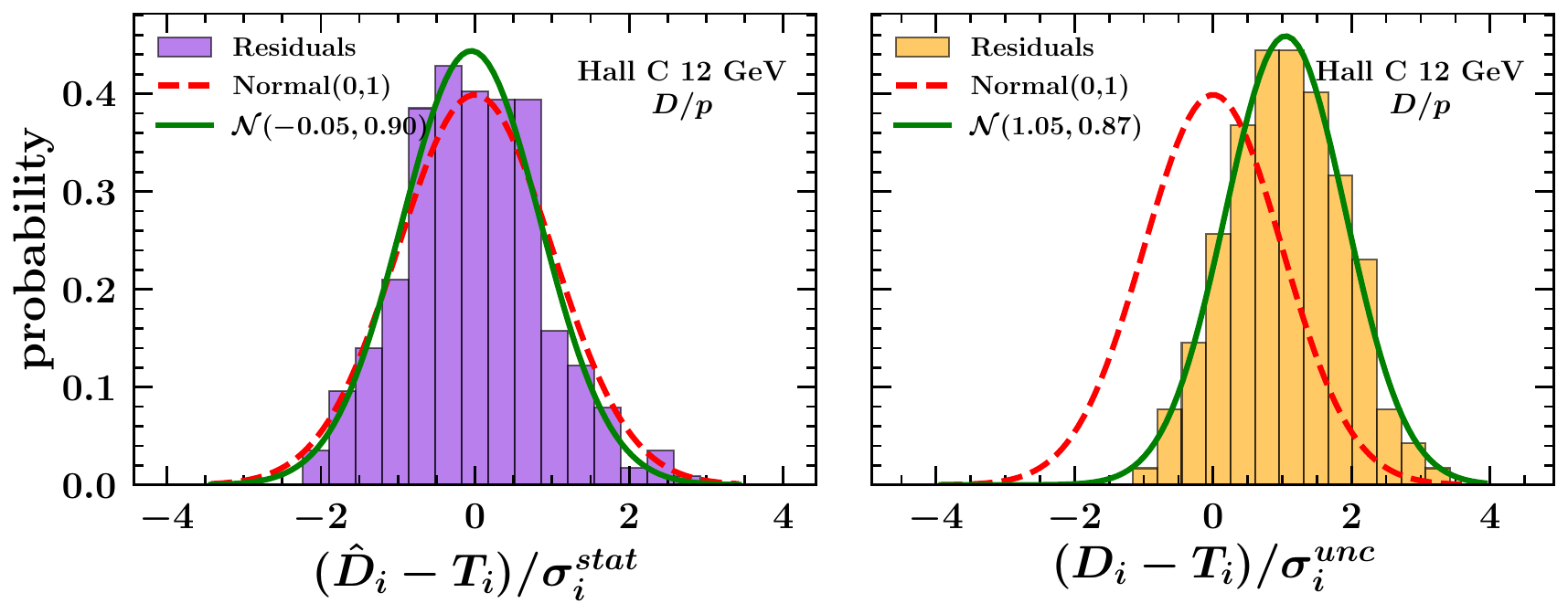}
    \includegraphics[width=1.0\linewidth]{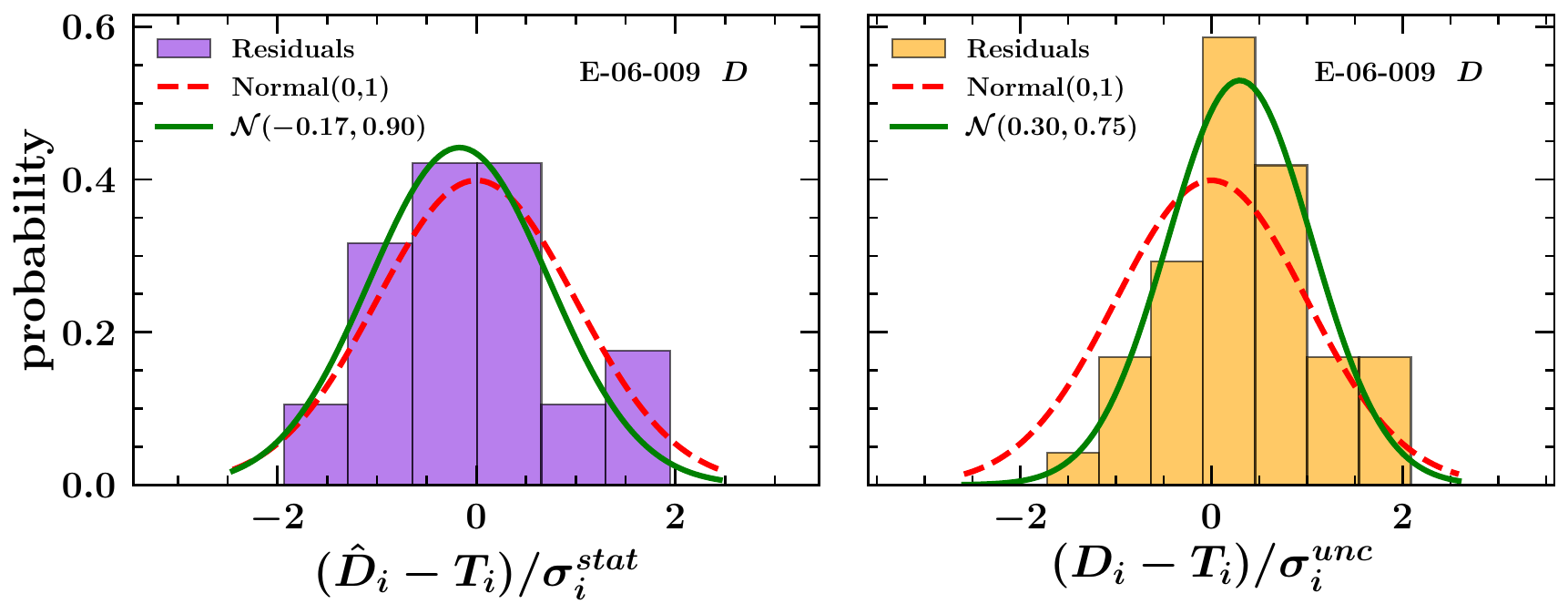}
    \caption{Same comparison and conventions as Fig.~\ref{f:pull_slac}, but for the Hall C 12 GeV ($D/p$) and E06-009 ($D$) datasets.}
    \label{f:pull_jlab}
\end{figure}

Whenever provided by the experimental collaborations, we explicitly include the correlated systematics in the CJ26 fit and examine their effect on the fit quality. To this end, in Figs.~\ref{f:pull_slac} and ~\ref{f:pull_jlab} we compare the pull distributions of selected datasets obtained when including correlations (left panels) and when decorrelating their systematic errors (right panels). Namely, on the left we plot $\Delta_i = (\hat D_i -T_i)/\sigma^{stat}_i$, same as in Fig.~\ref{f:pull_global}, where the data $\hat D_i$ include the fitted correlated shifts, the theoretical calculations $T_i$ are performed with the simultaneously fitted PDFs, and the denominator only includes the uncorrelated statistical uncertainties $\sigma^{stat}_i$ (see Ref.~\cite{Li:2023yda} for more detail). In the right panels, we decorrelate the systematic uncertainties as follows: the systematic shifts are removed from the numerator, $\hat{D}_i\rightarrow D_i$, the denominator includes all sources of errors added in quadrature, $\sigma^{stat}_i \to \sigma^{unc}_i=\sqrt{(\sigma^{stat}_i)^2+\sum_j(\sigma^{cor}_{i,j})^2}$, and the theory calculations remain the same. Each panel also displays a reference standard Gaussian distribution (red dashed line) and a Gaussian fit to the pull histogram (green solid line) for visual comparison.

%%%%%%%%%%%%%%%%%%
%%%%%%%%%%%%%%%%%%

The pull distributions of \textsc{SLAC} DIS measurements on proton ($p$) and deuteron ($D$) targets are displayed, respectively, in the upper and lower panels of Fig.~\ref{f:pull_slac}. For the proton target, the histogram in the left panel closely follows a standard Gaussian distribution. In the right panel, where the effect of correlated systematic shifts is removed, a deviation from the expected distribution is observed. The peak is shifted to the right by a small amount, and the narrower width indicates that the experimental uncertainties are indeed overestimated if the systematic error correlations are not taken into account.
This behavior is even more evident for the DIS dataset on deuteron target (lower panels), where the distribution is substantially shifted and 25\% narrower than a Gaussian distribution.

In Fig.~\ref{f:pull_jlab}, we turn to Jefferson Lab data, specifically the Hall C 12 GeV (DIS $D/p$ ratio) and the E06-009 (DIS $D$ target) datasets, for which a correlated systematic uncertainty analysis is available. To our knowledge this is the first study of the effect of systematic correlations performed on these data in the context of a global QCD analysis.
%%%%%%%%%%%%%%%%%%
%%%%%%%%%%%%%%%%%%
The E06-009($D$) measurement shows a peak shift and width narrowing similar to the SLAC($D$) data, although its interpretation is limited by the smaller number of data points.
But for the Hall C 12 GeV the deviations induced by the removal of the correlated systematic shifts are more pronounced. In the upper-right panel, the peak of the histogram is displaced by more than one standard deviation from the expectation, indicating that correlations play a significant role in the description of the Hall C data. In fact, given the relatively modest narrowing of the distribution width, neglecting the correlated systematics in the fit would put this otherwise very precise data in tension with the rest of the global dataset.

These results highlight the importance of treating systematic uncertainties in a consistent way. A proper account of correlated errors not only improves the statistical quality of the fit but also enhances the reliability of the physical quantities extracted from the data. This emphasizes the need for experimental collaborations to provide complete and transparent information on the quoted systematic uncertainties and their correlations, in order to facilitate accurate and consistent global analyses.

%%%%%
\subsection{Effect of implementation of higher twists}
\label{ss:syst-add-mult}

As discussed in detail in Ref.~\cite{Cerutti:2025yji} in the context of the \texttt{CJ22ht} fit, the implementation of the model for higher-twist corrections plays an important role in improving the stability of global QCD analyses focused on the large-$x$ region. As recommended in that study we allow for isospin dependence of the HT corrections ($p \neq n$) also in \texttt{CJ26} and perform the analysis under both the additive and multiplicative HT assumptions. Given the stability of our fits under these variations (see Tab.~\ref{tab:chi2_CJ26}), we take the envelope of the two results as the nominal CJ26 outcome.  
It is, however, instructive to investigate in more detail how the extracted quantities compare between the two approaches.

\begin{figure}[h!]
    \centering
    \includegraphics[width=1.0\linewidth]{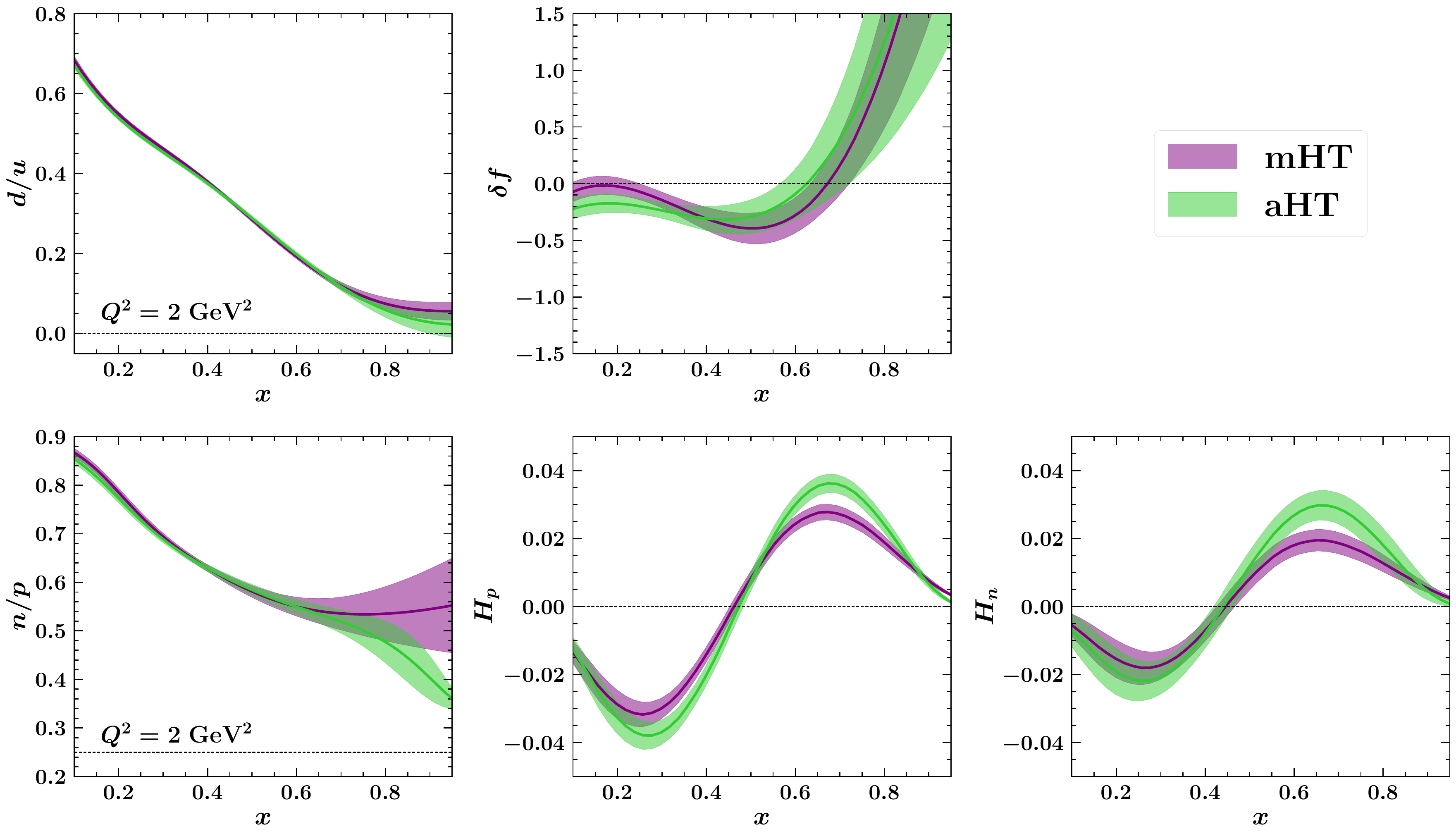}
    \caption{Comparison of the representative extracted quantities when implementing isospin-dependent additive (green band) or multiplicative (violet band) HT corrections. Upper row: $d/u$ ratio as a function of $x$ at $Q^2 = 2$ GeV$^2$ (left panel); off-shell function (right panel). Lower row: ratio $n/p$ of $F_2$ structure functions at $Q^2 = 2$ GeV$^2$ (left panel); proton higher-twist correction (central panel); neutron higher-twist correction (right panel). Bands represent $T^2$ = 2.7 uncertainties.}
    \label{f:mult_vs_add_general}
\end{figure}

In Fig.~\ref{f:mult_vs_add_general}, we show the comparison of a  selection of extracted quantities of interest in large $x$ studies for the multiplicative (violet bands) and additive (green bands) HT implementation scenarios.
In the upper row, we display the $d/u$ PDF ratio as a function of $x$ at $Q^2 = 2$ GeV$^2$ in the left panel, and the fitted off-shell function in the right panel. In the lower row, we display the $n/p$
ratio of $F_2$ structure functions at $Q^2 = 2$ GeV$^2$ in left panel, and the proton and neutron higher-twist term in the central and right panels. The error bands include a tolerance $T^2 = 2.7$ in the fit’s uncertainty~\cite{Owens:2012bv}. For the multiplicative HT analysis, we
plot the equivalent $\tilde{H}$ higher-twist function (see Eq.~\eqref{e:ht_tilde}).
%%%%%%%%%%%%%%%%%%
%%%%%%%%%%%%%%%%%%

The extracted leading-twist quantities in two fits present a very high degree of stability. The $d/u$ PDF ratios are compatible between the two fit configurations in the entire $x$ range, with the additive result showing a smaller tail compatible to 0 at $x>0.85$. Also the extracted off-shell functions are compatible with each other, showing a non-zero, negative effect around $x=0.4$, and growing positive at $x \gtrsim 0.7$. This result is in agreement with the AKP analysis of Ref.~\cite{Alekhin:2022tip}. The previously reported negatively curved \texttt{CJ22} result \cite{Cerutti:2024hrm}, was due to the insufficiently flexible second order polynomial parametrization of $\delta f$ adopted there. Here we use a third-order polynomial function, and have verified the absence of parametrization bias with a fourth-order polynomial fit, see Section~\ref{ss:syst-offshell}. 

The $n/p$ ratios obtained in the two configurations agree with each other up to $x=0.85$. Beyond this value the multiplicative curve exhibits a slightly larger tail compared to the additive one that decreases due to assumed vanishing of the $H_{p,n}$ HT functions, which vanishes as $x \to 1$ (see Eq.~\eqref{e:ht_add_par}), while the multiplicative $C_{p.n}$ function grow as in the pioneering study by Virchaux and Milsztajn (\cite{Virchaux:1991jc} and several subsequent others such as Refs.~\cite{Blumlein:2006be,Blumlein:2008kz,Accardi:2016qay}).
The difference in the extracted $n/p$ ratio largely reflects the distinct behavior of the two models in an extrapolation region, where the data no longer constrain the fit. This is confirmed by the correlation of the behavior of the large-$x$ $n/p$ tail and the fitted higher-twist functions that are generally consistent at small and intermediate values of $x$, but, in the additive case, need to increase at $x\sim 0.8$ to compensate for the parametrization-induced drop at larger $x$ value. The observed discrepancy has no significant phenomenological impact at present, but provides a useful indication of the model dependence associated with the higher-twist treatment at the kinematic limit. With more precision data expected from Jefferson lab in the near future it will become important to also address this remaining source of modeling bias.

\begin{figure}[h!]
    \centering
    \includegraphics[width=0.8\linewidth]{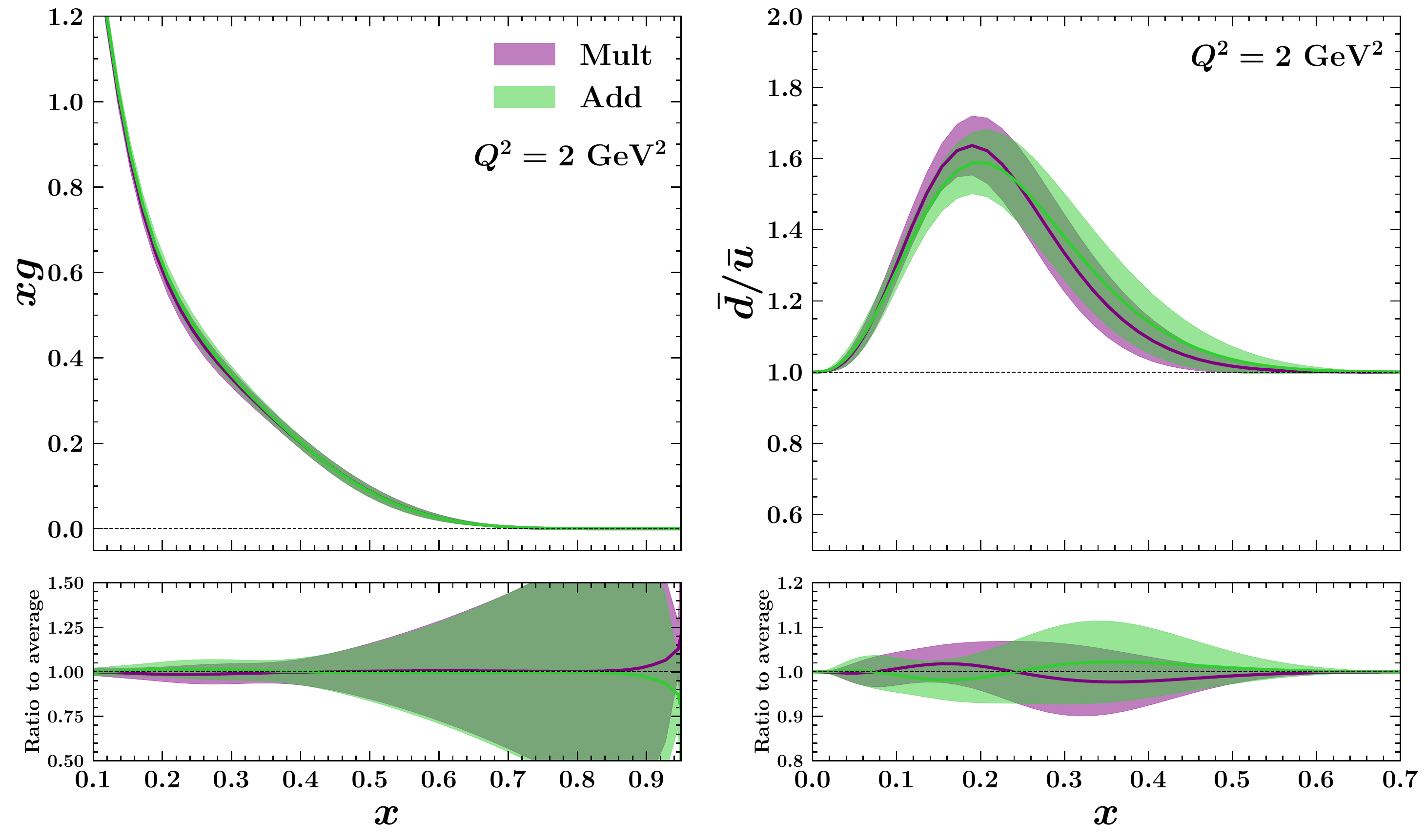}
    \caption{Comparison of the results when implementing isospin-dependent additive (green band) or multiplicative (violet band) HT corrections. Left panel: $xg$ gluon PDF as a function of $x$ at $Q^2 = 2$ GeV$^2$. Right panel: $\bar{d}/\bar{u}$ ratio. Bands represent $T^2$ = 2.7 uncertainties.}
    \label{f:mult_vs_add_g_dbub}
\end{figure}

In Fig.~\ref{f:mult_vs_add_g_dbub} we turn to the parton sea, and compare the gluon PDF and the $\bar{d}/\bar{u}$ PDF ratio obtained in the multiplicative and additive HT fits.
%%%%%%%%%%%%%%%%%%
%%%%%%%%%%%%%%%%%%
A high level of stability is observed between the two fit configurations also for these additional extracted quantities. In particular, no significant correlation is found between the HT implementation and the gluon distribution. The sea-quark ratios are likewise consistent, indicating that their behavior is largely independent of the specific treatment of higher-twist and nuclear corrections.

Overall, the comparison between additive and multiplicative higher-twist implementations confirms the robustness of the CJ26 analysis. While small differences appear in the extrapolation region at very large $x$, all physically relevant quantities (PDF ratios, off-shell functions, and higher-twist corrections) remain consistent within the fit uncertainties. This indicates that the choice of HT model has a limited impact on the extracted distributions, providing confidence in the reliability of the results across the explored kinematic range.

\subsection{Off-shell function parametrization}
\label{ss:syst-offshell}

Another source of (theoretical) systematic uncertainty in our analysis arises from the choice of parameterization for the off-shell function. This uncertainty can be assessed by varying the degree of the polynomial in Eq.~\eqref{e:off_poly}. In the nominal CJ26 fit, we used a polynomial of $3^{\rm rd}$ degree. Here, we examine the differences obtained when the offshell function is instead parameterized with a $2^{\rm nd}$- or $4^{\rm th}$-degree polynomial.

\begin{figure}[h!]
    \centering
    \includegraphics[width=1.0\linewidth]{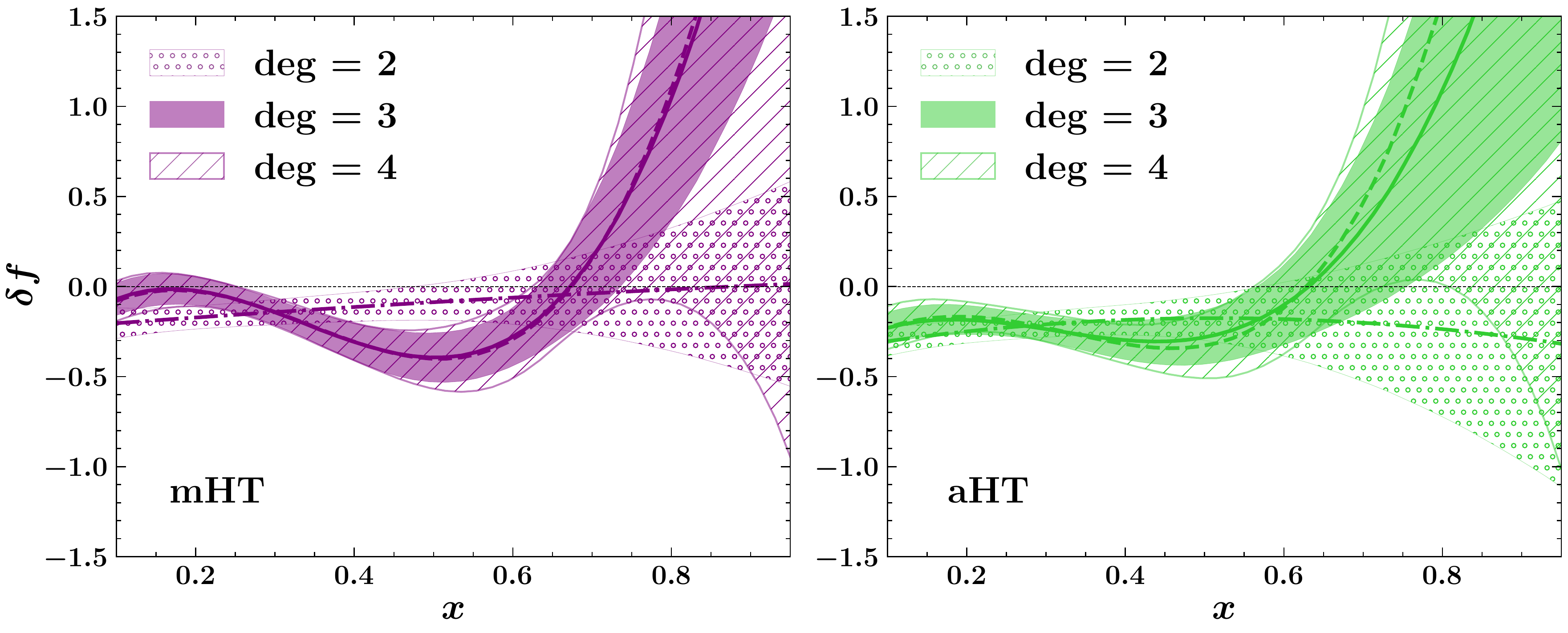}
    \caption{Comparison of the extracted off-shell function $\delta f$ for different polynomial parameterizations of the off-shell function in the CJ26 fit. Left panel: multiplicative HT implementation. Right panel: additive HT implementation. The dotted band corresponds to a second-degree polynomial parametrization of the HT functions, the solid band to the nominal \texttt{CJ26} third-degree polynomial fit, and the slashed band to a fourth-degree polynomial fit.}
    \label{f:syst_offs}
\end{figure}

In Fig.~\ref{f:syst_offs}, we show the extracted off-shell function $\delta f$ for different polynomial parameterizations, separately for the multiplicative (left panel) and additive (right panel) HT implementations. 
%%%%%%%%%%%%%%%%%%
%%%%%%%%%%%%%%%%%%
The second-order polynomial fits produce a flattish, sightly downward curve marginally compatible with 1, similarly to the \texttt{CJ22ht} fit. However, the third- and fourth-order polynomial fit show a qualitatively different off-shell function that changes curvature at $x\sim0.3$ where it is clearly negative, and then grows positive at larger $x$ values, in agreement with the AKP analysis \cite{Alekhin:2022uwc}. The obtained $\delta f$ shape is robust, as shown by the agreement between the $3^{\rm rd}$- and $4^{\rm th}$-order polynomial results at $x \lesssim 0.7$. As expected from a larger parametrization, the $4^{\rm th}$-order curve has an uncertainty band which is slightly larger than the $3^{\rm rd}$-order curve in the smaller-$x$ region, but grow rapidly at larger $x\gtrsim 0.7$ values. This indicates that the experimental data included in our analysis are not sensitive to the behavior of the off-shell function in this kinematic region. The upcoming BoNUS 12 tagged DIS data will help unraveling the off-shell function behavior in this region.

Although the fourth-degree polynomial parametrization provides a more conservative estimate of the $\delta f$ uncertainty bands and visually highlights the extrapolation region, in the nominal CJ26 fit we have adopted the third-degree polynomial to avoid instabilities from unconstrained parameters that do not improve the fit quality.  Nevertheless, we emphasize that the behavior of the extracted off-shell function beyond $x=0.75$, where we find that current data provide few constraints, should be interpreted with caution.

%%%%%%%%%%%%%%%%%%%%%%%%%%%%%%%%%%%%%%%%%%%%%%%%%%%%%%%
\section{Conclusions}
\label{sec:Conclusions}

In this work we have presented CJ26, a new global QCD analysis focused on the large-$x$ region of parton distribution functions, which includes for the first time in a global QCD analysis the complete set of Jefferson Lab 6~GeV inclusive and proton-tagged DIS measurements together with the first published data from the 12~GeV program. 
These measurements are the most precise DIS data currently available in this kinematic region and provide a unique lever arm in $Q^2$, with over 700 new data points compared to our previous fits.
The resulting global dataset exceeds 5000 experimental points, representing one of the most extensive and constraining inputs for large-$x$ studies to date. 

The CJ26 analysis is performed at next-to-leading order and includes a comprehensive treatment of power corrections, namely target mass corrections, nuclear effects in deuterium target (including smearing, Fermi motion, and off-shellness), and residual power corrections, commonly called Higher Twists. We explicitly propagate the theoretical systematic uncertainty associated with their implementation by comparing additive and multiplicative formulations, and incorporating them into the final uncertainty estimates.
We obtain a stable and robust fit, with a satisfactory description of the global dataset and no significant tensions introduced by the inclusion of the Jefferson Lab measurements. The extracted PDFs and related quantities are largely consistent with our previous determinations, confirming the overall reliability of the CJ framework. 
At the same time, the new JLab data allowed us to remove constraints previously placed on the PDF parameters and to enlarge the parametrization of the higher-twist and offshell correction functions, all the while improving the statistical precision of the extracted quantities. 

A key new result of this work is the first systematic study, within the CJ framework, of the impact of correlated experimental uncertainties. We find that their inclusion has a non-negligible effect on the consistency of the fit. We conclude that the availability of correlated systematic uncertainties is essential for precision QCD analyses, and their consistent release by experimental collaborations should be regarded as a priority.

We have also investigated the model dependence associated with the parametrization of the off-shell function. While the current dataset provides meaningful constraints up to $x \lesssim 0.7$, we find that the large-$x$ tail remains essentially unconstrained by data and is therefore driven by the assumed functional form. This highlights an intrinsic limitation of present extractions. Future measurements with extended kinematic coverage, like BONuS 12 GeV will be essential to properly constrain the large-$x$ behavior of this quantity.

A central result of the CJ26 analysis is the determination of the $d/u$ PDF ratio and of the neutron-to-proton structure function ratio $n/p$ in a framework that minimizes theoretical bias. Thanks to the simultaneous treatment of nuclear corrections, off-shell effects, and residual power corrections, together with the propagation of their associated systematic uncertainties, the CJ26 extraction provides one of the most unbiased determinations of these quantities currently available. We find that the $d/u$ ratio is stable with respect to previous CJ analyses, with a moderate reduction in its uncertainty driven by the inclusion of Jefferson Lab data and by the improved treatment of correlated systematics. For the $n/p$ ratio, we observe a more pronounced large-$x$ behavior, which can be traced back to the interplay of proton and neutron higher-twist corrections. These results illustrate the importance of a consistent and flexible treatment of subleading effects in extracting leading-twist observables at large $x$.

The impact of the Jefferson Lab data has been studied in detail by comparison with a baseline fit that excludes them. We find that their inclusion leads to a significant reduction of uncertainties in several key quantities, particularly in the neutron higher-twist contributions, the offshell function, and the $n/p$ structure function ratio. More importantly, the extended $Q^2$ coverage provided by the 12~GeV program enables a more effective disentangling of leading-twist dynamics from power corrections, which was only partially achievable in previous analyses. These results demonstrate that Jefferson Lab data play a central role in constraining the nonperturbative structure of the nucleon at large $x$.

Overall, CJ26 represents one of the most flexible and comprehensive global QCD analyses currently available in the large-$x$ region, with a realistic and systematic treatment of both experimental and theoretical uncertainties and an improved precision in the extraction of key leading-twist and higher-twist functions. Future data from dedicated large-$x$ measurements, such as BONuS12, inclusive DIS at JLab 12~GeV, SoLID, the planned JLab 22~GeV program, and ultimately the Electron-Ion Collider, will further refine these results. In particular, the increased $Q^2$ coverage and precision will allow for more stringent tests of correlations between the off-shell function, higher-twist corrections and the $d/u$ PDF ratio.

The PDF and structure function sets are available on the \mbox{CJ website}
\begin{center}
\url{https://www.jlab.org/theory/cj/pdfs}    
\end{center}
in LHAPDF~\cite{Buckley:2014ana} format and accompanied by the associated parameter values. 

%%%%%%%%%%%%%%%%%%%%%%%%%%%%%%%%%%%%%%%%%%%%%%%%%%%%%%%%%%
%% Acknowledgments  
%%%%%%%%%%%%%%%%%%%%%%%%%%%%%%%%%%%%%%%%%%%%%%%%%%%%%%%%%%

\begin{acknowledgments}
We gratefully acknowledge C.~Cocuzza, I.~Fernando, W.~Melnitcouk, T.~Je\v{z}o, K.~Kova\v{r}\'{i}k, A.~Kusina and F. Olness for informative discussions.

This work was supported in part by the  U.S. Department of Energy, Office of science contract DE-AC05-06OR23177, under which Jefferson Science Associates LLC manages and operates Jefferson Lab, and under contracts DE-SC0025004, DE-SC0010129, and DE-AC02-05CH11231.
The work of P.R. was also supported by the DOE Office of Science, Office of Nuclear Physics, within the framework of the Saturated Glue (SURGE) Topical Theory Collaboration.

\end{acknowledgments}

\bibliography{CJ22jlab}
\bibliographystyle{myrevtex}

%%%%%%%%%%%%%%%%%%%%%%%%%%%%%%%%%%%%%%%%%%%%%%%%%%%%%%
%%  APPENDICES
%%%%%%%%%%%%%%%%%%%%%%%%%%%%%%%%%%%%%%%%%%%%%%%%%%%%%%
\newpage
\appendix

%%%%%%%%%%%%%%%%%%%%%%%%%%%%%%%%%%%%%%%%%%%%%%%%%%%%%%

\section{Tables of fixed and fitted parameters}
\label{app:parameter_tables}
This appendix lists the best fit parameter values in the parametrization defined in \Cref{sec:formalism}. The brackets after the value give the Hessian $\Delta\chi^2=1$ uncertainty applied to be applied to the last two digits. Values without uncertainty are either determined through the sum rules or kept fixed. 

\subsection*{Multiplicative HT fit}
\begin{table}[h!]
\centering
\caption{\texttt{CJ26\_m} PDF shape parameters.}
\label{tab:CJ26mHT_pdf_params}
\begin{tabular}{|cccccc|}
\hline
      & $xu_v$ & $xd_v$ & $x(\bar{d}+\bar{u})$ & $x(\bar{d}-\bar{u})$ & $xg$ \\
\hline
    $a_0$ & $0.95961$ & $17.726$ & $0.1176(42)$ & $0.00713(48)$ & $64.618$ \\
    $a_1$ & $0.460(34)$ & $1.062(34)$ & $-0.2376(42)$ & $3.61(45)$ & $0.662(25)$ \\
    $a_2$ & $3.510(16)$ & $6.66(17)$ & $9.05(16)$ & $22.0(2.3)$ & $7.36(38)$ \\
    $a_3$ & $1.52(52)$ & $-3.701(75)$ & $0$ & $0$ & $-3.742(81)$ \\
    $a_4$ & $7.9(1.7)$ & $5.37(17)$ & $22.8(1.3)$ & $0$ & $4.08(20)$ \\
    $b$ & -- & $-0.00321(73)$ & -- & -- & -- \\
    $c$ & -- & $0.39119$ & -- & -- & -- \\
\hline
\end{tabular}
\end{table}

\begin{table}[h!]
\centering
\caption{\texttt{CJ26\_m} higher-twist and off-shell correction parameters.}
\label{tab:CJ26mHT_ht_off_params}
\begin{tabular}{|lc|lc|lc|}
\hline
      & HT (proton) &   & HT (neutron) &   & off-shell \\
\hline
    $\alpha_p$ & $-0.90(10)$ & $\alpha_n$ & $-4.88(93)$ & $a^0_{\texttt{off}}$ & $-0.408(88)$ \\
    $\beta_p$ & $1.593(98)$ & $\beta_n$ & $2.35(24)$ & $a^1_{\texttt{off}}$ & $5.2(1.1)$ \\
    $\gamma_p$ & $4.36(53)$ & $\gamma_n$ & $0$ & $a^2_{\texttt{off}}$ & $-20.7(4.4)$ \\
    $\delta_p$ & $-8.57(79)$ & $\delta_n$ & $-2.244(63)$ & $a^3_{\texttt{off}}$ & $20.5(4.4)$ \\
\hline
\end{tabular}
\end{table}

\clearpage

\subsection*{Additive HT fit}
\begin{table}[h!]
\centering
\caption{\texttt{CJ26\_a} PDF shape parameters.}
\label{tab:CJ26aHT_pdf_params}
\begin{tabular}{|cccccc|}
\hline
      & $xu_v$ & $xd_v$ & $x(\bar{d}+\bar{u})$ & $x(\bar{d}-\bar{u})$ & $xg$ \\
\hline
    $a_0$ & $1.9518$ & $14.462$ & $0.1339(42)$ & $0.00666(49)$ & $58.313$ \\
    $a_1$ & $0.576(34)$ & $1.026(40)$ & $-0.2172(35)$ & $3.17(46)$ & $0.637(24)$ \\
    $a_2$ & $3.635(28)$ & $6.23(22)$ & $9.02(18)$ & $19.4(2.3)$ & $7.31(39)$ \\
    $a_3$ & $0.05(39)$ & $-3.44(11)$ & $0$ & $0$ & $-3.713(88)$ \\
    $a_4$ & $4.96(69)$ & $4.99(19)$ & $18.9(1.2)$ & $0$ & $4.07(22)$ \\
    $b$ & -- & $0.0015(13)$ & -- & -- & -- \\
    $c$ & -- & $0.47902$ & -- & -- & -- \\
\hline
\end{tabular}
\end{table}

\begin{table}[h!]
\centering
\caption{\texttt{CJ26\_a} higher-twist and off-shell correction parameters.}
\label{tab:CJ26aHT_ht_off_params}
\begin{tabular}{|lc|lc|lc|}
\hline
      & HT (proton) &   & HT (neutron) &   & off-shell \\
\hline
    $\alpha_p$ & $-4.05(80)$ & $\alpha_n$ & $-3.2(1.8)$ & $a^0_{\texttt{off}}$ & $-0.474(90)$ \\
    $\beta_p$ & $2.29(15)$ & $\beta_n$ & $2.42(39)$ & $a^1_{\texttt{off}}$ & $3.9(1.3)$ \\
    $\gamma_p$ & $2.64(12)$ & $\gamma_n$ & $2.75(36)$  & $a^2_{\texttt{off}}$ & $-15.1(5.2)$ \\
    $\delta_p$ & $-2.112(27)$ & $\delta_n$ & $-2.269(69)$ & $a^3_{\texttt{off}}$ & $16.2(5.6)$ \\
\hline
\end{tabular}
\end{table}

\clearpage

\section{Impact of JLab data}
\label{app:JLab_impact}

In this Appendix we provide a detailed quantitative assessment of the impact of the JLab data sets on the global analysis, to add detail to and support the discussion in Section~\ref{s:Results-JLab}. To this end, we analyze a series of fits obtained by separately including different subsets of JLab measurements to our baseline fit that excludes the JLab data, and compare them with the full CJ26 fit.

A summary of the corresponding $\chi^2$ values for individual experiments, as well as for the total data set, is reported in Table~\ref{tab:chi2-JLab}. The intermediate fits allow one to disentangle the relative role of the various JLab inputs, namely, the tagged-DIS data from BoNUS6, the 6~GeV inclusive measurements, and the 12~GeV inclusive measurements.
\begin{table}[h!]
\renewcommand{\arraystretch}{0.4}
\begin{tabular}{|l|l|c|ccccc|}
\hline
Obs. & Experiment & \# Points & no JLab & +BoNUS6 & +6 GeV incl. & +12 GeV incl. & CJ26 \\
\hline
DIS & Hall C 12 GeV (D/p)${}^\dagger$ & 332 & -- & -- & -- & 275.9 & 281.1 \\
 & MARATHON12 (D/p) & 7 & -- & -- & -- & 3.1 & 3.0 \\
 & JLab E00-116 (p) & 91 & -- & -- & 84.1 & -- & 82.9 \\
 & JLab E00-116 (D) & 91 & -- & -- & 80.8 & -- & 81.2 \\
 & JLab E03-103 (p) & 32 & -- & -- & 25.4 & -- & 24.4 \\
 & JLab E03-103 (D) & 45 & -- & -- & 17.6 & -- & 17.6 \\
 & JLab E06-009 (D)${}^\dagger$ & 44 & -- & -- & 34.7 & -- & 34.5 \\
 & JLab E94-110 (p) & 46 & -- & -- & 44.5 & -- & 44.3 \\
 & JLab E99-118 (D) & 2 & -- & -- & 0.4 & -- & 0.5 \\
 & JLab JLCEE96 (p) & 100 & -- & -- & 95.6 & -- & 95.5 \\
 & JLab JLCEE96 (D) & 97 & -- & -- & 72.2 & -- & 71.5 \\
 & BoNUS6 (n/D) & 137 & -- & 154.2 & -- & -- & 158.4 \\
 & HERMES (p) & 37 & 40.9 & 38.8 & 43.1 & 41.2 & 42.4 \\
 & HERMES (D) & 37 & 33.0 & 35.1 & 36.0 & 32.4 & 36.6 \\
 & SLAC (p)${}^\dagger$ & 530 & 615.5 & 614.7 & 619.2 & 613.9 & 615.2 \\
 & SLAC (D)${}^\dagger$ & 541 & 546.0 & 553.1 & 550.5 & 545.8 & 556.7 \\
 & E140X (p) & 9 & 7.0 & 7.5 & 5.7 & 6.8 & 5.6 \\
 & E140X (D) & 13 & 5.6 & 5.8 & 5.4 & 5.9 & 5.6 \\
 & BCDMS (p)${}^\dagger$ & 351 & 438.8 & 440.7 & 435.9 & 440.1 & 441.4 \\
 & BCDMS (D)${}^\dagger$ & 254 & 299.3 & 294.2 & 296.4 & 300.4 & 291.6 \\
 & NMC (p)${}^\dagger$ & 275 & 405.6 & 407.1 & 403.8 & 405.5 & 404.7 \\
 & NMC (D/p)${}^\dagger$ & 189 & 173.4 & 172.5 & 172.1 & 172.9 & 169.9 \\
 & HERA (NC $e^-p$)${}^\dagger$ & 159 & 242.0 & 242.3 & 245.9 & 243.5 & 247.5 \\
 & HERA (NC $e^+p$)${}^\dagger$ & 945 & 1089.7 & 1090.4 & 1090.1 & 1091.3 & 1091.7 \\
 & HERA (CC $e^-p$)${}^\dagger$ & 42 & 46.2 & 46.0 & 47.4 & 46.4 & 47.4 \\
 & HERA (CC $e^+p$)${}^\dagger$ & 39 & 48.4 & 48.2 & 48.5 & 48.3 & 48.4 \\
LPP & E866 ($pp$) & 121 & 139.4 & 139.8 & 140.2 & 139.7 & 140.5 \\
 & E866 ($pd$) & 129 & 133.2 & 133.8 & 136.9 & 135.3 & 139.9 \\
 & SeaQuest ($d/p$) & 6 & 15.5 & 15.4 & 14.6 & 14.0 & 12.2 \\
$W$ & D\O ($e$)${}^\dagger$ & 13 & 18.7 & 19.8 & 17.7 & 17.7 & 18.2 \\
 & D\O ($\mu$)${}^\dagger$ & 10 & 16.3 & 16.2 & 16.4 & 16.4 & 16.2 \\
 & CDF ($W$) & 13 & 15.0 & 14.8 & 14.8 & 15.3 & 14.3 \\
 & D\O ($W$)${}^\dagger$ & 14 & 11.1 & 10.1 & 11.9 & 9.8 & 10.4 \\
 & STAR ($e^\pm$) & 9 & 22.3 & 21.8 & 22.9 & 22.7 & 22.9 \\
$Z$ & CDF (Z) & 28 & 26.7 & 28.1 & 26.5 & 26.5 & 28.9 \\
 & D\O (Z) & 28 & 16.2 & 16.3 & 16.2 & 16.2 & 16.4 \\
jet & CDF (jet)${}^\dagger$ & 72 & 92.4 & 92.3 & 95.6 & 93.6 & 97.0 \\
 & D\O (jet) ${}^\dagger$ & 110 & 99.9 & 99.9 & 101.8 & 100.0 & 102.4 \\
$\gamma+$jet & D\O (1, 2, 3, 4) & 44 & 50.5 & 50.8 & 50.5 & 50.3 & 50.7 \\
\hline
 & total        &  & 4661.3 & 4822.5 & 5134.2 & 4943.7 & 5582.8 \\
 & total points &  & 4030   & 4167   & 4578   & 4369   & 5054   \\
\hline
\end{tabular}
\caption{Comparison of total $\chi^2$ values across various inclusion levels of JLab data in the CJ26 analysis. Data sets marked with a dagger provide correlated uncertainties, which have been accounted for in the CJ26 analysis. 
}
\label{tab:chi2-JLab}
\end{table}

A few general features stand out. First of all, the inclusion of JLab data does not significantly deteriorate the description of the pre-existing DIS data sets. In particular, the $\chi^2$ values for SLAC, BCDMS, NMC, and HERA remain largely stable across the different fits, indicating a good level of consistency between the JLab measurements and the rest of the world data.

The BoNUS6 data set is well described already when included on its own, with a $\chi^2$ per data point close to unity, which does not significantly change in the full CJ26 fit. This indicates a good level of consistency with the global data set, with no obvious tensions.

The inclusion of the 6~GeV and 12~GeV data leads to moderate shifts in the $\chi^2$ of several DIS data sets, particularly for deuteron targets, reflecting the sensitivity of these measurements to nuclear and large-$x$ effects. Some of these shifts are also visible in hadronic data sets, such as E866 and SeaQuest, due to the nontrivial correlations between valence and light-sea quark ratios pointed out previously \cite{Accardi:2023gyr,Cerutti:2024hrm}.

The 12~GeV JLab data sets are generally well described, with $\chi^2$ values per data point close to unity. Their inclusion induces only moderate changes in the global $\chi^2$, indicating a good level of consistency with the rest of the data.

Overall, no single data set shows a dramatic degradation in $\chi^2$, and the changes observed across different experiments remain moderate. This supports the conclusion that the JLab data can be consistently incorporated into the global analysis, while providing additional constraints, especially in the large-$x$ region.

\begin{figure}[h!]
    \centering
    \includegraphics[width=1.0\linewidth]{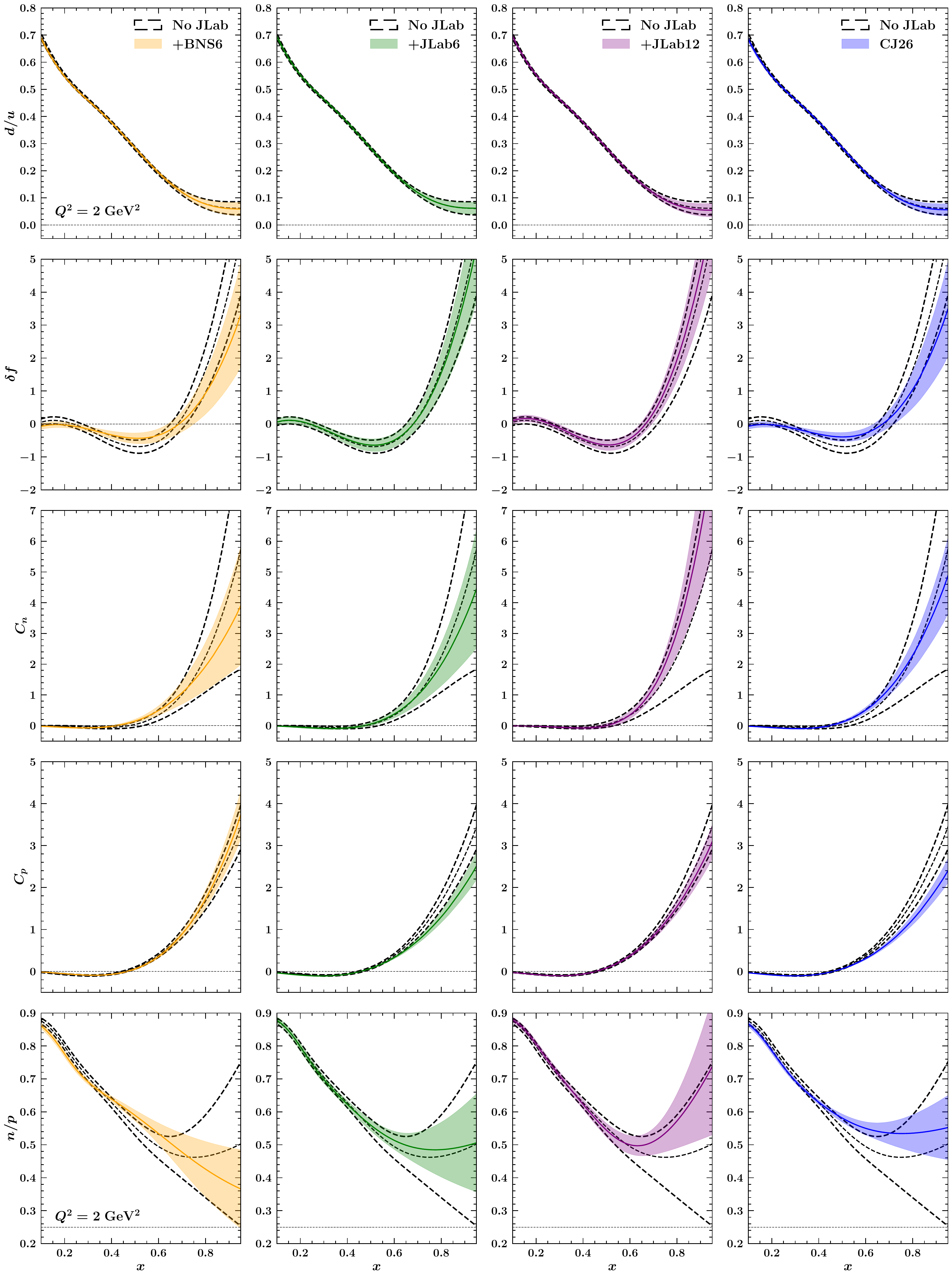}
    \caption{Impact of separately including new JLab data from left to right on a subset of extracted quantities.}
    \label{f:impact_sep}
\end{figure}

\begin{figure}[h!]
    \centering
    \includegraphics[width=1.0\linewidth]{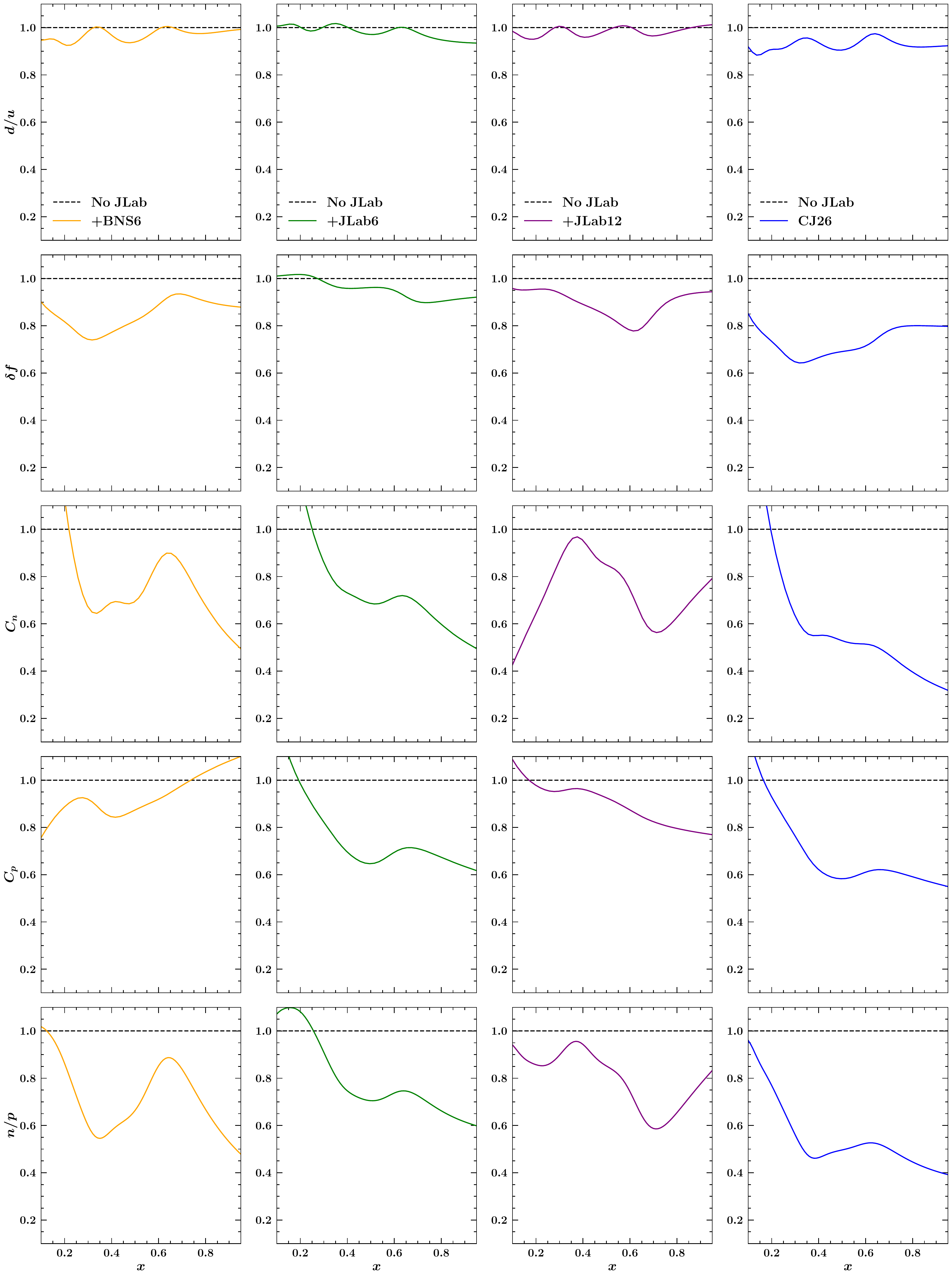}
    \caption{Same as Fig.~\ref{f:impact_sep} but for relative uncertainties.}
    \label{f:impact_ratio_sep}
\end{figure}

A more detailed view of the impact of the various JLab data sets is provided in Figs.~\ref{f:impact_sep} and \ref{f:impact_ratio_sep}, where we show the effect of including each subset of data separately on a representative set of extracted quantities. Figure~\ref{f:impact_sep} displays the corresponding shifts in the central values, while Fig.~\ref{f:impact_ratio_sep} shows the relative changes in the uncertainties.

None of the datasets alters significantly the central value of the $d/u$ ratio, either singly or combined, showing a remarkable agreement with the $W$ asymmetry data from the D\O{} experiment and testifying to the robustness of the nuclear and power corrections implemented in the CJ26 analysis. The statistical impact on the $d/u$ of the individual datasets is difficult to quantify, but is clear when combined in the full CJ26 fit.

The inclusion of the BoNUS6 data leads to a visible reduction of uncertainties, particularly for the off-shell function $\delta f$ (20\% smaller around $X\sim0.3$), and the neutron higher-twist term $C_n$ ($\sim$30\%). All central values remain compatible with the baseline fit within uncertainties, although $\delta f$ and $C_n$ tend to become smaller at large $x$, resulting in a slightly smaller tail in the $n/p$ ratio.

The 6~GeV JLab data primarily impact the higher-twist sector due to their relatively low, probed average $Q^2$. The central values of $C_n$ and $C_p$, and consequently of the $n/p$ ratio, are quite compatible with the baseline fit. At the same time, the uncertainties on these quantities are significantly decreased, with reductions at the level of $30$--$40\%$ in the relevantly probed $x=0.2-0.6$ range. The $x\gtrsim0.6$ region is unconstrained by this data, and the plots in Fig.~\ref{f:impact_ratio_sep} should be interpreted with caution in that extrapolation region. 

The impact of the 12~GeV data is qualitatively different, reflecting their extended kinematic reach in $x$, and even more importantly in $Q^2$. In addition to affecting higher-twist contributions, these data begin to constrain leading-twist quantities more directly. In particular, a reduction of uncertainties of the order of $5\%$ is observed for the $d/u$ ratio even at large $x$, together with a noticeable effect on the off-shell function in the region $x \simeq 0.6$--$0.7$ that was not covered by the JLab 6 data. This highlights the importance of the higher-energy JLab measurements in extending the sensitivity of the global analysis into previously less constrained kinematic regions.

\end{document}